\documentclass{aa}

\usepackage{graphicx}
\usepackage{txfonts}
\usepackage{amsmath}
\usepackage{xcolor}
\usepackage{subcaption}
\usepackage{lscape}
\usepackage{placeins}
\usepackage{hyperref}
\usepackage{setspace}

\AtBeginDocument{\renewcommand{\linenumbers}{}}

\newcommand{\dnds}{\mathrm{d}N/\mathrm{d}S}
\newcommand{\xiu}{\xi(u)}
\newcommand{\dxiu}{\Delta\xi(u)}
\newcommand{\sigc}{\sigma_{\mathrm{c}}}
\newcommand{\sigN}{\sigma_{\mathrm{N}}}
\newcommand{\sigtot}{\sigma_{\mathrm{tot}}}
\newcommand{\Scut}{S_{\mathrm{cut}}}
\newcommand{\Nbeam}{N_{\mathrm{beam}}}
\newcommand{\msun}{\mathrm{M}_{\odot}}
\newcommand{\mum}{\ensuremath{\mu\mathrm{m}}}
\newcommand{\mjb}{\mathrm{mJy\,beam}^{-1}}
\newcommand{\PD}{P(D)}
\newcommand{\Rcore}{\mathcal{R}_{\mathrm{core}}}
\newcommand{\Rtail}{\mathcal{R}_{\mathrm{tail}}}

\begin{document}

   \title{Probing submillimeter number counts below the confusion limit: extreme-value statistics of the P(D) distribution and its modulation by gravitational lensing}

   \titlerunning{Extreme-value statistics of the P(D) distribution}
   \authorrunning{Basu et al.}

   \author{Kaustuv Basu\inst{1}\corrauth{kbasu@uni-bonn.de}
        \and Andrea Guerrero\inst{1}\email{aguerrero@uni-bonn.de}
        \and Frank Bertoldi\inst{1}\email{bertoldi@uni-bonn.de}
        }

   \institute{Argelander-Institut f\"{u}r Astronomie, Universit\"{a}t Bonn, Auf dem H\"{u}gel 71, 53121 Bonn, Germany}

   \date{}

  \abstract
   {The shape of the submillimeter galaxy number counts below the confusion
    limit is a key record of the star formation history of the Universe, but
    it is accessible only statistically, through the one-point distribution
    of the map surface brightness, the $\PD$ distribution. Classical
    $\PD$ analysis compresses the counts into flux-integrated constraints
    and requires a complete forward model incorporating the instrument response.}
   {We introduce an extreme-value-theory analysis of the confusion $\PD$
    tail, in which exceedances above a
    threshold $u$ follow a generalized Pareto distribution (GPD). The GPD
    shape parameter $\xiu$ is a flux-resolved, normalization-free readout
    of the local logarithmic slope of the counts, and its lensing
    modulation $\dxiu$ probes their local curvature.} 
   {We derive analytic relations for $\xiu$ and $\dxiu$; validate these with end-to-end simulations including the
    instrument beam, source blending, declustering, and instrument noise;
    and apply the method to the \textit{Planck}
        857\,GHz maps of the cosmic infrared background (CIB) and the
    \textit{Herschel}/SPIRE 350\,\mum\ map of the GAMA-09 field, 
    with threshold correlations propagated into every statistic and 
    significances calibrated on simulated null skies.} 
   {At \textit{Planck}'s $5'$ resolution the tail reflects the bright and
    clustered sky rather than the faint counts, 
    but with \textit{Herschel}/SPIRE's finer resolution the method performs as designed: 
    the GPD shape rises with threshold, consistent
    with a strongly lensed bright population,  
    and the bright-masked map favors the
    Schechter count model. 
    The upcoming
    CCAT/FYST survey should separate the count models directly, though a
    cluster-lensing detection will need more $10^{15}\,\msun$ clusters than
    the sky contains.}
   {At current submillimeter-survey resolutions  
   the GPD tail statistic is a model-selection
    statistic: it discriminates the functional form of the counts below the individual-detection limit,
    while remaining insensitive to the map zero-level, gain, and count normalization,
    and robust against source clustering. 
    A differential cluster-lensing signal awaits
    deep, high-resolution surveys of massive clusters, for which we
    quantify the requirements.}


   \maketitle

\section{Introduction}
\label{sec:intro}

The differential number counts of galaxies at submillimeter wavelengths,
$\dnds$, encode the star-formation and galaxy-evolution history of the Universe. The bulk of
the cosmic star formation at $z \gtrsim 1$ takes place in dusty
environments: the ultraviolet light of young stars is absorbed by dust
grains and re-emitted in the rest-frame far-infrared, and the summed
emission of all such galaxies over cosmic time builds up the cosmic
infrared background \citep[CIB;][]{Puget1996, Fixsen1998, Hauser1998, Blain2002, Wright2004, Lagache2005, Kashlinsky2005,Odegard2019}, which
carries roughly half of the extragalactic radiative energy budget \citep[][]{HauserDwek2001}. The
number counts are the most basic statistic of this population, and
their \emph{shape} (the faint-end slope, the location and sharpness
of the break, and the steepness of the bright-end decline) encodes
the interplay between the luminosity function and its evolution in a
way that integrated quantities such as the background intensity cannot.
Measuring that shape, and in particular extending the measurement to
fluxes fainter than any individually detectable source, has therefore
been a central goal of every submillimeter survey
\citep[e.g.][]{Chary2001, Patanchon2009, Bethermin2010, Oliver2012, Geach2013, Magnelli2013, Carniani2015, Geach2017, GonzalezLopez2020, Bing2023}.

The obstacle is source confusion. At the $\sim\!20''$--$30''$ resolution of
single-dish submillimeter telescopes, the surface density of faint
sources is so high that many sources occupy each resolution element.
Below a flux limit set by the beam and the counts themselves, 
individual galaxies can no longer be extracted,
however deep the integration: for example, 
the confusion noise is $6.3 \pm 0.3\,\mjb$ for
\textit{Herschel}/SPIRE (Spectral and Photometric Imaging Receiver) at 350\,\mum\ \citep{Nguyen2010}, giving a
conventional $5\sigma$ extraction threshold of $\approx 31.5$\,mJy. 
The limiting ``noise'' is the sky itself. 
The information about the fainter population does not vanish, however;
it migrates into the statistics of the map. The one-point
distribution of the map surface brightness, historically called the
$\PD$ distribution, is the superposition of all sources, resolved and
unresolved, convolved with the beam, and its shape is a functional of
the counts. This insight is as old as radio astronomy itself
\citep{Scheuer1957, Condon1974}, and fitting a forward-modeled $\PD$
to the pixel histogram has yielded some of the deepest count
constraints available at submillimeter wavelengths
\citep{Maloney2005, Patanchon2009, Glenn2010, Scott2010, Berta2011, GonzalezLopez2020, Varnish2025}.

Powerful as it is, the classical $\PD$ method has two structural
limitations. First, it presupposes a parametric form for the
counts: one writes down a family $n(S;\theta)$, forward-models the full
histogram, and fits for $\theta$. The constraints are dominated by the
core of the distribution, where the central limit theorem has
compressed the counts into a few flux-integrated moments:
by
Campbell's theorem, the cumulants of the $\PD$ are beam-weighted
moments of $\dnds$ \citep{Condon1974}. As a result, the data constrain the
assumed family efficiently but test its functional form only weakly.
Second, the method requires a complete forward model of the
instrument: the absolute calibration, the noise level and its
decomposition from confusion, the beam profile, and the zero level of
the map all enter the predicted histogram and must be modeled or
marginalized. Both limitations bite hardest exactly where the science
interest is greatest: at the faint end, below the confusion limit,
where the instrument model is least certain and the assumed count
shape least tested.

In this paper we develop a different extraction of the same data,
focused on the tail of the $\PD$. The tail is where single sources still leave an individual imprint:
each is the brightest object in a beam, standing above the summed
confusion ``noise'' of all the others. This is the natural domain of
extreme-value theory (EVT).
Since this statistical framework is not part of the everyday toolkit
of most astronomers, we state what it provides in one paragraph. EVT occupies the same position for the extremes of a random
process that the central limit theorem occupies for its averages: just
as sums of many random variables converge to a Gaussian regardless of
the parent distribution, the exceedances of a random variable above a
high threshold likewise converge, again nearly regardless of the
parent distribution, to a single universal family, the generalized Pareto
distribution
\citep[GPD;][]{Pickands1975, BalkemaDeHaan1974,Davison1990}.

The GPD has one shape
parameter, $\xi$, which classifies the tail: positive for power-law
tails, zero for exponential, negative for tails with a hard endpoint.
Fitting the GPD to threshold exceedances, the ``peaks-over-threshold''
(POT) method, is therefore not a model choice but the tail
counterpart of assuming Gaussianity for a mean, and it is the standard
tool for tail inference in hydrology, insurance, and finance, among other domains
\citep{Rosbjerg1992, Embrechts1997, Coles2001, Choulakian2001, He2022}. EVT also has an established, if
scattered, astronomical lineage: the brightest cluster galaxies
\citep{BhavsarBarrow1985} or the number count of the most massive galaxy clusters \citep{Waizmann2012},
the maxima of the CMB temperature field
\citep{Coles1988, Mikelsons2009}, the solar activity cycle \citep{AsensioRamos2007} and
sunspot counts \citep{Acero2018}, and also BH spin statistics \citep{Silk2021}, and, more recently, the stellar/halo masses and luminosities of the brightest high-redshift galaxies revealed by \textit{JWST} \citep{Lovell2023,Heather2024}. 
The applications to source populations all fit the extremes of a catalog of discrete objects; the confusion-limited P(D), in which no catalog can be constructed at all, is a regime to which, to our knowledge, the formalism has not yet been applied.  

The key object in this paper is the GPD shape parameter measured as a function of the threshold, $\xiu$.  We show (Sect.~\ref{sec:theory})
that on the power-law shoulder of the confusion $\PD$,
$\xi(u) \approx 1/(\eta(S_u) - 1)$, where
$\eta \equiv -\mathrm{d}\ln(\dnds)/\mathrm{d}\ln S$ is the
local logarithmic slope of the counts at the flux $S_u$
selected by the threshold. Sweeping the threshold therefore scans the count slope flux by flux: in the ideal case a tomographic reconstruction of the shape of $\dnds$, as opposed to the flux-integrated moments of the classical analysis. Equally important is what $\xi$ does \emph{not} depend on:
as a pure shape descriptor of the exceedances, it is formally invariant under an additive offset of the map (the DC level, large-scale cirrus), under a multiplicative gain (absolute calibration), and under the overall normalization of the counts. A $\xiu$ analysis therefore sidesteps much of the instrument modeling that a full $\PD$ fit cannot avoid. The price is statistical efficiency: $\xiu$ is a lossy compression of the $\PD$, and we are explicit throughout about what is gained and lost relative to a full forward fit.

Gravitational lensing is the paper's other new theoretical element.  Behind a foreground mass concentration the counts are remapped as
$n_\mu(S) = \mu^{-2} n_0(S/\mu)$ \citep{Lima2010}, and a power law is the fixed point of this transformation: magnification changes only the amplitude of scale-free counts, never their slope. The lensing response of the tail therefore splits into two channels carrying different physics. The exceedance \emph{rate} shifts by the classical magnification-bias factor $\langle\mu^{\eta-2}\rangle$
\citep{Broadhurst1995}, while the \emph{shape} shifts only through the departure of the counts from a power law:
$\Delta\xi \propto \ln\mu \times \kappa$, with $\kappa$ the local
curvature of $\ln(\dnds)$ in $\ln S$. The differential observable $\dxiu$, measured between matched lensed and control fields, thus reads the count curvature with a known-$\mu$ prefactor, cancels common-mode systematics shared by the two fields, and, because a real count feature must shift in flux by the known factor $\mu$ while an instrumental artifact stays put, supplies both a \emph{ruler} that localizes count features and an \emph{authenticator} that certifies them as astrophysical.

One clarification of scope belongs up front. Since the magnification enters $\dxiu$ only logarithmically, inverting the measurement to infer cluster masses is intrinsically weak. 
The exceedance-rate above a threshold, which does carry mass as a power law but only in the combination $\langle\mu^{\eta-2}\rangle$, 
is degenerate with the count slope, so it amounts to a reformulation of classical magnification-bias mass estimation rather than a new capability. 
The architecture is therefore asymmetric: the magnification is supplied by external mass models of the foreground clusters (from weak lensing, X-ray, or Sunyaev--Zeldovich (SZ) observations), and the functional form of the counts is the target measured against that ruler
(Sect.~\ref{sec:theory:twosided}).

Three claims of novelty are central to what follows. The first is the formalism itself: to our knowledge this is the first application of the peaks-over-threshold / generalized-Pareto machinery to the $\PD$ distribution of a confusion-limited map, in which the threshold-resolved shape parameter $\xiu$ reads the local logarithmic slope of the counts at fluxes of order the threshold and discriminates their functional form below the individual-detection limit (Sect.~\ref{sec:theory}). Previous astronomical EVT applications fit catalogs of discrete objects or, for the CMB, the maxima of a Gaussian field; all previous $\PD$ analyses fit the full histogram under an assumed count parametrization. The estimator that results is invariant to the map mean, the gain and the count normalization by construction (under the conditions stated in Sect.~\ref{sec:theory:xi}), which decouples the count-shape measurement from the core-profile, calibration and noise-decomposition modeling a full $\PD$ fit cannot avoid.

The second is the treatment of gravitational lensing, which has not previously been developed systematically for $\PD$ analysis, with or without extreme-value statistics: the two-channel decomposition of the lensing response into rate and shape, the master relation connecting the shape channel to the count curvature, and the use of a known magnification as both ruler and authenticator of count features (Sect.~\ref{sec:theory:lensing}).

The third is empirical. We build and validate an end-to-end pipeline (beam and declustering formalism, matched filtering, covariance-correct combination of correlated thresholds, simulation-calibrated nulls, and a quantitative demonstration that the tail statistic is protected against source clustering) and
apply it to the two best available datasets and to one forthcoming survey. What each arm delivers is summarized in the roadmap below and
itemized in Sect.~\ref{sec:discussion:conclusions}.

We proceed in three stages that ascend in angular resolution. After establishing the formalism (Sect.~\ref{sec:theory}) and validating it
with end-to-end simulations (Sect.~\ref{sec:sims}), we apply it to the \textit{Planck} 857\,GHz CIB maps of \citet{Lenz2019}
(Sect.~\ref{sec:planck}) for the methodology demonstration, the quantitative angular-resolution requirement, and an absolute-background constraint on the count models that differential measurements cannot access.
 We then analyze the \textit{Herschel}/SPIRE 350\,\mum\ map of the GAMA-09 field (Sect.~\ref{sec:herschel}), where the method operates as designed.
Finally (Sect.~\ref{sec:ccat}) we forecast the performance of the upcoming 850\,GHz survey of the CCAT Observatory's Fred Young Submillimeter Telescope (CCAT/FYST), where the count-model discrimination becomes decisive and the requirements for a first
detection of the lensing modulation can be stated concretely.
Sect.~\ref{sec:discussion} summarizes the method's place among existing techniques and closes with an itemized list of conclusions.
Throughout, the technical underpinnings are collected in six appendices: the count-model fits (Appendix~\ref{app:counts}), the beam and declustering formalism (Appendix~\ref{app:beam}), exact peak statistics of Gaussian random fields (Appendix~\ref{app:gaussianpeaks}),
the matched-filter treatment of the SPIRE data
(Appendix~\ref{app:matchedfilter}), the statistical methodology for
correlated thresholds (Appendix~\ref{app:stats}), and the numerical
clustering tests (Appendix~\ref{app:clustering}).

\section{Theory: the GPD shape of the P(D) tail and its lensing modulation}
\label{sec:theory}

\subsection{The P(D) distribution and the extreme-value view of its tail}
\label{sec:theory:pd}

Consider a map of the sky observed with a beam $B(\theta)$, populated by
point sources drawn from differential counts $n_0(S) = \dnds$. The
brightness in a map pixel, denoted $D$, is the sum of the beam-weighted fluxes of all
sources near that line of sight, plus instrument noise. The probability
distribution of this quantity, the $\PD$, was first computed by
\citet{Scheuer1957} for radio surveys and given its modern form by
\citet{Condon1974}: for Poisson-distributed sources, the
log-characteristic function of the noiseless $\PD$, as a function of the Fourier variable $\omega$ conjugate to the pixel brightness $D$ (with $S$ the source flux density, as above), is
\begin{equation}
\ln\psi(\omega) \;=\; \int \mathrm{d}^2\theta \int \mathrm{d}S\;
n_0(S)\,\Bigl[\mathrm{e}^{\,\mathrm{i}\omega S B(\theta)} - 1\Bigr],
\label{eq:condon}
\end{equation}
so that the cumulants $c_k$ of the $\PD$ are beam-weighted moments of the counts, $c_k \propto \int B^k \mathrm{d}\theta \int S^k n_0\,
\mathrm{d}S$. (The generalization to clustered sources, which adds correlation-function terms to Eq.~(\ref{eq:condon}), is taken up in
Sect.~\ref{sec:sims:clustering}.) 

The morphology of the resulting distribution divides into three regimes, and this division organizes everything that follows. (i) A quasi-Gaussian core of width
$\sigtot = (\sigc^2 + \sigN^2)^{1/2}$, where $\sigc$ is the confusion
noise from the summed faint population and $\sigN$ the instrument
noise: here many sources contribute comparably to each beam and the
central limit theorem has erased their individual identities. (ii) A
power-law shoulder, where the brightest single source in a beam
dominates the pixel value: here the $\PD$ tail inherits the local
shape of the counts directly, because the probability of a pixel
exceeding a high value is, to leading order, the probability of one
sufficiently bright source landing in it. (iii) A ceiling at
the flux where sources become individually detectable and are masked
or cataloged, beyond which the map tail is emptied by construction.

The shoulder is the scientifically valuable regime (it is where the
counts below the detection limit imprint their local shape), and it
is by definition a tail regime. This motivates trading the
full-histogram fit for a method designed for tails. The
peaks-over-threshold formalism rests on a limit theorem
\citep{Pickands1975, BalkemaDeHaan1974} of the same character as the
central limit theorem: for essentially any parent distribution $F$,
the distribution of the exceedances $Y = D - u$, conditioned on
$D > u$, converges as the threshold $u$ grows to the generalized
Pareto distribution, stated below for the realized value $y$ of $Y$ once a pixel brightness $D$ is observed,
\begin{equation}
G_{\xi,\sigma}(y) \;=\; 1 - \Bigl(1 + \frac{\xi y}{\sigma}\Bigr)^{-1/\xi},
\qquad y \ge 0,
\label{eq:gpd}
\end{equation}
 with a scale parameter $\sigma > 0$ and a shape parameter $\xi$ that
classifies the tail: $\xi > 0$ for power-law (heavy) tails, $\xi = 0$
for exponential decay (the limit of Eq.~(\ref{eq:gpd}) as
$\xi \to 0$ is $1 - \mathrm{e}^{-y/\sigma}$), and $\xi < 0$ for tails
with a finite endpoint. For a parent tail
$\bar F(x) \propto x^{-a}$ the shape converges to $\xi = 1/a$. The
force of the theorem is that the GPD is not a model choice but
the universal limiting form: the analyst decides only the threshold
and the estimator, not the family.

The GPD fixes the exceedance \emph{shape}; the exceedance \emph{rate}
is the second half of the description and we retain it throughout. In
the point-process formulation of POT \citep[ch.~7]{Coles2001} the
expected number of exceedances per unit area above a threshold $u$
follows from the same two parameters,
\begin{equation}
\lambda(u) \;=\; \lambda(u_0)
\Bigl[1 + \frac{\xi\,(u-u_0)}{\sigma_{u_0}}\Bigr]^{-1/\xi},
\label{eq:rate}
\end{equation}
anchored at any reference threshold $u_0$ in the valid range. The
pair $\{\lambda(u), \xi(u)\}$ is the complete observable of this
paper: $\xi$ carries the count \emph{shape}, $\lambda$ the count
amplitude  (equivalently, in point-process language, the \emph{exceedance rate} of Eq.~(\ref{eq:rate}); we use ``amplitude'' for its role in the count normalization and ``rate'' for its POT meaning),  
and the lensing response splits cleanly along the
same seam (Sect.~\ref{sec:theory:lensing}).

Both decisions have standard diagnostics, which we apply throughout
and record here because they recur in every data section. The
mean-excess function, $e(u) = \mathrm{E}[D - u \,|\, D > u]$,
is linear in $u$ exactly when the exceedances are GPD, so a
mean-excess plot identifies the threshold above which the asymptotic
regime has been reached (Fig.~\ref{fig:planckbright}a, a \textit{Planck} example introduced in Sect.~\ref{sec:planck}).  The parameter-stability plot (the
fitted $\hat\xi$ and the reparametrized scale
$\sigma^* = \hat\sigma - \hat\xi u$ as functions of $u$) must
plateau over the valid threshold range, since above a valid threshold
the GPD family is closed under threshold raising
\citep{Coles2001, Scarrott2012}. Rather than selecting a single
``optimal'' threshold, we elevate the stability plot itself to the
observable: we fit the GPD at a ladder of thresholds and treat the
curve $\xiu$ as the measurement. As the following subsection shows,
the curve carries strictly more science than any single point on it:
its plateaus, bends, and steps are the fingerprint of the count
shape. The threshold-to-threshold correlations that this
introduces are handled with the full threshold covariance in Appendix~\ref{app:stats}. The
estimator throughout is maximum likelihood on declustered peak
exceedances (Sect.~\ref{sec:theory:beam}), with uncertainties from a
bootstrap over the largest exchangeable data units
(Appendix~\ref{app:stats}).

\subsection{The shape parameter as the local slope of the counts}
\label{sec:theory:xi}

Define the local logarithmic slope and curvature of the counts,
\begin{equation}
\eta(S) \equiv -\frac{\mathrm{d}\ln n_0}{\mathrm{d}\ln S},
\qquad
\kappa(S) \equiv \frac{\mathrm{d}^2\ln n_0}{\mathrm{d}(\ln S)^2}.
\label{eq:etakappa}
\end{equation}
On the shoulder, where the tail of the $\PD$ is set by the single
brightest source per beam, the exceedance distribution inherits the
local power-law index of the counts, and the derivation is short
enough to give in full. An isolated source of flux $S$ seen through a
Gaussian beam of solid angle $\Omega_{\rm beam}$ raises an area
$\Omega_{\rm beam}\ln(S/u)$ of the map above the level $u$, so the
expected area above threshold is
\begin{equation}
A({>}u) \;=\; \Omega_{\rm beam}\!\int_{u}^{\infty}\!
n_0(S)\,\ln\frac{S}{u}\;\mathrm{d}S .
\label{eq:areaabove}
\end{equation}
For locally power-law counts $n_0 = C\,S^{-\eta}$\footnote{By \emph{local power-law} we mean the approximation to the (otherwise general) counts at the flux scale near $u$, with $C$ and $\eta$ evaluated there, not a global functional form assumed for $n_0$.} the integral is
elementary, substituting $S = ut$ and using
$\int_1^\infty t^{-\eta}\ln t\;\mathrm{d}t = (\eta-1)^{-2}$,
and gives
\begin{equation}
A({>}u) \;=\; \frac{\Omega_{\rm beam}\,C}{(\eta-1)^{2}}\;
u^{-(\eta-1)} ,
\label{eq:areapl}
\end{equation}
a Pareto tail of index $\eta - 1$. Since a Pareto tail
$\bar F \propto u^{-a}$ has GPD shape $\xi = 1/a$
(Sect.~\ref{sec:theory:pd}), matching gives, to leading order,
\begin{equation}
\boxed{\;\xi(u) \;\approx\; \frac{1}{\eta(S_u) - 1}\;}
\label{eq:xislope}
\end{equation}
where $S_u$ is the intrinsic source flux that the threshold $u$
preferentially samples (set by the beam and the confusion level;
Appendix~\ref{app:beam}). The factor $(\eta-1)^{-2}$ left behind in
Eq.~(\ref{eq:areapl}) is not incidental: it is the derivative
$\mathrm{d}\xi/\mathrm{d}\eta$, and it reappears as the lever arm of
the lensing response in Sect.~\ref{sec:theory:lensing}.

Two points about Eq.~(\ref{eq:xislope}) deserve
emphasis. First, $\xi$ measures a logarithmic derivative of
the counts at one flux, not a moment. The distinction matters because
the classical connection between the
$\PD$ and the counts runs through the cumulants, which are
flux-integrated moments, whereas $\xi$ is a tail parameter
that responds to the count slope locally, at the flux the
threshold selects. Where the cumulant analysis compresses $\dnds$
into a handful of integrated numbers, the curve $\xiu$ scans
$\eta(S)$ point by point; the appropriate description is a local
Taylor expansion of $\ln n_0$ in $\ln S$, with $\xiu$ reading the
first coefficient and (through the lensing modulation below) the
second. Second, the relation is leading-order on the shoulder; the
model curves actually confronted with data in this paper are computed
by evaluating the exact $\PD$, or by direct simulation at the level
of declustered peaks (Sect.~\ref{sec:sims}), with
Eq.~(\ref{eq:xislope}) serving as the interpretive backbone rather
than the fitting function.

The invariances of $\xi$ are worth listing explicitly, because they
define the method's practical usefulness, and each corresponds to a
nuisance parameter that a full $\PD$ fit must account for. (i)
\emph{Location.} An additive offset of the map (a DC level, a
large-scale gradient, Zodiacal light and Galactic cirrus, or the summed mean of the
faint sources) leaves the exceedances $Y = D-u$ unchanged, provided
the threshold is anchored to the distribution itself rather than to
an absolute level. (ii) \emph{Scale.} A multiplicative gain error
$D \to aD$ rescales $u$ and $\sigma$ but not $\xi$, since a power-law
index is scale-free: $\xi$ is immune to absolute-calibration error.
(iii) \emph{Normalization.} On the power-law shoulder the overall
source density enters only the exceedance \emph{rate} $\lambda(u)$,
not the exceedance shape (nearer the core it also moves the core width
through $\sigc \propto \sqrt{q_2}$, which is one reason the \textit{Planck} threshold ladder is anchored
to the measured core, and the simulated \textit{Herschel} peaks to the
measured core centroid):
$\xi$ measures the count slope decoupled from the count amplitude,
which is precisely the slope--normalization degeneracy that most afflicts
direct faint-end count fits, where completeness corrections and
amplitude are strongly covariant. The practical consequence is that a
$\xiu$ analysis never needs the exact Gaussian-core profile, the
decomposition of the central variance into confusion versus
instrument noise, the absolute flux calibration, or the map zero
level. The complementary price is that all amplitude information
lives in $\lambda(u)$, which we retain alongside $\xi$ through the
point-process formulation of POT \citep[ch.~7]{Coles2001}, and that
robustness is bought with statistical efficiency: $\xi$ is robust to
the normalization because it discards it.

\begin{figure}
\centering
\includegraphics[width=\hsize]{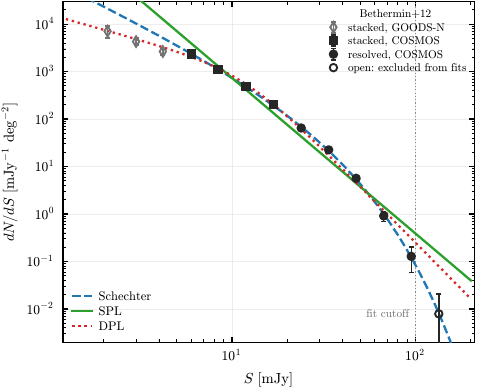}
\caption{The 350\,\mum\ differential number counts of
\citet{Bethermin2012} (their Table~3, ``All'' column) together with the
three count models used
throughout this paper: a Schechter function (blue dashed), a single power law (SPL; green solid)
and a smoothly broken double power law (DPL; red dotted). The data are
split by method, as in the legend: stacked GOODS-N (gray diamonds),
stacked COSMOS (filled squares) and resolved COSMOS (filled circles);
the gray dotted vertical line marks the 100\,mJy fit cutoff. Open symbols (the
three GOODS-N points below 6\,mJy and the point above the 100\,mJy
fit cutoff) are excluded from the fits. These are working descriptions of the
measured 6--94.6\,mJy counts, chosen to span the range of plausible
functional forms so that the tail statistic has something to
discriminate between; they are not offered as the best available
determination of the submillimeter counts. Functional forms, fitted
parameters, goodness of fit and methodology are all given in
Appendix~\ref{app:counts}.}
\label{fig:counts}
\end{figure}

\begin{figure}
\centering
\includegraphics[width=\hsize]{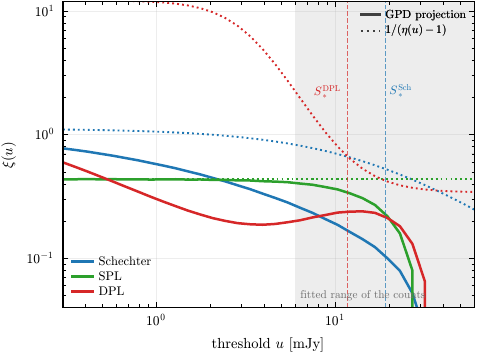}
\caption{The $\xiu$ fingerprints of the three count models in the
ideal case: no beam, no instrument noise, and no declustering. Dotted
curves are the asymptotic power-law limit $1/(\eta(u)-1)$; solid
curves are the exact GPD projection of the source-flux distribution,
so the gap between them is the finite-threshold correction alone,
with no instrumental content. In the asymptotic (dotted) curves, a pure power law produces a flat
plateau, a Schechter cutoff a smooth decline toward zero, and the
smoothly broken double power law a broad ramp
($\simeq 0.7$\,dex wide) down onto its bright-end plateau
$1/(\beta-1) = 0.34$ (the DPL $\beta$ of Table~\ref{tab:countfits}); the solid projections carry the finite-threshold correction on top of these. The counts are truncated at
$\Scut = 100$\,mJy, as throughout the paper (Sect.~\ref{sec:theory:beam});
the downturn of all three solid curves above $u \approx 30$\,mJy is
the finite-threshold approach to that endpoint, not a feature of the
counts.
Shading marks the
$6$--$94.6$\,mJy range over which the counts were fitted; the
color-matched dashed vertical lines mark the characteristic fluxes $S_*^{\rm DPL} =
11.7$\,mJy and $S_*^{\rm Sch} = 19.0$\,mJy at which each model bends.
}
\label{fig:fingerprints}
\end{figure}

\begin{figure}
\centering
\includegraphics[width=\hsize]{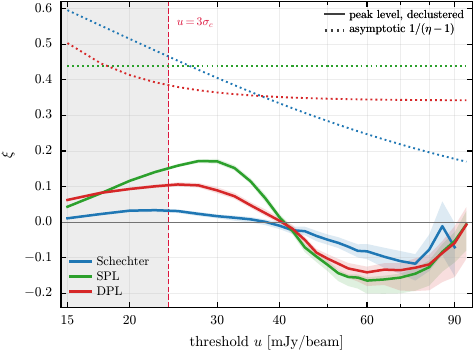}
\caption{The same three count models after the beam, the confusion
core, the instrument noise and the declustering step: solid curves
are peak-level Monte Carlo simulations at SPIRE resolution
($25.15''$, $\sigma_N = 5.40\,\mjb$), with bootstrap bands; dotted
curves repeat the asymptotic $1/(\eta-1)$ of
Fig.~\ref{fig:fingerprints} for reference. The vertical gap between
dotted and solid, $0.3$--$0.5$ in $\xi$, is the full
degradation budget of the measurement, and is the reason no analytic
curve is confronted with data anywhere in this paper
(Sect.~\ref{sec:theory:beam}).
At $u = 25\,\mjb$ the asymptotic values run
Schechter $>$ SPL $>$ DPL, while the measurable peak-level values run
SPL $>$ DPL $>$ Schechter; above $u \approx 40\,\mjb$ all three cross
into $\xi < 0$ as the $\Scut$ truncation takes hold. This inversion
is a further reason why only simulated peak-level curves are
confronted with data. Shading and the labeled red dashed vertical line mark $u < 3\,\sigc$, below which
$\hat\xi$ is no longer count-dominated; the horizontal line is $\xi = 0$.
}
\label{fig:fingerprintspeak}
\end{figure}

\begin{table}
\caption{What $\xi$ reports for common parent tails. The domain of
attraction fixes the asymptotic shape; the last column is the
signature in a measured $\hat\xi(u)$ ladder. A measured
$\hat\xi < 0$ diagnoses a finite endpoint; in our maps that endpoint
is the bright-source truncation at $\Scut$
(Sect.~\ref{sec:theory:pd}(iii)), never the exponential cutoff of the
counts, which is Gumbel and approaches zero from above. Column~1
gives the survival function $\bar F(x) = \Pr(X > x)$ of the parent,
with $a, b > 0$ tail exponents, $x_*$ an exponential scale and
$x_{\max}$ a finite endpoint; column~3 is the resulting GPD shape.}
\label{tab:parents}
\centering
\begin{tabular}{l l r l}
\hline\hline
parent tail $\bar F(x)$ & domain & $\xi$ & signature \\
\hline
power law $x^{-a}$              & Fr\'echet & $1/a$   & flat plateau \\
Schechter $x^{\alpha}\mathrm{e}^{-x/x_*}$ & Gumbel & $0^{+}$ & decline toward 0 \\
exponential                     & Gumbel   & $0$     & flat at zero \\
lognormal                       & Gumbel   & $0^{+}$ & very slow decline \\
endpoint $(x_{\max}\!-\!x)^{b}$ & Weibull  & $-1/b$  & decline through 0 \\
\hline
\end{tabular}
\tablefoot{
The \emph{domain of attraction} is the classical Fisher--Tippett--Gnedenko classification of a distribution's tail into one of three universal limiting families for its \emph{maximum}: Fr\'echet (power-law tails), Gumbel (exponential-class tails), and Weibull (tails with a finite endpoint); \citep[see e.g.][ch.~3]{Coles2001}. The GPD shape parameter $\xi$ is the same classification derived from the exceedances rather than the block maximum, so a parent's domain of attraction fixes the sign of $\xi$, and through it, what a measured $\hat\xi(u)$ can and cannot distinguish.} 
\end{table}

The power of reading the counts through Eq.~(\ref{eq:xislope}) is
that the \emph{functional form} of $\dnds$ becomes a morphological
statement about the curve $\xiu$ (Fig.~\ref{fig:fingerprints}), with
Table~\ref{tab:parents} as the dictionary. For a
pure power law, $\eta$ is constant and $\xiu$ is a flat plateau. For
a Schechter function the local slope $\eta(S) = -\alpha + S/S_*$
steepens smoothly toward the exponential cutoff, so $\xiu$ declines
monotonically toward zero: for our fitted parameters
$\eta - 1 = 0.890 + S/S_*$ is positive at every flux, so the
untruncated Schechter tail lies in the Gumbel domain and $\xi$
approaches $0^{+}$ without ever crossing it. 
For the smoothly broken double power law adopted here the local slope
$\eta_{\rm DPL}(S)$ of Eq.~(\ref{eq:etamodels}) is a logistic in
$\ln S$ interpolating from $\alpha$ to $\beta$ over
$1.91/(\beta - \alpha) \simeq 0.7$\,dex (a broad ramp rather than a
step), after which $\xiu$ settles onto the bright-end plateau
$1/(\beta - 1) = 0.34$. 

The long-standing ambiguity between a Schechter cutoff and a broken
power law in the submillimeter counts, both of which fit the binned
counts comparably well over limited flux ranges, is thus recast as a
question about how the measured curves end: over the accessible window the Schechter
declines through zero as its endpoint is approached, while the broken
power law flattens onto a positive plateau. It becomes a
model-selection problem on the joint observables
$\{\lambda(u_i), \xi(u_i)\}$, evaluated across the threshold ladder $\{u_i\}$ of Sect.~\ref{sec:theory:pd} (``we fit the GPD at a ladder of thresholds''), rather than a
parameter-fitting problem inside an assumed family. We adopt as our
three reference models the Schechter, single power-law (SPL), and double power-law (DPL) fits to the
\citet{Bethermin2012} 350\,\mum\ counts described in
Appendix~\ref{app:counts} (Fig.~\ref{fig:counts}); their local slopes
at the fluxes relevant here differ enough that the fingerprints are
separated by $0.1$--$0.5$ in $\xi$ in the ideal, beam-free case
(Fig.~\ref{fig:fingerprints}) and by $0.1$--$0.2$ at peak level over
the accessible window (Fig.~\ref{fig:fingerprintspeak}),
setting the precision target for the measurements that follow.

\subsection{Beam, declustering, and instrument noise}
\label{sec:theory:beam}

Three instrumental effects stand between Eq.~(\ref{eq:xislope}) and a
measurement on a real map. Each is treated at full length in
Appendix~\ref{app:beam}; here we describe what each one does and the 
implementation framework that comes from those. 
The effects are illustrated concretely below by the three instruments this paper analyzes (\textit{Planck}, \textit{Herschel}/SPIRE, and CCAT/FYST), with full configuration in Table~\ref{tab:ladder}.

\begin{figure}
\centering
\includegraphics[width=\hsize]{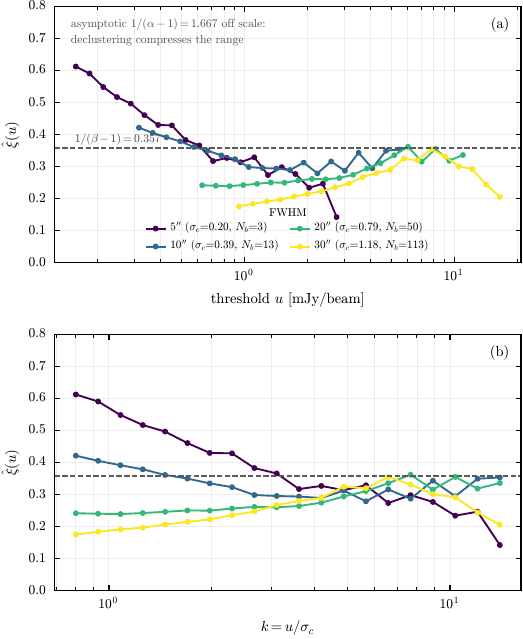}
\caption{What the beam does to the measurable tail statistic, at peak
level. Both panels are declustered Monte Carlo simulations of the
same illustrative double power law ($\alpha = 1.6$, $\beta = 3.8$,
$S_* = 4$\,mJy), chosen for its well-separated asymptotes and its
break placed low enough to leave a long bright-end plateau. $\sigN = 0$ throughout, so the panels
isolate the beam. 
\emph{(a):} $\hat\xi(u)$ against absolute threshold
for four beam sizes, color-coded as in the legend, which also
gives each beam's confusion noise $\sigc$ (in $\mjb$) and sources
per beam $\Nbeam$ (labeled $N_b$); the dashed horizontal line in both
panels is the bright-end asymptote $1/(\beta-1) = 0.357$, and the
downturn of the curves at the highest thresholds is the truncation
depression calibrated in Table~\ref{tab:truncbias}, not a failure
to converge. The beam decides the sign of the trend:
at $5''$ ($\Nbeam = 3$) sources are near-isolated and $\hat\xi$
falls from $0.61$ toward the bright-end asymptote
$1/(\beta-1) = 0.357$, whereas at $30''$ ($\Nbeam = 113$) confusion
suppresses the low-$u$ end and $\hat\xi$ rises from $0.18$ to
meet the same value. 
\emph{(b):} The same curves against
$k = u/\sigc$: they converge over $k \approx 4$--$10$, which is the
statement that the confusion core, not the beam size as such, sets
the scale of the low-$u$ suppression. The asymptotic faint-end value
$1/(\alpha-1) = 1.667$ lies off scale in both panels: declustering
compresses the dynamic range relative to the pixel $\PD$, by the
mechanism quantified in Appendix~\ref{app:beam}.}
\label{fig:beamxi}
\end{figure}

\paragraph{The beam sets the dynamic range.} Convolution with the
beam fixes the confusion noise: for a Gaussian beam,
$\sigc^2 = \tfrac12\,\Omega_{\rm beam}\int S^2 n_0(S)\,\mathrm{d}S$,
so $\sigc$ (the rms confusion noise) scales linearly with the beam FWHM at fixed counts. With
it are fixed the number of sources per beam, $\Nbeam$, and the
skewness of the $\PD$. The single number that determines whether the
shoulder regime exists at all is the ratio of the brightest
admissible source flux (the bright-source truncation $\Scut$ of the count models, $100$\,mJy throughout this paper) to the confusion noise, $\Scut/\sigc$: a
positive-$\xi$ power-law shoulder can only rise above the Gaussian
core if single sources deposit several $\sigc$ in one beam.
Figure~\ref{fig:beamxi} shows how completely the beam controls the
measurable statistic: at fixed counts it decides even the sign
of the trend of $\hat\xi$ with threshold, and the curves for
different beams approximately collapse onto one another over the shoulder when plotted against
$u/\sigc$ rather than against absolute flux.
For the same intrinsic 350\,\mum\ counts truncated at
$\Scut = 100$\,mJy, this ratio is $1.05$ at \textit{Planck}'s $300''$
beam, $12.5$ at SPIRE's $25''$, and $21$ at CCAT's $15''$. The 
$S_{\rm cut}/\sigma_c$ ratio is the 
quantitative spine of the three-instrument ladder this paper climbs,
and the reason the three instrument result sections read so differently. 

The full configuration of the three instruments is collected in
Table~\ref{tab:ladder}, alongside the simulation setup that generates
their model curves (the simulation pipeline itself is introduced fully in Sect.~\ref{sec:sims}; forward-referenced here only for the instrument configuration table); we refer back to it throughout.

Two threshold windows follow, and they are not equivalent.
The \emph{interpretive} one is the range over which
Eq.~(\ref{eq:xislope}) may be read quantitatively,
\begin{equation}
3\,\sigc \;\lesssim\; u \;\lesssim\; \Scut/10 ,
\label{eq:window}
\end{equation}
bounded below by the Gaussian ``penultimate'' regime and above by the
level at which the bright truncation depresses $\hat\xi$ as much as
the count models differ (Appendix~\ref{app:beam},
Table~\ref{tab:truncbias}). It is non-empty only for
$\Scut/\sigc \gtrsim 30$, a condition none of our three
instruments meets ($1.05$, $12.5$, $21.0$;
Table~\ref{tab:ladder}). That is the quantitative reason behind a
rule invoked throughout this paper: the analytic relations orient the
interpretation but never supply a number, and every model curve is
generated by simulation instead.

The \emph{measurement} window is the one that represents data, and
it is wider because model comparison does not require
Eq.~(\ref{eq:xislope}) to hold: the bright mask is applied
identically to map and models (Appendix~\ref{app:matchedfilter:forward}),
so it is forward-modeled rather than interpreted and largely cancels
between them. It runs from the same
$3\,\sigc$ up to the exceedance floor, and is non-empty for the far
weaker $\Scut/\sigc \gtrsim 3$. Hence the \textit{Herschel} ladder
reaches $75\,\mjb$ and the CCAT ladder $55\,\mjb$ at
$\Scut = 100$\,mJy without contradicting Eq.~(\ref{eq:window}); hence
also the bright-masked \textit{Herschel} variant is compared with the
models only up to $50\,\mjb$, and the unmasked curve above that level reports the
bright lensed population rather than the intrinsic counts
(Sect.~\ref{sec:herschel:rise}). The two criteria bracket the ladder
this paper climbs: at \textit{Planck} both windows are void and no
sky area recovers either; at a $40''$ beam the interpretive window is
already empty while the measurement window survives as the half
decade above $3\,\sigc = 38\,\mjb$; at SPIRE and CCAT the measurement
window is comfortably open, which is why those two arms deliver
count-model comparisons and \textit{Planck} does not.

\paragraph{Declustering: one peak per beam.} Map pixels are
correlated on the beam scale, so pixel exceedances violate the
independence that the GPD likelihood assumes, and a single bright
source contributes many contiguous pixels, the classic clustered-%
exceedances problem of the POT literature, for which the standard
remedy is declustering \citep{Coles2001}. We retain local maxima
separated by at least one beam FWHM, reducing every exceedance
cluster to its peak. Two properties of that operation are used
repeatedly below and derived in Appendix~\ref{app:beam}: it is a rank
operation, so it commutes with any monotonic rescaling of the flux
axis and is therefore common-mode between lensed and control fields;
and it is an order statistic rather than a thinning, so the peak
marginal is a reweighting of the pixel marginal
(Eq.~\ref{eq:peakmarginal}) that crushes the core and leaves the far
tail merely rescaled.

The consequence that bears on every figure in this paper is that
peaks and pixels are different distributions with different fitted
shapes. Analytic $\xiu$ curves computed from the \emph{pixel} $\PD$
differ from the peak-level measurement by a median $0.03$--$0.06$,
reaching $0.22$ at the ends of the threshold ladder (comparable
to, or larger than, the count-model separations themselves), and
the asymptotic $1/(\eta-1)$ limit overshoots the peak-level curve by
$0.3$--$0.5$ (Fig.~\ref{fig:fingerprintspeak}). Accordingly, no analytic curve is fitted directly to data or to Monte Carlo anywhere in this paper: model predictions are instead generated at peak level by the simulation pipeline detailed in Sect.~\ref{sec:sims}. The simulations
use beam-correlated rather than white noise throughout, a convention
forced by the declustering step and calibrated in
Appendix~\ref{app:beam} (Fig.~\ref{fig:noisedecl} there).

\paragraph{Noise moves the threshold, not the tail.} Adding Gaussian
instrument noise convolves the $\PD$ but leaves the power-law decay
of the far tail intact: a Gaussian tail dies faster than any power
law, so sufficiently far out the noise cannot manufacture or destroy
exceedances. Its effect is to raise the flux at which the asymptotic
regime begins (to push the usable window upward), not to change
the limiting $\xi$. Confusion itself acts as a second, irreducible
noise floor of the same character, present even at $\sigN = 0$.
Because the noise is common to a lensed field and its control, this
threshold shift is largely common-mode in the differential
measurement (Sect.~\ref{sec:theory:lensing}): a shared noise level
biases $\xi_{\rm lensed}$ and $\xi_{\rm control}$ in the same
direction and drops out of the difference, costing sensitivity but
not fidelity.

\subsection{The lensing modulation of the tail}
\label{sec:theory:lensing}

The magnification field is the external input when adding in the lensing effect, 
so we outline all its components. 
For an axially symmetric lens the convergence and tangential shear are
\begin{equation}
\kappa_{\rm l}(\theta) = \frac{\Sigma(\theta)}{\Sigma_{\rm cr}},
\qquad
\gamma(\theta) = \bar\kappa_{\rm l}({<}\theta) - \kappa_{\rm l}(\theta),
\label{eq:kappagamma}
\end{equation}
with $\bar\kappa_{\rm l}({<}\theta) = (2/\theta^{2})\int_0^{\theta}
\kappa_{\rm l}(\theta')\,\theta'\,\mathrm{d}\theta'$, $\Sigma(\theta)$
the projected mass density and $\Sigma_{\rm cr}$ the critical surface
density, with $\theta$ the angular distance from the cluster center; the magnification is the inverse Jacobian of the lens
mapping,
\begin{equation}
\mu(\theta) \;=\;
\bigl[\,(1-\kappa_{\rm l})^{2} - \gamma^{2}\,\bigr]^{-1} .
\label{eq:mu}
\end{equation}
For a truncated Navarro--Frenk--White (NFW; \citealt{Navarro1997}) halo of virial mass $M_{\rm vir}$, radius
$r_{\rm vir}$ and concentration $c$ both quantities are analytic
\citep[][and references therein]{Lima2010},
\begin{equation}
\begin{pmatrix}\kappa_{\rm l}\\[2pt]\gamma\end{pmatrix}\!(\theta)
= \frac{M_{\rm vir}\,f\,c^{2}}
       {2\pi r_{\rm vir}^{2}\,\Sigma_{\rm cr}}
\begin{pmatrix}F\\[2pt] G\end{pmatrix}\!
\Bigl(\frac{c\,\theta}{\theta_{\rm vir}}\Bigr),
\qquad
f = \Bigl[\ln(1+c) - \tfrac{c}{1+c}\Bigr]^{-1},
\label{eq:nfwkg}
\end{equation}
with the closed forms of $F$ and $G$ given by Eqs.~(11) and (16) of
\citet{Lima2010}. Aperture averages throughout are image-plane and
area-weighted,
\begin{equation}
\bigl\langle X \bigr\rangle_{<R} \;=\; \frac{2}{R^{2}}
\int_{0}^{R} X\bigl(\mu(\theta)\bigr)\,\theta\,\mathrm{d}\theta ,
\label{eq:apav}
\end{equation}
which is the correct weighting for a statistic accumulated over map
pixels; the corresponding source-plane average carries an extra
factor $\mu^{-1}$ \citep[cf.\ Eq.~(40) of][]{Lima2010}. The
magnification entering these averages is $|\mu|$, clipped at
$\mu_{\max} = 100$ near the critical curves of the halo (for the
benchmark lens of Sect.~\ref{sec:ccat} they lie at $0.02$ and
$0.06\,\theta_{500}$, inside every aperture, enclosing $1.3\%$ of the
innermost one). The quantity that drives $\dxiu$, $\langle\ln\mu\rangle$,
is insensitive to the clip ($0.691$, $0.693$ and $0.694$ for
$\mu_{\max} = 100$, $300$ and $1000$ at $r < 0.5\,\theta_{500}$),
whereas $\langle\mu\rangle$ grows logarithmically with it ($3.07$ to
$3.90$ over the same range); the simulated $\hat\xi$ changes by less
than $0.005$ between $\mu_{\max} = 5$ and $100$. We take
$c(M,z)$ from \citet{Duffy2008} and the virial overdensity from
\citet{BryanNorman1998}.

Behind such a lens, the counts per unit \emph{image-plane}
solid angle at magnification $\mu$ are
\begin{equation}
n_\mu(S) \;=\; \mu^{-2}\, n_0(S/\mu),
\label{eq:lensedcounts}
\end{equation}
each patch gaining because it samples the more abundant counts at the
demagnified flux $S/\mu$, and losing the factor $\mu^{-2}$ of
solid-angle dilution \citep[Eq.~31 of][]{Lima2010}. Inserting a pure power law
$n_0 \propto S^{-\eta} $ (again the local approximation at the relevant flux, as in Sect.~\ref{sec:theory:xi} above, not a global assumption), the slope passes through the transformation
untouched and only the amplitude is rescaled, by $\mu^{\eta-2}$. This is an exact statement of symmetry: magnification rescales the
flux axis, a power law is the unique scale-invariant function, and a
scale transformation maps a self-similar function onto itself. It
cleanly splits the lensing response of the $\PD$ tail into two
channels.

\textbf{The amplitude channel.} The exceedance rate shifts by the
classical magnification-bias factor,
\begin{equation}
\frac{\lambda_{\rm lensed}(u)}{\lambda_{\rm control}(u)}
\;=\; \bigl\langle \mu^{\,\eta - 2} \bigr\rangle ,
\label{eq:ampchannel}
\end{equation}
\citep{Broadhurst1995}, where the average runs over the magnification
distribution within the selected aperture. The fixed point at
$\eta = 2$ separates two regimes with opposite signs. For the shallow
slopes of the faint submillimeter counts ($\eta < 2$; $1.1$--$1.9$
for the faint-end slopes of the Schechter and DPL fits of Appendix~\ref{app:counts}; the SPL has $\eta = 3.28$ at every flux)
the factor is below unity, a depletion: the faint tail thins
behind the cluster, because the solid-angle dilution wins against the
flux boost. For the steep counts near the cutoff the factor grows
without bound, becoming the strong enhancement that selects the
well-known strongly lensed submillimeter population
\citep{Blain1996, Negrello2010}. Equation~(\ref{eq:ampchannel}) is a
single number blending the count slope with a moment of the
magnification distribution, which is why we treat it as a
consistency channel rather than a primary observable: on its own it
is degenerate between mass and slope.

\textbf{The shape channel.} Under a single magnification the constant
prefactor $\mu^{-2}$ drops out of the logarithmic derivative, so the
lensed local slope at observed flux $S$ is simply the unlensed slope
at the demagnified flux, $\eta_\mu(S) = \eta(S/\mu)$. Expanding for
small $\ln\mu$,
$\eta(S/\mu) \simeq \eta(S) - (\ln\mu)\,\mathrm{d}\eta/\mathrm{d}\ln S$,
using the definition $\kappa \equiv -\mathrm{d}\eta/\mathrm{d}\ln S$ of Eq.~(\ref{eq:etakappa}), and differentiating Eq.~(\ref{eq:xislope}), we have
\begin{equation}
\begin{aligned}
\Delta\eta &= \eta(S/\mu) - \eta(S) \simeq \kappa(S)\,\ln\mu \;,\\
\Delta\xi  &\equiv \frac{\partial\xi}{\partial\eta}\,\Delta\eta
          = -\,\frac{\Delta\eta}{(\eta-1)^{2}} \;,
\end{aligned}
\label{eq:mastersteps}
\end{equation}
so that composing the two gives the master equation of this paper:
\begin{equation}
\boxed{\;\Delta\xi(u) \;\approx\;
-\,\frac{\ln\mu}{\bigl(\eta - 1\bigr)^{2}}\;\kappa(S_u)\;}.
\label{eq:master}
\end{equation}

This equation shows the lensing-induced change of the GPD shape as proportional to the
curvature of the log-counts, times $\ln\mu$. The
pure-power-law invariance is now simply the statement that a straight
line has zero curvature, and the three count models map onto three
distinct lensing signatures: no shape response at all (SPL, all
information in the amplitude channel); a smooth positive $\dxiu$
growing toward the cutoff (Schechter: magnification sends
$S_* \to \mu S_*$, so at fixed threshold the lensed tail sits further
from the cutoff, in shallower counts, and is heavier); and a
localized response that tracks the break as it shifts to $\mu S_*$
(DPL). Far from being the weakness it might first appear, the
power-law invariance is the statement that the shape channel switches
on exactly in proportion to the departure of the counts from
scale-free behavior, which is precisely the functional-form
question the method aims to answer.

This last property is what elevates the differential signal beyond a
statistics convenience. A genuine feature of the counts, a cutoff
or a break, must move in threshold by the known factor $\mu$
between control and lensed fields, whereas an instrumental kink knows
nothing of the foreground cluster and stays put. With the
magnification supplied by external mass models, the predicted shift
acts simultaneously as a \emph{ruler}, localizing the feature in
intrinsic flux, and as an \emph{authenticator}, certifying it as
astrophysical. A further consistency channel comes for free: the
unlensed curve already carries the count curvature as its own
threshold derivative, $\mathrm{d}\xi/\mathrm{d}u \propto \kappa$,
while $\Delta\xi \propto \ln\mu\,\kappa$ carries the same
$\kappa$ with a known-$\mu$ prefactor and different systematics:
the former requires the beam- and confusion-dependent
threshold-to-flux mapping, the latter does not. These are two routes
to one physical quantity, and their agreement is a systematics
cross-check, not a second independent observable, since both derive
from the same counts. 

In addition, because the lensed and control fields
share the instrument, the additive noise, and the
beam, the dimensionless machinery of the measurement (declustering
conventions, peak-selection shifts, filter choices) cancels in
$\dxiu$ to the 1--3\% level (Appendix~\ref{app:beam:lensing}). A calibration difference between
fields observed in separate runs does not enter either, since $\xi$
is gain-invariant (Sect.~\ref{sec:theory:xi}); it would only displace
the flux axis of a count feature by the relative gain error, which is
at the percent level for the instruments considered here. What does
\emph{not} cancel is the flux scale itself. Substituting Eq.~(\ref{eq:lensedcounts}) into the
confusion variance of Sect.~\ref{sec:theory:beam}  and changing
variable to $S' = S/\mu$,
\begin{equation}
\begin{split}
\sigc^{2}(\mu) &= \tfrac12\,\Omega_{\rm beam}\!\int\! S^{2}\,
   n_\mu(S)\,\mathrm{d}S \\
 &= \mu\;\tfrac12\,\Omega_{\rm beam}\!\int\! S'^{2}\,
   n_0(S')\,\mathrm{d}S'
 \;=\; \mu\,\sigc^{2}(1) ,
\end{split}
\label{eq:sigcmu}
\end{equation}
so that $\sigc(\mu) = \sqrt{\mu}\,\sigc(1)$,
exactly for untruncated counts, and to a few percent with $\Scut$
held fixed in the observed frame. The usable window therefore sits at
different absolute fluxes in the two arms, and the threshold ladder
of a differential measurement must be constructed on the lensed arm
(Appendix~\ref{app:beam}).

Where does the signal live spatially? Not, as intuition might
suggest, in the strongly magnified cusp near the cluster core, where
sources are ``brought out of the confusion''. Quantitatively that is
not what happens, and getting this correct matters for the
aperture design of Sects.~\ref{sec:planck}--\ref{sec:ccat}. At the
fluxes sampled by the shoulder the counts are only moderately steep
($\eta \approx 2.5$--$3.3$ near the Schechter knee), so the per-area
enhancement $\mu^{\eta-2}$ of a strongly magnified zone is of order
unity (the flux boost and the solid-angle dilution nearly cancel),
and the area at $\mu \gtrsim 5$ around even a massive halo is a
few percent of the innermost aperture and a fraction of a percent
beyond $r \approx \theta_{500}$, the angle subtended by $R_{500}$ at the lens redshift. The measurable $\dxiu$ is
instead carried by the large moderate-magnification region
($\mu \sim 1.3$--$2$), which is also where the weak-$\mu$
linearization of Eq.~(\ref{eq:master}) is accurate. The strongly
magnified caustic regime is different physics: there the image-plane
counts flatten to the universal $S^{-3}$ behavior set by the
magnification distribution $P(\mu) \propto \mu^{-3}$ near a caustic
\citep{SchneiderEhlersFalco1992}, and $\xiu$ would read $P(\mu)$
rather than the intrinsic counts. That regime, the selection
mechanism behind the strongly lensed submillimeter population, is
deliberately excluded by our aperture choices; it is a different
observable, not a contaminant we failed to model.

Finally we write down the object from which every quantitative
lensing statement in this paper is actually computed, since
Eq.~(\ref{eq:master}) is never used for prediction
(see Sect.~\ref{sec:sims:validation}). Substituting $n_\mu$ of
Eq.~(\ref{eq:lensedcounts}) into the Condon integral
Eq.~(\ref{eq:condon}) gives the local, single-$\mu$ distribution
$P(D\,|\,\mu)$ through
\begin{equation}
\ln\psi_{\mu}(\omega) = \int\!\mathrm{d}^{2}\theta' \int\!\mathrm{d}S\;
n_{\mu}(S)\,\bigl[\mathrm{e}^{\,\mathrm{i}\omega S B(\theta')}-1\bigr] ,
\label{eq:condonlensed}
\end{equation}
and the distribution actually fitted (the pooled pixel histogram
of an aperture of radius $R$) is the area-weighted \emph{mixture}
of these local distributions,

\begin{equation}
P_{R}(D) \;=\; \frac{2}{R^{2}}\int_{0}^{R}
P\bigl(D \,\big|\, \mu(\theta)\bigr)\;\theta\,\mathrm{d}\theta
\label{eq:mumixture}
\end{equation}
with $\mu(\theta)$ from Eq.~(\ref{eq:mu}). Equation~%
(\ref{eq:mumixture}) is what we refer to throughout as the exact
$\mu$-mixture $\PD$. It makes explicit a point that governs every
aperture choice in our analysis: \emph{the aperture, not the map, defines the
magnification distribution the measurement responds to}. A
cluster-centered map always contains the high-$\mu$ cusp, but the
fitted $\hat\xi$ sees it only in so far as cusp pixels enter the
histogram, and a whole-map fit is diluted to nothing by the
$\mu \approx 1$ periphery.

Equation~(\ref{eq:mumixture}) differs from the
superficially similar construction of \citet{Hezaveh2013}, who
compute the flux distribution behind a cluster by decomposing the
aperture into annuli of constant $\mu$ and convolving the
per-annulus Poisson distributions. That is the correct object for
their observable, the total flux integrated over an aperture. Ours is
a one-point histogram of the pixels within the aperture, for which
the local distributions mix rather than add. The two coincide only in
the limit of a single pixel per aperture.

\subsection{The suppression of \texorpdfstring{$\dxiu$}{Delta xi(u)},
and the mass-determination leverage}
\label{sec:theory:twosided}

The relations of Sect.~\ref{sec:theory:lensing} invite an optimistic
estimate that is worth working through explicitly, because it is
wrong by more than an order of magnitude and the reasons are
physical rather than technical. Inside a small aperture on a massive
cluster the mean magnification is substantial: for the benchmark
halo used throughout this paper ($M_{500} = 10^{15}\,\msun$ at
$z_{\rm l} = 0.5$, $z_{\rm s} = 2$), $\langle\mu\rangle = 3.07$ and
$\langle\ln\mu\rangle = 0.69$ within $r < 0.5\,\theta_{500}$. With
the lever arm of Eq.~(\ref{eq:master}) evaluated for the Schechter
counts at a CCAT-like threshold $u = 15\,\mjb$, where
$\eta = 2.679$ and the log-curvature is $\kappa = -0.789$,
\begin{equation}
\frac{\kappa}{(\eta-1)^2} = -0.280 ,
\label{eq:leverarm}
\end{equation}
so Eq.~(\ref{eq:master}) predicts $\Delta\xi = +0.193$, comfortably
larger than the $0.1$--$0.2$ separations between the count models
themselves, and apparently a routine detection. It is not. Four
effects are at work: the first is already contained in that number
and bounds how large it could ever be, and the remaining three stand
between it and any measurement.

\emph{First, $\dxiu$ is a curvature effect, not a magnification
effect.} A pure power law is scale-invariant, so magnification maps
it onto itself with a changed amplitude and $\Delta\xi$ vanishes
identically (Sect.~\ref{sec:theory:lensing}); for the single
power law Eq.~(\ref{eq:leverarm}) is exactly zero. Magnification slides
the observer along the counts; only the bending of the counts
registers in the shape. Over the accessible decade the submillimeter
counts are close enough to a power law that the lever arm
(Eq.~\ref{eq:leverarm}) is modest even at its best, and the
$+0.193$ is a ceiling set by the curvature of the Schechter counts
at that flux.

\emph{Second, the $\PD$ suppresses what remains by a further factor
$2$--$14$.} Table~\ref{tab:dxisuppression} compares
Eq.~(\ref{eq:master}) against the exact area-weighted
$\mu$-mixture $\PD$ for the benchmark lens. The suppression is
strongest precisely where the magnification is largest, for two
compounding reasons, both traceable to
Sect.~\ref{sec:theory:beam}. A map value $u$ is a sum over
$\Nbeam$ sources rather than a single source of flux $S_u$, so the
local curvature that Eq.~(\ref{eq:master}) evaluates at one flux is
in reality averaged over a broad range and partly cancels; and
$\sigc$ is itself lensed, $\sigc(\mu) \simeq \sqrt{\mu}\,\sigc$,
so the magnified field carries a wider confusion core that pushes the
measurement to a lower effective $k = u/\sigc$ and undoes part of the
shape shift. Both scale with $\mu$: the more magnification one buys,
the larger the fraction of it that is eaten.

\emph{Third, the exact $\Delta\xi$ barely depends on the aperture at
all.} The fifth column of Table~\ref{tab:dxisuppression} varies by
less than a factor three across a factor six in radius, and
peaks near $r \simeq \theta_{500}$ rather than at the center.
Shrinking the aperture to chase magnification does not increase the
signal; below $r \sim \theta_{500}$ it slightly decreases it.

\emph{Fourth, the aperture trend is roughly scale-invariant, so it cannot be
optimized away.} The single-cluster significance goes as
$|\Delta\xi|\sqrt{N_{\rm pk}}$. Since $N_{\rm pk} \propto r^2$ while
$\langle\ln\mu\rangle$ falls steeply (roughly as $r^{-1}$ to
$r^{-2}$) through the outer NFW profile, the two factors nearly
cancel: the final column of
Table~\ref{tab:dxisuppression} is flat to a factor $3.6$ over the
whole scan, with a broad, shallow optimum near
$r \simeq 1.5$--$2\,\theta_{500}$. This is why the $N_{3\sigma}$
requirement of Sect.~\ref{sec:ccat:lensing} varies so little between
apertures, and it is the quantitative form of the
dilution-versus-statistics pincer that the \textit{Planck} analysis
meets in data (Sect.~\ref{sec:planck:lensed}).

\begin{table*}
\caption{Why the lensing modulation is small. Benchmark lens
($M_{500} = 10^{15}\,\msun$, $z_{\rm l} = 0.5$, $z_{\rm s} = 2$), Schechter
counts, CCAT configuration at $u = 15\,\mjb$.}
\label{tab:dxisuppression}
\centering
\begin{tabular}{r r r r r r r r}
\hline\hline
$r/\theta_{500}$ & $\langle\mu\rangle$ & $\langle\ln\mu\rangle$ &
Eq.~(\ref{eq:master}) & exact $\mu$-mixture $\PD$ & ratio &
$N_{\rm pk}$/cluster & $|\Delta\xi|\sqrt{N_{\rm pk}}$ \\
\hline
0.5 & 3.071 & 0.690 & $+0.193$ & $+0.0130$ & 0.07 &   32.5 & 0.074 \\
1.0 & 1.657 & 0.299 & $+0.084$ & $+0.0188$ & 0.22 &  130.1 & 0.214 \\
1.5 & 1.328 & 0.168 & $+0.047$ & $+0.0150$ & 0.32 &  292.8 & 0.258 \\
2.0 & 1.192 & 0.102 & $+0.028$ & $+0.0115$ & 0.41 &  520.5 & 0.263 \\
3.0 & 1.086 & 0.046 & $+0.013$ & $+0.0066$ & 0.52 & 1171.0 & 0.227 \\
\hline
\end{tabular}
\tablefoot{Columns 4 and 5 are $\Delta\xi$ from the linearized
master relation and from the exact $\mu$-mixture $\PD$ respectively,
and column 6 is their ratio (exact/linearized); $\langle\mu\rangle$
and $\langle\ln\mu\rangle$ are area-weighted means within the
aperture. The exact column is what the peak-level Monte Carlo
reproduces (Sect.~\ref{sec:sims:validation}).
$N_{\rm pk}$ is the declustered peak yield at one peak per four beam
solid angles. The last column is the per-cluster significance up to
the $(1+\xi)$ prefactor common to all rows.}
\end{table*}

The four effects share a single root: \emph{lensing changes how
many peaks are seen far more than it changes their shape.} The
amplitude channel of Eq.~(\ref{eq:ampchannel}) responds as
$\langle\mu^{\eta-2}\rangle$, a power law: a factor $1.85$ over
the $r < 0.5\,\theta_{500}$ aperture ($\langle\mu\rangle = 3.07$) for
$\eta = 2.679$, an $85\%$ effect. The
shape channel responds as $\ln\mu$, logarithmically, and is then
reduced again by curvature and by confusion. One is attempting to
read a logarithmic response in the presence of a power-law one.

That asymmetry also settles the direction of inference.
Equations~(\ref{eq:ampchannel}) and (\ref{eq:master}) are formally
invertible: with the counts $n_0(S)$ known to sufficient accuracy,
the same observables constrain the magnification, and hence the mass
of the foreground lens. But the asymmetry just
established runs in the same direction for mass as it does for
detection, and for the same reason. A mass inferred through $\xi$
inverts a logarithm, so it carries the exponential of the measurement
error, and the exact column of Table~\ref{tab:dxisuppression} shows
the response is a further order of magnitude below what
Eq.~(\ref{eq:master}) would suggest, so that exponential is taken of
a large number. The amplitude channel does carry mass as a power law,
but Eq.~(\ref{eq:ampchannel}) is a single number degenerate between
$\langle\mu^{\eta-2}\rangle$ and the local slope $\eta$: breaking the
degeneracy requires exactly the external information (a known slope,
or a known mass) that one hoped to infer. It is, in effect, a
reformulation of classical magnification-bias mass estimation, with
the same degeneracy structure and no new information content. We
therefore adopt the asymmetric architecture throughout this paper:
the magnification is an external input, supplied by mass models of
the foreground clusters from weak lensing, X-ray, or SZ data, and the
functional form of $\dnds$ is the target measured against that ruler.
Mass is not a rival target but the calibration that makes the shape
measurement work. The reverse mode does become attractive in a future
regime (counts pinned down to high accuracy, large homogeneous
cluster ensembles), and we return to it briefly in
Sect.~\ref{sec:discussion}.

\subsection{What fluxes the measurement reaches}
\label{sec:theory:reach}

A flux-boundary criterion is central to this paper, and we state
it before any data are presented. The threshold $u$ must clear the
confusion core ($u \ge 3\,\sigc$; Sect.~\ref{sec:theory:beam}), so
$\xiu$ reads the counts at fluxes of
order the threshold itself: for the instruments considered here, the
accessible band begins at $\approx 25\,\mjb$ (SPIRE) or
$\approx 15\,\mjb$ (CCAT) and extends to the bright mask. Its lower
end is genuinely below the individual-detection limit
($\approx 31.5$\,mJy at SPIRE resolution; \citealt{Nguyen2010}), so
the method reaches sources that cannot be cataloged. But it is one
to two decades above any plausible faint-end turn-over of the
counts, and the distinction between those two statements must be kept
sharp. The faint population enters $\xiu$ only through its
contribution to $\sigc$, a channel worth at most a few percent
because $\sigc^2 = \tfrac12\,\Omega_{\rm beam}\,q_2$ (Sect.~\ref{sec:theory:beam}) is set by the flux second moment $q_2 \equiv \int S^2 n_0\,\mathrm{d}S$, which converges at the faint end: for our fiducial counts, 96\% of $q_2$ is
contributed by sources above 1\,mJy, and a full third by the
1--10\,mJy decade alone. We have verified quantitatively that a
faint-end turn-over is structurally invisible to the statistic: even
a turn-over strong enough to reconcile the counts with the measured
background (below) changes $\xiu$ by at most $0.007$ (against
measured errors of $0.009$--$0.099$) and does so as a nearly
uniform vertical shift of the whole curve, the one mode that the
threshold-covariance analysis of Appendix~\ref{app:stats} identifies
as carrying most of the measurement noise. A dedicated faint-end
search with this statistic is therefore not worth designing.

The CIB background monopole $q_1 = \int S\,n_0\,\mathrm{d}S$ is the
mirror image in this respect: it is spread nearly evenly across every decade of flux,
converges only slowly toward the faint end for the Schechter slope
(and diverges for the SPL), and is measured to a few percent \citep{Fixsen1998,
Odegard2019}. The two probes are therefore nearly orthogonal in flux,
and their combination, not either alone, is the strongest available
constraint architecture (Sect.~\ref{sec:discussion}). To state the
flux reach precisely: the statistic is \emph{locally} sensitive at
source fluxes of order the threshold, $S_u \approx u$ (the
identification used throughout, e.g.\ in Appendix~\ref{app:clustering}),
i.e.\ over $25$--$75\,\mjb$ at SPIRE resolution, a band that
straddles the individual-detection limit. The 1--10\,mJy decade is
constrained by the measurement only through the functional forms of
the count models, which tie their behavior in that decade to their
slopes at $S_u$, and directly only through $\sigc$. In that sense the
decade of scientific interest is 1--10\,mJy: it contributes a third of
the confusion variance, it is where our three count models genuinely
diverge, and it is where, as Sect.~\ref{sec:planck} shows, two of the three models
already conflict with the measured background. One further caveat on
the word ``flux-resolved'': a GPD fit at threshold $u$ uses all
exceedances above $u$, so $\hat\xi(u)$ is not strictly local in flux;
Sect.~\ref{sec:herschel:unlensed} quantifies this, masking the
$S > 100$\,mJy sources moving $\hat\xi(25\,\mjb)$ by $0.1$, and it is
the reason the bright-masked variant is the one compared with the
models, which are forward-modeled through the same mask.

\section{Simulation framework and validation}
\label{sec:sims}

Every model curve confronted with data in this paper, and every
forecast, is generated by an end-to-end simulation pipeline that
mirrors the measurement step by step: sources are injected on a pixel
grid, convolved with the instrument beam, superposed with noise,
declustered to one peak per beam, thresholded, and fitted with the
identical GPD estimator used on the real maps. This section
records the design of the pipeline, the validation of the analytic
relations against it, and (at somewhat greater length, because the
question has acquired fresh urgency from recent literature) the
treatment of source clustering.

\begin{table}
\caption{The three-instrument ladder: configurations and derived
quantities (Schechter counts, $\Scut = 100$\,mJy,
$S_{\min} = 1$\,mJy). }
\label{tab:ladder}
\centering
\setlength{\tabcolsep}{4pt}
\begin{tabular}{l r r r}
\hline\hline
 & \textit{Planck} & \textit{Herschel} & CCAT \\
\hline
band & 857\,GHz & 350\,\mum & 850\,GHz \\
beam FWHM & $300''$ & $25.15''$ & $15''$ \\
map pixel & $60''$ & $8''$ & $3''$ \\
$\sigN$ [$\mjb$] & 30.4 & 5.40 & 3.1 \\
$\sigc$ [$\mjb$] & 95.1 & 7.98 & 4.76 \\
$\sigtot$ [$\mjb$]$^{(a)}$ & 187.9 & 9.63 & 5.68 \\
measured core [$\mjb$] & 185.4 & 9.00 & -- \\
$\Scut/\sigc$ & \textbf{1.05} & \textbf{12.5} & \textbf{21.0} \\
$\Nbeam$ & 680 & 4.8 & 1.7 \\
$\PD$ skewness & 0.15 & 1.77 & 2.97 \\
$u$ window [$\mjb$] & 271--1012 & 25--75 & 15--55 \\
analysis area [deg$^2$] & 1820 & 25.8 & 9.1 \\
\hline
\end{tabular}
\tablefoot{$^{(a)}$ For \textit{Planck}, $\sigtot$ includes the
measured non-Poisson (clustered CIB + cirrus) Gaussian term of
$159.2\,\mjb$ (Sect.~\ref{sec:planck:data}); the simulated core then
reproduces the measured one to 1.4\%. The measured
\textit{Herschel} core is the matched-filtered value; CCAT is a
forecast. The CCAT window floor is the first ladder point above the
$3\sigc = 14.3\,\mjb$ validity criterion
of Sect.~\ref{sec:ccat:lensing}; the \textit{Planck} ladder is
anchored to the measured core width rather than to $\sigc$
(Sects.~\ref{sec:planck:data} and \ref{sec:herschel:data}) and its floor is not a $3\sigc$ criterion. Analysis areas are the cutout
totals actually used, not the full survey footprints. Beam FWHM and $\sigN$ are instrument/survey properties taken from the cited map data (Sects.~\ref{sec:planck:data}, \ref{sec:herschel:data}, \ref{sec:ccat}); $\sigc$, $\Scut/\sigc$, $\Nbeam$, the $\PD$ skewness, and the $u$ window are derived from the fitted count models via the beam formalism of Sect.~\ref{sec:theory:beam} and Appendix~\ref{app:beam}, not separately measured.}
\end{table}

\subsection{Sky and lens simulation}
\label{sec:sims:setup}

Sources are drawn as an inhomogeneous Poisson process from each count
model and injected individually down to $S_{\min} = 1$\,mJy wherever
the source density above that flux is below 200 per beam, which is
the case at the \textit{Herschel} and CCAT beams. At \textit{Planck}'s
beam it is not: there, sources below a split flux $S_{\rm split}$
(2--4\,mJy, set by $\Nbeam({>}S_{\rm split}) = 200$) are not injected
individually; their summed contribution is replaced by a Gaussian
field with the exact analytic mean and variance of the sub-split
population. The substitution is justified by the central limit
theorem (the sub-split population contributes hundreds of sources
per beam, so its beam convolution is Gaussian to high accuracy), and
it is validated rather than assumed: across all nine (instrument,
count model) pairs of this paper, the simulated pixel rms reproduces
the analytic $\sigtot$ to better than 0.5\%. The counts are truncated at
$\Scut = 100$\,mJy, above which the observed 350\,\mum\ counts are
dominated by strong lensing rather than the intrinsic population
\citep{Lima2010}, and at $S_{\min} = 1$\,mJy at the faint end; the
motivation for both truncations, and the sensitivity of the results
to them, are discussed in Appendix~\ref{app:counts}.

Lensing is implemented at the flux level with the resolved NFW
magnification profile of Eqs.~(\ref{eq:mu})--(\ref{eq:nfwkg}): the
local counts at image-plane position $\theta$ are
$n_{\mu(\theta)}(S)$, and the pooled aperture histogram is the
mixture Eq.~(\ref{eq:mumixture}). Aperture selection is therefore an
explicit design element of every lensed analysis in this paper, not
an afterthought. Our benchmark
lens throughout is a cluster of $M_{500} = 10^{15}\,\msun$ at
$z_{\rm l} = 0.5$ with sources at $z_{\rm s} = 2.0$, for which
$\theta_{500} = 3.43'$ and the aperture-averaged magnifications run
from $\langle\mu\rangle = 3.07$ ($r < 0.5\,\theta_{500}$) to $1.19$
($r < 2\,\theta_{500}$).
We do not address the sensitivity of the lensing predictions to the assumed source redshift here; our results are built on a fixed $z_{\rm s} = 2.0$ choice throughout.
However, one property of magnification mixtures deserves explicit note:
pooling apertures over clusters of different masses
and redshifts inflates the fitted $\hat\xi$ beyond the naive sample
average of single-cluster values, because the pooled histogram is a
mixture of differently stretched tails. Predictions for
heterogeneous samples are therefore always computed as mixtures over
the actual sample (as in
Sects.~\ref{sec:planck:lensed} and \ref{sec:herschel:lensed}), and
narrow magnification bins are the recommended design wherever sample
size permits.

\subsection{Validation of the analytic relations}
\label{sec:sims:validation}

\begin{figure}
\centering
\includegraphics[width=\hsize]{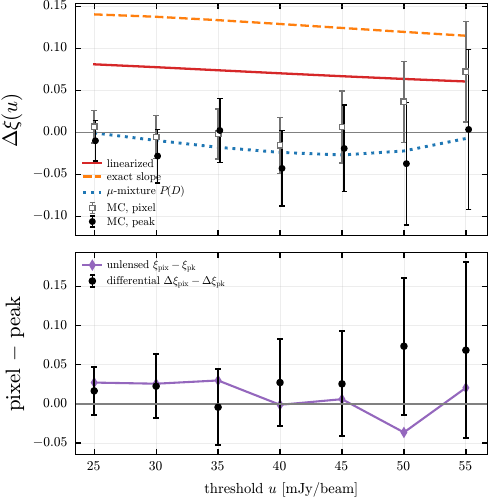}
\caption{Regime of validity of the analytic $\dxiu$ relations, for
the benchmark lens at $r < 1.0\,\theta_{500}$
($\langle\mu\rangle = 1.66$), Schechter counts. \emph{Top:} three
analytic routes, namely the linearized master relation
(Eq.~\ref{eq:master}; red solid), the exact slope evaluation
$1/(\eta(S_u/\mu)-1) - 1/(\eta(S_u)-1)$ without expansion in
$\ln\mu$ (orange dashed), and the exact $\mu$-mixture $\PD$ (blue dotted), against Monte Carlo
measured in both the pixel (open squares) and the declustered-peak (filled circles) convention, with bootstrap errors. The
two Monte Carlo conventions agree with each other and with the
$\mu$-mixture $\PD$; both disagree with the two shoulder-based
relations, which assume a one-to-one map between a threshold $u$ and
a source flux $S_u$ that does not hold at $\Nbeam \approx 5$.
\emph{Bottom:} the pixel-minus-peak convention offset, which reaches
$|0.036|$ in the unlensed $\xi$ (purple diamonds) but is consistent with zero in the
differential $\Delta\xi$ (black circles), the cancellation that justifies the
differential design.}
\label{fig:dxivalidation}
\end{figure}

For a first-in-field formalism the analytic relations must be
validated against exact computation, and the outcome of that
exercise is a statement about their domain of validity. Figure~\ref{fig:dxivalidation} compares four routes to
$\dxiu$: the linearized master relation (Eq.~\ref{eq:master}), the
exact single-$\mu$ slope evaluation without expansion in $\ln\mu$,
the exact area-weighted $\mu$-mixture $\PD$, and the spatially
resolved Monte Carlo in both estimator conventions.

The two Monte Carlo conventions agree with each other and with the
$\mu$-mixture $\PD$ (between $-0.001$ and $-0.027$ over
$u = 25$--$55\,\mjb$). Both differ from the linearized relation
($+0.081$ to $+0.061$) and from the exact slope evaluation ($+0.141$
to $+0.115$): a discrepancy larger than the signal itself, and of
opposite sign. The cause is not the linearization: the unexpanded
slope evaluation is further from the truth than the linearized one.
Nor is it the spread of $\mu$ across the aperture, which accounts for
only $\approx 0.02$ of the gap, nor proximity to the noise core,
since the gap survives with $\sigN$ set to zero and at
$u/\sigtot \approx 10$. It is the shoulder relation
Eq.~(\ref{eq:xislope}) itself: both analytic routes descend from
$\xi = 1/(\eta(S_u)-1)$, which presumes that a map value $u$ is
produced by a single source of flux $S_u$. In a confusion-limited map
with $\Nbeam \approx 5$ that value is a sum over several sources, the
correspondence $u \leftrightarrow S_u$ blurs, and a relation built on
it cannot be quantitative.

What survives is the structure of Eq.~(\ref{eq:master}), and there 
it survives exactly: $\Delta\xi$ is proportional to $\ln\mu$, it is
proportional to the curvature of the log-counts, and a pure power law
has zero curvature and therefore identically zero $\Delta\xi$ in the
master relation (the non-zero SPL response of
Fig.~\ref{fig:ccatlensedtheory} is the endpoint and lensed-core effect
that the relation does not contain). The
master relation is therefore retained throughout this paper as the
interpretive statement that it is, and it is never used to predict a
number. Every forecast in Sect.~\ref{sec:ccat} is computed from the exact
$\mu$-mixture $\PD$, which the Monte Carlo validates directly. The
physical consequences of that suppression (how large $\dxiu$
actually is, and why no aperture choice recovers it) are set out
in Sect.~\ref{sec:theory:twosided}.

Two systematic offsets between analytic and simulated curves are
understood, quantified, and inform how every result in this paper is
presented. The first is the \emph{pixel-versus-peak} offset of
Sect.~\ref{sec:theory:beam}, which the simulations quantify (a median
of $0.03$--$0.06$ in the fitted shape, reaching $0.09$--$0.22$ at the
ends of the ladder) and which the lower panel of
Fig.~\ref{fig:dxivalidation} shows to cancel in the lensed-minus-control
difference, both arms being declustered identically: the convention
offset is $0.02$--$0.03$ in the unlensed $\xi$, changing sign along
the ladder and reaching $-0.036$ at its top, and consistent with zero
in $\Delta\xi$. This is the origin of the standing
rule that every model curve shown against data is simulated at peak
level. The second is a small-sample
estimator effect: at low exceedance counts the GPD
maximum-likelihood $\hat\xi$ departs from its asymptotic sampling
distribution, Eq.~(\ref{eq:gpdvar}), the standard large-$n$ result $n\,\mathrm{Var}(\hat\xi) \to (1+\xi)^2$ \citep{Smith1984}. Wherever the exceedance count is
small (most acutely in the \textit{Planck} cluster apertures), we
therefore calibrate the estimator with end-to-end null Monte Carlos
rather than relying on asymptotics (Appendix~\ref{app:stats}).

\subsection{Source clustering}
\label{sec:sims:clustering}

The fiducial simulations place sources at random, uncorrelated
positions, while the real CIB is clustered. Whether this matters has
recently been examined for classical $\PD$ count inference by
\citet{Wang2026}, who find that clustering biases $\PD$-derived
counts in the 500\,\mum\ SPIRE band by a factor $1.63 \pm 0.27$
(with $1.25 \pm 0.38$ at 350\,\mum, a non-detection, and
$1.12 \pm 0.14$ at 250\,\mum), building on the SIDES simulation of
\citet{Bethermin2017}, which first showed that clustering distorts
the flux histogram at large beams. Whether the statistic proposed here inherits the same bias is a
natural question given those results, and the answer is instructive
precisely because it splits: the shape parameter $\xi$ is protected,
while simulation-derived error bars are not. We summarize the
argument here, and the full numerical tests are given in 
Appendix~\ref{app:clustering}.

One tempting argument must be disposed of first: that a one-point
statistic cannot respond to clustering at all. Writing the
$\PD$ for a general (non-Poisson) point process via the probability
generating functional \citep{Peebles1980}, Eq.~(\ref{eq:condon})
acquires an additive term for every connected correlation function
of the source field, and already the second cumulant picks up the
familiar shot-plus-clustering decomposition,
$c_2 = \tfrac12 q_2\,\Omega_{\rm beam} + q_1^2 W_{11}$, with
$W_{11}$ the beam-filtered integral of the angular power spectrum.
The one-point distribution is a functional of all the
correlation functions; the variance of a CIB map is demonstrably not
clustering-free. The correct argument is specific to the
tail, and it has two parts.

First, a rigorous part. The map value in a beam is a random sum of
source responses, and for responses with a regularly varying
(power-law) tail, precisely the regime $\xi > 0$ in which the GPD
analysis operates, the classical subexponential limit theorem
\citep[Thm.~1.3.9 of][]{Embrechts1997} states that
$P(\sum_k x_k > u) \sim \mathrm{E}[K]\,\bar F_x(u)$ as
$u \to \infty$: the asymptotic tail depends on the source count $K$
in a beam only through its mean. This is the
``single big jump'' principle (a far-tail exceedance is achieved by
one bright source, not by a conspiracy of many), and clustering,
which is over-dispersion of $K$ at fixed mean, drops out of the
limit entirely. The one-point intensity $\dnds$ is unchanged by
clustering by definition, and it alone fixes the tail.

Second, a quantitative part. The clustering corrections to the core
variance and to the tail carry different powers of the source
density, and their ratio is independent of the clustering amplitude
altogether: the unknown correlation strength cancels, leaving
Eq.~(\ref{eq:coretail}), which evaluates to two to four orders of
magnitude of protection across the \textit{Herschel} fitting window
(Appendix~\ref{app:clustering}). The tail is protected not because
clustering is weak but because the tail statistic is structurally
insensitive to it; the core, by contrast, is genuinely inflated, by
15\% in variance at SPIRE's $25''$ beam and by a factor $\approx 3.9$ at
\textit{Planck}'s $5'$ beam.

Direct simulation with lognormal-clustered source fields confirms
both statements (Appendix~\ref{app:clustering}). The measured $\hat\xi$
of clustered and Poisson-simulated images differs by $-0.02$ near the confusion
core, a shift decaying through zero by $u \approx 3\,\sigc$ and
unresolvable above it: the predicted pattern, and safely below
every fitting window used here. What clustering does \emph{not}
leave intact is the realization-to-realization scatter of the
exceedance counts, which is inflated by a factor $\approx 2$ in rms
at the floor of the fitting window (falling to $\approx 1.5$ at its top; Appendix~\ref{app:clustering}). The
consequence is that
all measured error bars of
this paper are cutout-level bootstraps on the real, clustered sky and
are correct as measured, while any uncertainty derived from a Poisson
simulation -- including the forecasts of Sect.~\ref{sec:ccat} --
must carry the corresponding factor-4 variance inflation, and does.
The inflation factor is calibrated at a $25''$-class beam and applied
globally; it is appropriate at \textit{Herschel}'s resolution, likely
conservative at CCAT's $15''$, and optimistic at \textit{Planck}'s
(Appendix~\ref{app:clustering}).

This also reconciles our approach with \citet{Wang2026}. Their quoted
500\,\mum\ distortion peaks at $1.5\,\sigc$, the
\emph{peak} of the histogram, exactly where the $\Nbeam$-weighted
clustering term lives and where classical $\PD$ inference draws its
constraining power. The GPD estimator deliberately discards that
region: it pays in statistical power and is repaid in insensitivity of the
shape to the two-point function. The two approaches are not in tension; they read
different parts of the same distribution, with different exposure to
clustering. The limiting assumption of our own treatment, shared with
theirs, is luminosity-independent clustering
(Appendix~\ref{app:clustering}).

\section{\textit{Planck} 857\,GHz: the resolution floor}
\label{sec:planck}

The \textit{Planck} arm of this work asks three questions of the best
available all-sky CIB data: does the POT/GPD machinery operate on a
real confusion-limited $\PD$; is the cluster-lensing modulation
measurable at $5'$ resolution; and, if not, are the data at least
consistent with the NFW $\mu$-mixture prediction computed from known cluster
masses? The answers are yes, no, and yes, each a
quantitative result rather than a qualitative reading. 
This section additionally delivers one
constraint that no differential measurement can access: an
absolute-background test of the count models. We present the
\textit{Planck} analysis first not because it is the most sensitive (it is the least) but
because it is where the statistical
methodology of the whole paper was developed and most stringently tested, and
because its null results define, quantitatively, what the higher
resolution of the later sections demonstrates.

\subsection{Data}
\label{sec:planck:data}

We use the 857\,GHz CIB maps of \citet{Lenz2019}, constructed from
\textit{Planck} data with SMICA CMB subtraction, an
\ion{H}{i}-based Galactic dust model removed, and point sources
masked above the \textit{Planck} detection limit
($\approx 600$--$700$\,mJy at 857\,GHz). Among the delivered mask
variants we adopt
$N_{\ion{H}{i}} < 4\times10^{20}\,\mathrm{cm}^{-2}$ with a 40\%
Galactic-plane cut, retaining $f_{\rm sky} = 0.338$
(13\,955\,deg$^2$): the looser column-density cut buys sky at the
cost of residual cirrus, a trade that is acceptable here precisely
because the shape statistic is location-invariant and the
non-Gaussian residuals are measured rather than assumed (below). All
fluxes are quoted in $\mjb$ with
$\Omega_{\rm beam} = 1.133\,(5')^2$, the map's delivered resolution after \citet{Lenz2019} smoothed it to match the SMICA CMB template used for the CMB subtraction above, not the native \textit{Planck} 857\,GHz instrumental beam ($\approx\!4.3'$); we work throughout with the map as delivered; the shape parameter $\xi$ is
invariant under this or any gain convention.

The noise budget of the masked map is measured, not modeled, and
its decomposition is critical for everything that follows. Half-difference
(odd--even survey) maps give an instrument noise of
$\sigN = 30.4\,\mjb$. The post-baseline Gaussian core of the map,
however, has width $\sigma_{\rm core} = 185.4\,\mjb$, and the
Poisson confusion prediction of the adopted counts is only
$\sigc = 95.1\,\mjb$. The bulk of the core, an effective Gaussian of
$159.2\,\mjb$ (the value that makes the simulated core reproduce
the measured one; Table~\ref{tab:ladder}), is therefore non-Poisson: clustered CIB and residual
cirrus, which at a $5'$ beam dominate the confusion budget (this is
the $\Rcore \approx 2.9$ regime, a factor $1 + \Rcore \approx 3.9$
in variance, where $\Rcore$ is the ratio of the clustered to the Poisson
contribution to the core variance defined in Appendix~\ref{app:clustering}, anticipated in
Sect.~\ref{sec:sims:clustering}). This measured decomposition is
carried explicitly into every simulation and null of this section.

The foreground cluster sample comprises the 772 PSZ2 clusters of the
\citet{Erler2018} compilation with $(RA, Dec, z, R_{500})$, of which
496 lie on the adopted mask (median $M_{500} =
4.5\times10^{14}\,\msun$, median $\theta_{500} = 4.8'$, redshifts
$0.05$--$0.89$). NFW magnification profiles are derived per cluster
from the catalog $R_{500}$, with the same mass--concentration
conversion as in Sect.~\ref{sec:sims:setup}. The identity of this
sample with that of \citet{Erler2018} is deliberate and pays off in
Sect.~\ref{sec:planck:dust}.

\subsection{The beam-dilution statement}
\label{sec:planck:dilution}

With the adopted Schechter counts truncated at $\Scut = 100$\,mJy,
the $5'$ beam gives $\sigc = 95.1\,\mjb$,
so $\Scut/\sigc = 1.05$, against $12.5$ at SPIRE (Table~\ref{tab:ladder}),
and this ratio frames the entire section. A \textit{Planck} beam
contains several hundred sources above 1\,mJy, so the central limit
theorem has all but Gaussianized the confusion $\PD$; the brightest
source the intrinsic counts allow cannot raise a single beam by even
one confusion sigma. The power-law shoulder on which
Eq.~(\ref{eq:xislope}) operates is not merely undetected at this
resolution; it is structurally absent, and no amount of sky area or
statistical care can recover it. There is a second, sharper way to
state the same limitation, by contrast with the \textit{Herschel}
arm. At \textit{Herschel} one can mask
the map above the model truncation flux, pass the models through the
same mask, and compare map and model on a common support. At \textit{Planck} that is impossible:
$\Scut = 100$\,mJy is $0.54\,\sigma_{\rm core}$, so masking the map
at the model's truncation would remove most of the sky; \emph{there
exists no flux at which map and model can be brought onto common
support by masking, because the model's bright cut lies inside the
noise}. The comparison must instead run the other way, extending the
model upward, which is how the null tests below are constructed. A
null result at \textit{Planck} is a measurement of beam dilution, not
a failure of the method.

\subsection{The unlensed \texorpdfstring{$\hat\xi(u)$}{xi(u)}: a
clean EVT characterization, not a counts measurement}
\label{sec:planck:unlensed}

\begin{figure}
\centering
\includegraphics[width=\hsize]{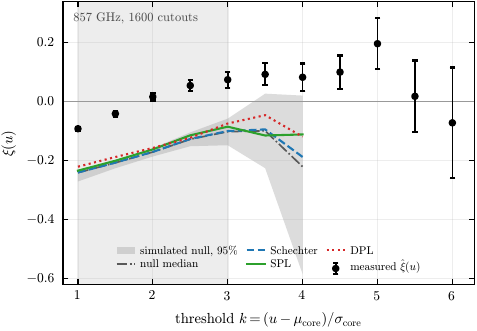}
\caption{Threshold-stability scan of $\hat\xi$ for the \textit{Planck}
857\,GHz $\PD$: the measured $\hat\xi(u)$ climbs out of the Gaussian
floor, crosses zero at $k \approx 1.9$, and reaches $+0.093\pm0.037$ at
$k=3.5$, where $u = \mu_{\rm core} + k\,\sigma_{\rm core}$,
with $\mu_{\rm core}$ and $\sigma_{\rm core}$ the mean and width of the
fitted Gaussian core ($\mu_{\rm core}$ is unrelated to the
magnification $\mu$). This core-standardized $k$, used throughout
Sect.~\ref{sec:planck}, differs from the confusion-scaled
$k = u/\sigc$ of Sect.~\ref{sec:theory:beam}. Black points with
bootstrap errors are the measurement; the gray band and the
dash-dotted line are the 95\% interval and the median of the
truncated-counts null Monte Carlo; the three colored curves are the
peak-level simulations of the Schechter (blue dashed), SPL (green
solid) and DPL (red dotted) count models; the vertical gray shading
marks the sub-window core $k < 3$, and the in-panel label gives the
band and the number of cutouts. The
threshold morphology (floor climb, plateau onset, mask ceiling) is the
first EVT characterization of a real confusion-limited $\PD$. The
companion mean-excess diagnostic is shown in
Fig.~\ref{fig:planckbright}, panel (a).}
\label{fig:planckstability}
\end{figure}

From 1600 cluster-avoiding $64'$ cutouts (1820\,deg$^2$, 231k beams,
53\,977 declustered peaks), the measured $\hat\xi(u)$ exhibits
exactly the threshold morphology the formalism predicts
(Fig.~\ref{fig:planckstability}): deep in the core the fitted shape
is negative, as it must be for a Gaussian-dominated sample;
it climbs monotonically as the threshold rises, crosses zero at
$k \approx 1.9$ (here and below $k$ counts core widths above the core
mean, $u = \mu_{\rm core} + k\,\sigma_{\rm core}$; Fig.~\ref{fig:planckstability}); and it reaches a positive excursion of
$\hat\xi = +0.093 \pm 0.037$ at $k = 3.5$ before the point-source
mask empties the tail. This is, as far as we are aware, the first
extreme-value characterization of a real confusion-limited $\PD$,
and every feature of it (the floor climb, the crossing, the
plateau onset, the ceiling) appears where the formalism says it
should. The machinery works on real data.

\begin{figure}
\centering
\includegraphics[width=\hsize]{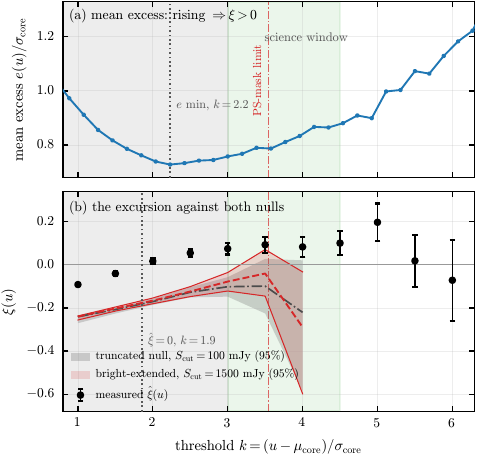}
\caption{Diagnosis of the positive $\hat\xi$ excursion, on a shared
threshold axis. \emph{(a)} The mean-excess function $e(u)$
(Sect.~\ref{sec:theory:pd}), normalized by $\sigma_{\rm core}$, of the same
declustered peaks. Above a valid threshold its slope is
$\mathrm{d}e/\mathrm{d}u = \xi/(1-\xi)$, so the turnover locates the
sign change of $\xi$ without fitting a GPD anywhere: it falls at $k = 2.2$, against
$k = 1.9$ for the fitted scan of Fig.~\ref{fig:planckstability}, and
the curve rises monotonically thereafter. \emph{(b)} The measured
$\hat\xi(u)$ against two simulated nulls: the truncated-counts null
from which the significance is quoted, and a bright-extended null in
which the simulated counts run to $\Scut = 1500$\,mJy. The extension
raises the null without lifting it to the measurement. The two
nulls are drawn as their median (gray dash-dotted and red dashed
lines) and 95\% band (gray and pink shading); the dotted vertical
lines are the two independent locators of the sign change, the $e$
minimum at $k = 2.2$ in (a) and the fitted $\hat\xi = 0$ crossing at
$k = 1.9$ in (b), annotated in each panel. Gray vertical shading
marks the sub-window core, green the science window
$k \in [3, 4.5]$, and the red dash-dotted vertical line the point-source mask
limit. The rise of $e(u)$ at the highest thresholds lies beyond both
the mask limit and the window, and is not used.}
\label{fig:planckbright}
\end{figure}

The interpretation of the positive excursion requires the null
calibration, and the answer is more modest than the raw
significance suggests. Against a
null simulated through the identical pipeline (sky realizations in
which sources are placed at random from each of the three count
models, truncated at $\Scut$, on top of the measured Gaussian budget
of Sect.~\ref{sec:planck:data}, i.e.\ instrument noise plus the
non-Poisson term, and then extracted, declustered, thresholded and
bootstrapped exactly as the data; the three models bound the range
quoted), the
excursion is formally significant at $6.0$--$7.1\sigma$
(covariance-corrected over the science window $k \in [3, 4.5]$;
Appendix~\ref{app:stats}). But the excursion is \emph{not} a
measurement of the sub-100\,mJy counts, and two diagnostics
establish what it is instead (Fig.~\ref{fig:planckbright}).

The first excludes the estimator. The mean-excess function
$e(u)$ of Sect.~\ref{sec:theory:pd} of the same
declustered peaks turns over at $k = 2.2$ and rises through the
science window, locating the sign change of $\xi$ within
$\Delta k \approx 0.4$ of the fitted scan's $k \approx 1.9$ and doing
so with no GPD fit involved. The positive tail is therefore in the
data, not in the small-sample bias of the maximum-likelihood
$\hat\xi$ (Appendix~\ref{app:stats}).

The second identifies its origin. Extending the simulated counts to
1500\,mJy (the bright, lensing-dominated population that the
truncated models exclude but the real sky contains below the
\textit{Planck} point-source mask) raises the null without lifting
it to the measurement: the measured-minus-null offset shrinks by
14\% at $k=3.0$ and 30\% at $k=3.5$, and the excursion clears the
bright-extended band at every threshold in the window. That the map
holds such a population is direct: above $u \simeq 650$\,mJy
($k = 3.5$, mid-window) it carries 1303 declustered peaks over
1820\,deg$^2$, where the double power law -- extrapolated above its
truncation -- predicts 64. These are map peaks rather than
identified sources. What remains after the bright extension is
attributable to clustered CIB and cirrus non-Gaussianity, which the
Poisson null does not contain and which the measured non-Poisson core
of Sect.~\ref{sec:planck:data} independently demonstrates to be
present. The correct reading is that the \textit{Planck}
$\hat\xi(u)$ measures the bright and clustered sky. We therefore
quote no count-model ranking from this arm: a ranking computed
mechanically from these data would reflect the bright tail and the
cirrus, not $\dnds$ below 100\,mJy, and would mislead.

This arm forced a new statistical machinery into our analysis:
correlated thresholds, the \citet{Hartlap2007} correction, and
end-to-end null Monte Carlos calibrating the small-sample bias of the
GPD estimator. It is set out in Appendix~\ref{app:stats} and applied
uniformly to both data arms. One number shows why it cannot be
skipped: with only $n_{\rm eff} \approx 2.2$--$2.8$ independent
thresholds out of the 10 sampled here, naive quadrature combination
across the \textit{Planck} science window would overstate this
excursion's significance by a factor $\sim 1.4$.

One constraint available at \textit{Planck} and nowhere else in this
paper deserves a sentence here, although its details belong with the
count models (Appendix~\ref{app:counts}). A differential tail measurement
is blind to the map mean, but \textit{Planck} measures the absolute
background: extrapolated to $S_{\min} = 1$\,mJy the single power law
over-produces the 857\,GHz monopole by a factor $5.4$ and is excluded
as a global count model outright, while the Schechter over-produces
it by $1.4\times$ using its $S > 1$\,mJy sources alone, placing
that tension in the 1--10\,mJy decade, which is precisely the decade
the tail statistic targets (Sect.~\ref{sec:theory:reach}). Absolute
photometry and the tail statistic are near-orthogonal in flux, and
this arm is the only one that supplies the former.

\subsection{The lensed \texorpdfstring{$\dxiu$}{Delta xi(u)}: an
upper limit, and a consistency test}
\label{sec:planck:lensed}

\begin{figure}
\centering
\includegraphics[width=\hsize]{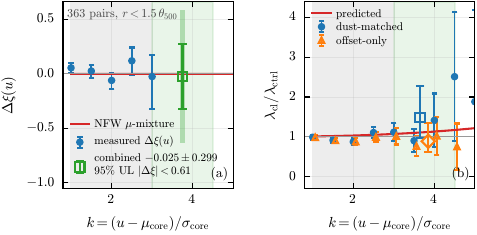}
\caption{The two channels of the \textit{Planck} cluster/control
differential at the default $r < 1.5\,\theta_{500}$ aperture, from
363 PSZ2 cluster/control pairs (primary sample, 100\% aperture
coverage). (a) The shape channel: the measured $\dxiu$ against the
NFW mixture prediction (three orders of magnitude inside the
errors). The open square is the window-combined value over
$k \in [3, 4.5]$ with its $1\sigma$ error (dark) and 95\% interval
(light band); because the GPD fit
requires a minimum number of exceedances in both arms, $\dxiu$ is
available only up to $k = 3$, whereas the amplitude ratios of panel (b) extend
over the full window. The remaining apertures
are shown in Fig.~\ref{fig:planckaperturedust}a and discussed in
Sect.~\ref{sec:planck:lensed}. (b) The amplitude channel: the
exceedance-rate ratio $\lambda_{\rm cl}/\lambda_{\rm ctrl}$ of
Eq.~(\ref{eq:ampchannel}) with dust-matched (circles) and
offset-only (triangles) controls, Poisson errors per threshold,
against the exceedance-rate ratio of the NFW $\mu$-mixture $\PD$
(red line; $1.07$--$1.17$ across the science window). The open square (dust-matched) and diamond (offset-only)
are the window-combined ratios with their 95\% intervals, $1.49$
$[0.98, 2.29]$ and $0.89$ $[0.63, 1.34]$: these two 
 straddle unity (Sect.~\ref{sec:planck:lensed}).
Thresholds at which either pool is empty are omitted. In both
panels the gray band is the sub-window core and the green band the
$k \in [3, 4.5]$ science window; the horizontal lines mark
$\Delta\xi = 0$ in (a) and a ratio of unity in (b).}
\label{fig:planckdxi}
\end{figure}

For the differential measurement, each cluster is paired with local
control fields drawn from an annulus $1\degr$--$3\degr$ away and
matched to the cluster field in the \citet{Lenz2019} dust-model column
density (``dust-matched'' controls; a variant with the offset alone and
no matching is carried as a check),
and the bootstrap is performed over pairs, so that sky
systematics cancel within every replicate rather than merely on
average. Aperture acceptance requires 100\% valid pixels on both
sides of every pair, a criterion adopted after the corresponding
null test (mask balance between cluster and control apertures)
revealed that the \textit{Planck} point-source mask preferentially
removes pixels near clusters, and that the affected apertures were
not harmless (see below). This yields 363 accepted pairs; the sample is
not materially reshaped relative to the looser 97\% criterion
(median $\theta_{500}$, $z$, and $M_{500}$ all essentially
unchanged), and the mask-balance null closes exactly, at ratio
1.000, by construction.

The result is a null at every measurable aperture
(Fig.~\ref{fig:planckdxi}): $\Delta\xi = -0.025 \pm 0.299$ at
$r < 1.5\,\theta_{500}$ (in practice the $k = 3$ threshold alone,
the only one in the science window at which both arms retain enough
exceedances for a GPD fit), $-0.124 \pm 0.204$ at $r < 2.0$, and
$-0.312 \pm 0.143$ at $r < 2.5$, with 95\% upper limits
$|\Delta\xi| < 0.61$, $0.52$, and $0.59$ respectively. The NFW
mixture prediction at $r < 1.5\,\theta_{500}$ for this sample is
$-3.2\times10^{-4}$: consistency, three orders of magnitude inside
the error. The two smallest apertures fail the exceedance floor
outright: at a $5'$ beam, an aperture of $r < 1.0\,\theta_{500}$
around a median PSZ2 cluster contains about two beams and, on
average, less than one declustered peak. This is the beam-dilution
statement of
Sect.~\ref{sec:planck:dilution} in its most severe form.

Two robustness findings on the lensing-modulated signal deserve to be highlighted. 
First, on the looser 97\%-coverage sample (378 pairs; hereafter
``the looser sample'') the same
measurement gives $\Delta\xi = -0.378 \pm 0.296$ at
$r < 1.5\,\theta_{500}$: removing the 4\% of clusters with masked
pixels inside the aperture moves the central value by $0.35$,
far more than two nearly identical samples should differ. The
$r<1.5$ result of the looser sample was carried by a handful of
masked-aperture clusters and was not robust; we quote the
100\%-coverage sample as primary and report the looser sample once,
as the robustness check that motivated the cut. 
Second, the mildly significant negative offset that persists at the largest
aperture ($-2.2\sigma$ at $r < 2.5\,\theta_{500}$) was investigated
with sample splits by mass, redshift, and angular size. In an
SZ-selected sample these three variables select roughly the same
clusters (the low-mass, low-$z$, and angularly large halves overlap
at 78--94\%), so the splits constitute one test, not three; the
variable with a mechanism attached is $\theta_{500}$, which sets the
sky area the aperture admits. The offset is carried entirely by the
angularly large half ($\Delta\xi = -0.52 \pm 0.19$, against
$+0.19 \pm 0.20$ for the compact half on the baseline sample), and
its sign is wrong for a lensing origin: the high-mass half, which
has the largest $\langle\ln\mu\rangle$, is the one consistent with
zero. It is also the wrong sign for cluster dust, which likewise
scales with mass. It is the right sign, and the right ordering, for residual
cirrus and clustered-CIB contamination growing with the admitted sky
area, which for these clusters reaches aperture radii of
$\sim 18'$. We therefore describe this offset as a sky-systematic
residual associated with large angular apertures, not as an
unexplained excursion, and in any case the predicted signal is
three orders of magnitude smaller.

In the amplitude channel
(Eq.~\ref{eq:ampchannel}; Fig.~\ref{fig:planckdxi}b), the
exceedance-rate ratio on the primary sample is $1.49$ with 95\% interval $[0.98, 2.29]$ using
dust-matched controls, and $0.89$ $[0.63, 1.34]$ using offset-only
controls: the two control constructions straddle unity, and we
conclude that no robust amplitude excess or deficit is detected. On
the looser sample the dust-matched ratio had appeared elevated, and
the disagreement between the two control constructions on the primary
sample is the reason we do not press any amplitude-channel claim at
\textit{Planck}.

\subsection{Cluster dust: a closure test against an external
measurement}
\label{sec:planck:dust}

\begin{figure}
\centering
\includegraphics[width=\hsize]{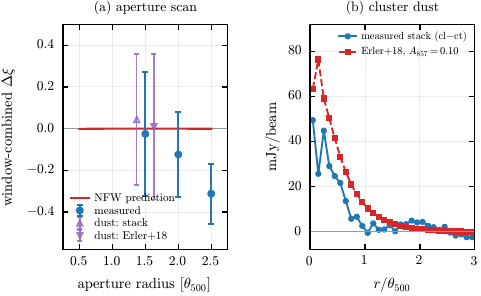}
\caption{\emph{(a)} Aperture dependence of the window-combined
\textit{Planck} $\dxiu$, against the NFW $\mu$-mixture prediction
(flat at $\sim\!10^{-4}$ on this scale). Filled triangles show the two
dust-corrected variants at the default $r < 1.5\,\theta_{500}$
aperture; all three treatments agree with one another and with zero.
\emph{(b)} The cluster-dust closure test: the measured stacked
cluster-minus-control profile of the primary (100\%-coverage) sample
(blue circles, in $\mjb$) against the \citet{Erler2018} GNFW
template at their published amplitude $A_{857} = 0.10$\,MJy\,sr$^{-1}$
(red squares, dashed) for the identical PSZ2
sample, forward-modeled through the beam and the baseline
subtraction (Sect.~\ref{sec:planck:dust}).}
\label{fig:planckaperturedust}
\end{figure}

Thermal dust emission from the clusters themselves~\citep{MontierGiard2005,PlanckXXIII2016,PlanckXLIII2016,McDonald2016,Melin2018}
is a foreground
to any cluster-differential measurement at 857\,GHz, and our sample
choice turns it from a nuisance into a validation. The 772 PSZ2
clusters used here are the identical sample whose stacked
submillimeter spectrum was measured by \citet{Erler2018}, so the
expected dust contamination of the cluster-side apertures is fixed
by an external measurement, with no free parameters and no
self-referential template built from the same data. Correcting the
cluster apertures with the \citet{Erler2018} amplitude changes the
measured $\dxiu$ by an amount consistent with that spectrum,
about a fifth of the (null) negative offset of the looser sample at
the default aperture, and on the primary sample the dust-corrected
variants agree with
the uncorrected one and with zero
(Fig.~\ref{fig:planckaperturedust}a). 
Figure~\ref{fig:planckaperturedust}b
shows the stacked cluster-minus-control profile of the primary sample
against the \citet{Erler2018} template forward-modeled through the
beam and the baseline: the measured excess runs at $60$--$75\%$ of
the template inside $\theta_{500}$, an implied
$A_{857} \approx 0.06$\,MJy\,sr$^{-1}$ against the published
$0.10 \pm 0.02$, while the looser sample gives $0.085$. The
difference between the two samples is the point-source mask at work
(the 100\% criterion preferentially removes clusters with compact
bright emission, i.e.\ the dustiest cores), and it bounds the sample
variance of the stack at the same level as the residual offset from
the template. Within that bounded sample variance the stack is
consistent with the template, and the test validates both the
pairing machinery and the dust treatment that the
\textit{Herschel} arm inherits, and we are not aware of a previous
use of a stacked cluster SED as an external dust control in a
lensing-differential context (cf.\ \citealt{Melin2018}).

\section{\textit{Herschel}/SPIRE 350 \texorpdfstring{\mum}{um}: the measurement}
\label{sec:herschel}

At SPIRE's $25''$ resolution the dynamic range of the shoulder opens for detection
($\Scut/\sigc = 12.5$), and the method operates as designed. This
section presents the first count-sensitive measurement of the GPD tail
statistic of a confusion-limited submillimeter map, its confrontation with the count
models, the physical origin of its most prominent feature, and the
first calibrated upper limits on the cluster-lensing modulation.

\subsection{Data and the matched filter}
\label{sec:herschel:data}

We use the \textit{Herschel} Extragalactic Legacy Project
\citep[HELP;][]{Shirley2019} homogenized SPIRE 350\,\mum\ map of the
GAMA-09 field: 53.44\,deg$^2$ of usable coverage
($9.7\times10^5$ beams) at $8''$ pixels, of which the unlensed
analysis uses the 25.8\,deg$^2$ covered by 103 non-overlapping $30'$
cutouts placed at random, at least $5'$ from any cluster of the eFEDS sample introduced below and
with their full padded filter footprints unmasked (the packing
fraction, $\sim\!55$--$60\%$ of the area left eligible by the
cluster-exclusion zones and the footprint requirement, is the jamming
limit of random non-overlapping placement, not a data cut), with an
effective point-source FWHM of $25.15''$ (the Condon-equivalent width
of the delivered matched-filter kernel, quoted as $25''$ elsewhere in
this paper; Appendix~\ref{app:matchedfilter}).
The choice of field is dictated by the foreground sample: GAMA-09
overlaps the eROSITA Final Equatorial Depth Survey
\citep[eFEDS;][]{Brunner2022, Liu2022}, the performance-verification
survey of \textit{SRG}/eROSITA, whose cluster catalog carries
weak-lensing-calibrated masses from the HSC-SSP analysis of
\citet{Chiu2022} (median $M_{500} \approx 10^{14}\,h^{-1}\msun$,
redshifts up to $\sim 1.3$). The overlap of a wide, homogeneous SPIRE
mosaic with a uniformly selected, mass-calibrated X-ray cluster
sample is what makes the differential measurement of
Sect.~\ref{sec:herschel:lensed} possible at all.

The raw SPIRE map cannot be thresholded directly: its declustered
peak population is a mixture of sky peaks and correlated
instrument-noise structure. The diagnostic is the lag-one
autocorrelation of the map pixels, the correlation coefficient
between adjacent pixels (Appendix~\ref{app:matchedfilter}): only $0.25$ for the raw map, in
which white instrument noise is superposed on the beam-correlated sky,
rising to $0.72$ once the matched filter described below is applied
($0.87$ for a pure $25.15''$ beam-correlated field; the forward model of
Appendix~\ref{app:matchedfilter:forward}, which passes simulated sky and
white noise through the delivered kernel, gives $0.78$). We
therefore apply the confusion-weighted matched filter delivered with
the HELP products (the inverse-variance-weighted filter of
\citealt{Chapin2011}, in which the kernel is built from the beam and
the confusion power spectrum, applied with the per-pixel noise as
weight), reproducing its operator from the delivered kernel
and applying it ourselves to padded cutouts of the primary signal map
(Appendix~\ref{app:matchedfilter}), which guarantees that one fixed
linear operator of identical form acts on cluster and control fields
alike, its only field dependence entering through the delivered
per-pixel error map (Appendix~\ref{app:matchedfilter}), and is
therefore common-mode in every differential statistic below. What the filter buys is not noise suppression but
the removal of the noise correlation structure, which is what
makes one-peak-per-beam declustering (the same one-FWHM
maximum-filter rule as in the \textit{Planck} analysis;
Appendix~\ref{app:beam}) applicable at all; the filtered
map retains $\sigN = 5.40\,\mjb$ against $\sigc = 7.98\,\mjb$
($\sigN/\sigc = 0.68$), so a Gaussian instrument floor is carried
explicitly in every forward model. Throughout, $\sigc$ denotes the
beam confusion width: the Condon prediction for a
$25.15''$ Gaussian beam, tabulated in Table~\ref{tab:ladder}, used
for the $3\,\sigc$ floor and reproduced by every simulation. The
distinct robust (median-absolute-deviation, MAD) confusion widths of the unfiltered and
filtered maps, $6.5$ and $5.2\,\mjb$, enter only the noise
budget of Appendix~\ref{app:matchedfilter}.

Thresholds in this section are specified in absolute flux
($\mjb$) rather than as multiples of the measured core width. The
reason is Eq.~(\ref{eq:xislope}): $\xiu$ reads the counts at the
flux the threshold selects, so an absolute-flux ladder makes the
same threshold mean the same intrinsic flux in every cutout, every
map variant, and every simulation leg, whereas a $k\sigma$ ladder
would re-map fluxes whenever the local noise varies. (The
core-anchored convention is retained at \textit{Planck}, where the
non-Poisson core dominates and no absolute count scale is
accessible anyway.)

\subsection{The unlensed \texorpdfstring{$\hat\xi(u)$}{xi(u)} and the
count models}
\label{sec:herschel:unlensed}

\begin{figure}
\centering
\includegraphics[width=\hsize]{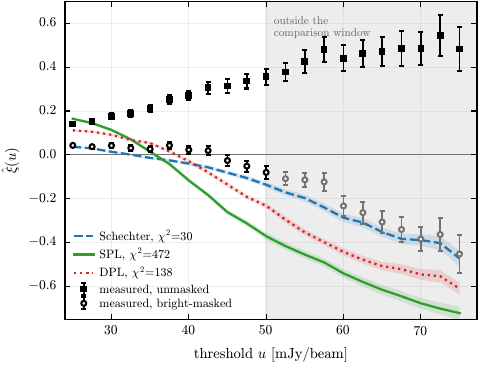}
\caption{The measured $\hat\xi(u)$ of the \textit{Herschel}/SPIRE
GAMA-09 350\,\mum\ map over
$u = 25$--$75\,\mjb$, in the bright-masked ($S > 100$\,mJy sources
removed; open circles, gray above the comparison window) and unmasked
(filled squares) variants. The three count models (Schechter blue
dashed, SPL green solid, DPL red dotted) are forward-modeled through
the same operations as the masked data: the intrinsic model below
$\Scut$ plus the observed bright population above it, the
matched-filter response of sky and noise, and the identical baseline
removal, bright mask, declustering and core anchoring
(Appendix~\ref{app:matchedfilter:forward}). The bands are the curves'
own map-bootstrap errors (400 maps, 100\,deg$^2$ per model); the
covariance-correct $\chi^2$ against the eleven bright-masked points
of the comparison window $u \le 50\,\mjb$ (unshaded) is given in the
legend; the horizontal line is $\xi = 0$. The unmasked
variant lies above every model because it contains the population
the mask removes (Sect.~\ref{sec:herschel:rise}).}
\label{fig:herschelxi}
\end{figure}

From 103 non-overlapping $30'$ cutouts spanning the field
(25.8\,deg$^2$), with cutout-level bootstrap
errors, the measured $\hat\xi(u)$ rises from $0.141 \pm 0.009$ at
$u = 25\,\mjb$ to $0.48 \pm 0.10$ at $75\,\mjb$
(Fig.~\ref{fig:herschelxi}): a positive, heavy-tailed shape
parameter, measured at percent-level precision deep in the
confusion regime (the first count-sensitive measurement of its kind) and an
$8.1\sigma$ departure from flatness whose interpretation is taken up
in
Sect.~\ref{sec:herschel:rise}. Because a handful of the brightest
objects in the field dominate the highest thresholds, we quote the
measurement in two variants: with the map masked above $\Scut = 100$\,mJy
(pixels above $100\,\mjb$ after baseline removal, grown by 3 pixels,
one FWHM: 70 sources and 0.10\% of the area) and unmasked. The masked
variant is the one compared with the count models, which are
forward-modeled through the identical filter, mask, declustering and
anchoring (Appendix~\ref{app:matchedfilter:forward}: a map-level mask
is not equivalent to truncating the injected counts, and the models
must carry the same mask); the unmasked variant contains the
population above $\Scut$ and is the subject of
Sect.~\ref{sec:herschel:rise}. As a systematic
check, repeating the analysis on the independently processed
nebular-filtered variant of the map changes no conclusion, and the
like-for-like comparison of predicted and measured confusion widths
agrees to 7--9\%, inside the count fits' own normalization
uncertainty (Appendix~\ref{app:matchedfilter}).

Confronting the bright-masked measurement with the three count
models (as simulated peak-level curves processed through the
identical pipeline, with no free parameters and no renormalization)
yields two results. First, the forward simulation of the
Schechter model tracks the level and run of the measurement (though
not to the formal $\chi^2$, below): $+0.037 \pm 0.004$ simulated
against $+0.043 \pm 0.009$ measured at $u = 25\,\mjb$, $-0.137 \pm
0.013$ against $-0.080 \pm 0.037$ at $50\,\mjb$, and, beyond the
comparison window, the steep decline of the masked measurement to
$-0.45 \pm 0.10$ at $75\,\mjb$, which is the map-level mask acting
as a hard endpoint (Appendix~\ref{app:matchedfilter:forward}),
reproduced as $-0.47 \pm 0.04$. A parameter-free forward model
reproducing the absolute level and run of a first-of-its-kind
statistic is among the strongest results of this paper. Second, with the full threshold
covariance, the model ranking is Schechter, then
DPL, then SPL: over the eleven thresholds at which the bright-masked
variant is compared, $\chi^2 = 30.1$, $138.4$ and $472.4$
respectively, with a Hartlap-corrected inverse covariance estimated
from 400 bootstrap replicates over the 103 cutouts (the correction
factor evaluated for the 103 independent cutouts rather than the 400
replicates, $\alpha_{\rm H} = 0.88$ instead of $0.97$, scales all
three to $27.4$, $125.9$ and $429.7$ and changes nothing below). The
model curves are treated as exact in this $\chi^2$; their own Monte
Carlo uncertainty is $0.004$ at $u = 25\,\mjb$, rising to $0.013$ at
$50\,\mjb$ (bands in Fig.~\ref{fig:herschelxi}), comparable to
the data errors, and the ranking is preserved at every $\pm1\sigma$
corner of the count-fit parameters of Table~\ref{tab:countfits}
(Schechter $\chi^2 \le 64$ against DPL $\ge 83$ when the analytic
pixel-level shifts are applied to the peak-level curves). The
comparison window ends at $u = 50\,\mjb$, half the mask level, so
that the threshold stays several exceedance scales below the mask
ceiling and the fitted shape reads the counts rather than the
endpoint; over all 21 thresholds of the masked ladder, where the
upper ten test the mask forward model of
Appendix~\ref{app:matchedfilter:forward} rather than the counts, the
$\chi^2$ are $74$, $199$ and $628$ for 21 degrees of freedom, with
the same ranking. Two
remarks place this ranking. It is a ranking of three pre-fitted
functional forms, not a measurement of $\eta$ with an error bar,
because the interpretive window of Sect.~\ref{sec:theory:beam} is
empty at this resolution; and since the SPL is already excluded by the
background monopole (Appendix~\ref{app:counts:monopole}), the live
comparison is Schechter against DPL. That the binned counts of
\citet{Bethermin2012} also prefer the Schechter form
(Table~\ref{tab:countfits}) is not what the tail adds: the models were
fitted to COSMOS and GOODS-N, the tail is measured on GAMA-09, and the
two measurements share neither sky area nor extraction method, so the tail
supplies an independent confirmation of the same preference (with the
regime caveat of Appendix~\ref{app:counts} in mind: the DPL's own
counts put it at $\Nbeam \approx 2$ at this beam, which the peak-level
simulation forward-models rather than assumes away). Generating the model curves at peak
level rather than from the pixel $\PD$ changes these numbers
substantially (the same comparison against pixel-level curves of
the truncated counts gives $51.3$, $132.2$ and $158.8$), and it
changes them in the direction that matters: the favored model's
$\chi^2/\nu$ moves from $4.7$ to $2.7$ ($2.5$ with the cutout-based
Hartlap factor), still formally a poor fit (the count-fit parameter
uncertainty of Appendix~\ref{app:counts}, worth $0.02$--$0.04$ in
$\xi$, is not in the $\chi^2$), and the $\Delta\chi^2$ separating it
from the SPL rises from $108$ to $442$. The convention error was
understating both the quality of the preferred fit and the
discrimination between models; the ranking itself is unaffected.
 Following Appendix~\ref{app:stats}, we quote
the full-covariance values together with $n_{\rm eff}$, and require
the ranking they give to be visible in the per-threshold comparison of
Fig.~\ref{fig:herschelxi} as well, which it is.

\subsection{The rise of \texorpdfstring{$\hat\xi(u)$}{xi(u)}:
the imprint of the lensed bright population in the tail}
\label{sec:herschel:rise}

\begin{figure}
\centering
\includegraphics[width=\hsize]{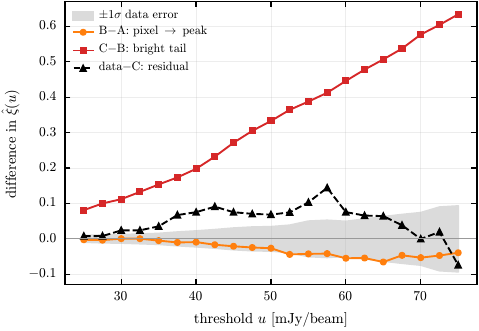}
\caption{Forward-simulation decomposition of the measured rise of
$\hat\xi(u)$, shown as the successive differences between simulation
legs: A is the pixel-level simulation of the truncated counts, B
the same at declustered-peak level, and C is B with the bright
($S > \Scut$) lensed tail added. The dominant term is the bright, strongly lensed population
above the intrinsic-counts truncation (C$-$B, red squares, contributing $+0.0102$\,mJy$^{-1}$ to
the slope $\mathrm{d}\hat\xi/\mathrm{d}u$); the pixel-versus-peak effect (B$-$A, orange circles) is small and of the wrong sign;
a residual (data$-$C, black triangles, dashed) of slope $+0.0022$\,mJy$^{-1}$
(all slopes fitted over the comparison window $u \le 50\,\mjb$) remains,
unattributed (Sect.~\ref{sec:herschel:rise}). The baseline of the
decomposition is the Gaussian-beam analytic model of the truncated
counts; its $-0.0001$\,mJy$^{-1}$ offset from leg A (text) is not drawn. The gray band is the $\pm1\sigma$ error of the
measurement, and the horizontal line marks zero. Legs A--C are
unmasked simulations (500 maps per leg) through the matched-filter
forward model of Appendix~\ref{app:matchedfilter:forward}; they are
compared with the unmasked measurement, and leg B is not the masked
Schechter curve of Fig.~\ref{fig:herschelxi}.}
\label{fig:h1c}
\end{figure}

The unmasked $\hat\xi(u)$ does something no intrinsic count model
can produce: it rises steeply with threshold, while all three
models, whose local slopes steepen toward their cutoffs, predict
a falling curve. Before interpreting this astrophysically, we
excluded the mundane explanations by direct simulation: the
pixel-versus-peak distinction contributes with the wrong sign (peaks give a slightly
lower $\xi$ than pixels here, slope $-0.0012$\,mJy$^{-1}$), and
threshold-anchoring and map-variant systematics are excluded at the
same level. The resolution is astrophysical, and it is the single
most instructive result of this arm. A forward simulation in which
the intrinsic (truncated) counts are supplemented by a bright tail
with the amplitude and slope of the observed $S \gtrsim 100$\,mJy
population, which at 350\,\mum\ is dominated by strongly lensed
galaxies \citep{Negrello2010, Wardlow2013, GonzalezNuevo2012, Lima2010}, reproduces the measured
rise quantitatively (Fig.~\ref{fig:h1c}). The measured slope over the comparison window ($u \le 50\,\mjb$) is
$\mathrm{d}\hat\xi/\mathrm{d}u \approx +0.0088$\,mJy$^{-1}$, i.e.\
$+0.011$ above the $-0.0025$ of the truncated analytic model; of
that excess the lensed bright population contributes $+0.0102$, the
pixel-versus-peak distinction $-0.0012$, and the
filter-and-simulator leg against the Gaussian-beam analytic model
$-0.0001$ (its level shift against that analytic pixel model, $-0.01$
to $-0.04$ inside the window and reversing above it, carries no slope), leaving a $+0.0022$
residual that we do not attribute: source clustering is not detected as a
contributor across this window at the precision of
Appendix~\ref{app:clustering} ($|\Delta\hat\xi| \lesssim 0.01$ above
$3\,\sigc$ in 96 simulated maps, and the window begins at
$3.1\,\sigc$), and the remaining candidates are the $4\%$ flux-scale
stretch of the matched filter (Appendix~\ref{app:matchedfilter}, worth
$|\Delta\xi| \lesssim 0.013$) and the luminosity dependence of
clustering that the Poisson-versus-clustered tests do not model. In
other words, $\xiu$ is reading the transition from the intrinsic
counts to the lensing-dominated bright tail: the imprint of the
lensed bright population is detected in the $\PD$ tail of this map.
The identification of that population as lensing-dominated rests on
the $500\,\mum$-selected samples \citep{Negrello2010, Wardlow2013} and
on the count models \citep{Lima2010, Bethermin2017}; at $350\,\mum$
the unmasked map also contains local galaxies above $100$\,mJy, which
the statistic does not distinguish from lensed ones. It is not the
differential cluster signal that the formalism of
Sect.~\ref{sec:theory:lensing} targets; it is the integrated
lensing effect of all the cosmic structures along all sightlines, appearing in
the unlensed channel. But it is a demonstration that the bright population, which the
$500\,\mum$ samples identify as lensing-dominated, shapes the
extreme-value statistics of the submillimeter sky at an amplitude our
statistic measures at high significance. It is also consistent with
the choice $\Scut \approx 100$\,mJy as the flux above which the
observed counts no longer trace the intrinsic population.

\subsection{The lensed \texorpdfstring{$\dxiu$}{Delta xi(u)} behind
eFEDS clusters: calibrated upper limits}
\label{sec:herschel:lensed}

\begin{figure*}
\centering
\includegraphics[width=\hsize]{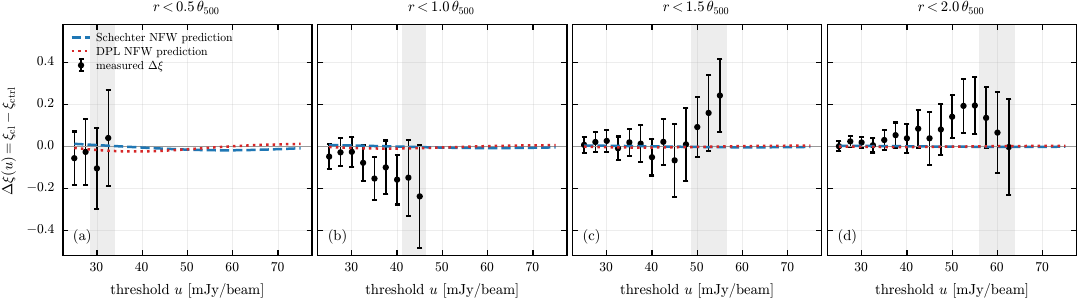}
\caption{The measured $\dxiu$ over 110 eFEDS cluster/control pairs in
four apertures \emph{(a)}--\emph{(d)} (black circles with bootstrap
errors), with the NFW predictions from
the \citet{Chiu2022} weak-lensing masses for the Schechter (blue
dashed) and DPL (red dotted) count models, the two not excluded by
the background monopole (Appendix~\ref{app:counts:monopole}); the
horizontal line marks zero (the predictions, of either sign and
below $2.5\times10^{-2}$ in magnitude, are indistinguishable from it
on this scale). The four apertures contain
$1258/1273$, $5104/5181$, $11449/11596$ and $20551/20624$ declustered
peaks at all levels in the cluster and control arms respectively; this
total peak density is set by the beam, not by lensing, so the near-equal
counts show that the control apertures sample the same peak density
as the cluster apertures, as the control placement is designed to
ensure (the exceedance counts above threshold, which lensing does
modulate, are discussed in the text). All apertures are consistent
with zero; the 95\% upper limits are $|\Delta\xi| < 0.37$, $0.20$,
$0.094$, and $0.067$ for $r < 0.5$, $1.0$, $1.5$, $2.0\,\theta_{500}$
respectively. Every plotted threshold (at least 50 exceedances in
both arms) enters the inverse-variance-weighted combination, with the
error propagated through the bootstrap replicates; the shaded
thresholds have fewer than 100 cluster-arm exceedances and carry
$31\%$, $4\%$, $4\%$ and $1\%$ of the weight in \emph{(a)}--\emph{(d)}.}
\label{fig:herscheldxi}
\end{figure*}

For the differential measurement we select, from the 542 clusters
and groups of the eFEDS catalog, the 110 with a \citet{Chiu2022}
weak-lensing mass and a redshift
whose padded filter footprints are 100\% unmasked (132 and 140
would survive at $\ge 99\%$ and $\ge 95\%$ coverage), the same
criterion whose importance the \textit{Planck} arm demonstrated
after the fact (Sect.~\ref{sec:planck:lensed}); here it was imposed
from the start, because masked pixels inside the kernel footprint
distort the matched filter in a non-common-mode way. Each cluster is
paired with a control field displaced by $1\degr$ in a random
direction, sharing its local depth and cirrus environment, and both
arms are processed identically; a dust-matched second control
construction of the \textit{Planck} kind (Sect.~\ref{sec:planck:lensed}) is
not used here, because at this high-latitude field and $25''$
resolution cirrus enters only through the common-mode baseline, and
the control re-randomization null below tests the construction
directly, with the threshold ladder
constructed on the lensed arm
(Sect.~\ref{sec:theory:lensing}). Cluster masses come from the
core-excised, $R > 0.5\,h^{-1}$Mpc weak-lensing calibration of
\citet{Chiu2022}, the variant least sensitive to cluster baryonic
physics.

No flux cut is applied in the differential measurement: both arms are
unmasked, so the observed bright population is common-mode, whereas the
$\mu$-mixture predictions of Sect.~\ref{sec:ccat} truncate the
intrinsic counts at $\Scut$, magnified to $\mu\Scut$ in the lensed arm.
The result (Fig.~\ref{fig:herscheldxi}) is a clean null at every
aperture, with 95\% upper limits $|\Delta\xi| < 0.37$, $0.20$,
$0.094$, and $0.067$ at $r < 0.5$, $1.0$, $1.5$, and
$2.0\,\theta_{500}$, and all null tests (control-versus-control,
mask balance between arms, and control re-randomization) passing. The null
is expected, and we knew it going into this analysis: for this cluster sample the
aperture-averaged $\langle\ln\mu\rangle$ runs from $0.25$
($r < 0.5\,\theta_{500}$) to $0.04$ ($r < 2\,\theta_{500}$), so the
NFW predictions ($3\times10^{-3}$ to $2.5\times10^{-2}$ in magnitude,
of either sign depending on aperture and model) lie an order of
magnitude below the achieved errors. 

The amplitude channel of Sect.~\ref{sec:theory:lensing}, which
at \textit{Planck} was measured through the exceedance-rate ratio, is
not analyzed separately here beyond a consistency check: the
cluster-to-control ratio of exceedance counts above $u = 25\,\mjb$
is $1.09 \pm 0.12$, $1.03 \pm 0.06$, $1.04 \pm 0.04$ and
$1.03 \pm 0.03$ at $r < 0.5$, $1.0$, $1.5$ and $2.0\,\theta_{500}$
(Poisson errors, to be inflated by the clustering factor of
Sect.~\ref{sec:sims:clustering}), positive at every aperture and at
every threshold from $25$ to $45\,\mjb$, and consistent both with unity and
with the few-percent to $\sim\!30\%$ magnification bias that
$\langle\mu^{\eta-2}\rangle$ implies at leading order for these
$\langle\ln\mu\rangle$ and $\eta(S_u) \approx 3$; the sample cannot distinguish the two, and a
calibrated amplitude measurement would require the dust-matched
control construction of Sect.~\ref{sec:planck:lensed}, which was not
built for this field.

The value of the
exercise is methodological and deliberate: it establishes the full
differential pipeline (pairing, common-mode filtering,
declustering, lensed-arm threshold construction, and
covariance-correct combination) as a calibrated instrument on real
data, its error budget anchors the forecast of
Sect.~\ref{sec:ccat}, and its limits are the first observational
constraints we know of on the lensing modulation of the
$\PD$ tail.

\section{CCAT/FYST forecast: where the regime opens}
\label{sec:ccat}

The Fred Young Submillimeter Telescope (FYST) of the CCAT Observatory
will survey $\sim\!100$\,deg$^2$ (with deeper 4--16\,deg$^2$ tiers)
at 850\,GHz with the Prime-Cam instrument \citep{CCAT2023}. We
emphasize for clarity that this is the FYST 850\,GHz module
($353$\,\mum, beam FWHM $= 15''$, the band matched to our 350\,\mum\
count models) and not the Prime-Cam channel labeled ``350\,GHz''
(which is 857\,\mum, with a $37''$ beam). We adopt the projected
first-quartile weather noise of $\sigN = 3.1\,\mjb$
\citep[Table~1 of][]{CCAT2023}, with the median-weather value
($4.9\,\mjb$) as a degraded leg. At $15''$ the confusion noise of
the Schechter counts is $\sigc = 4.76\,\mjb$, giving
$\sigtot = 5.68\,\mjb$ and $\Scut/\sigc = 21$: the widest shoulder
dynamic range of the three instruments we consider. The simulated survey area is
9.1\,deg$^2$, representative of a deep tier; all forecast statements
below scale with area in the usual $\sqrt{A}$ manner.

\subsection{Count-model discrimination: the near-term deliverable}
\label{sec:ccat:models}

\begin{figure}
\centering
\includegraphics[width=\hsize]{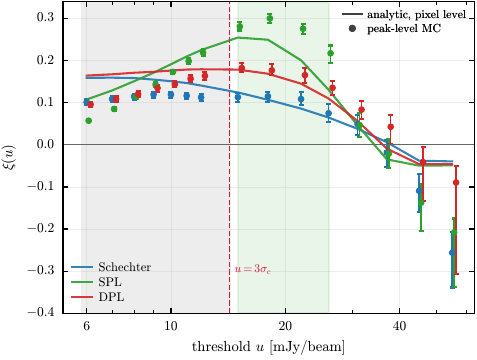}
\caption{Simulated CCAT/FYST unlensed $\hat\xi(u)$ for the three
count models (peak-level Monte Carlo, $15''$ beam,
$\sigN = 3.1\,\mjb$ first-quartile noise, 9.1\,deg$^2$) on the
extended threshold ladder reaching $u = 6\,\mjb$, with
bootstrap errors. The three models converge and cross at
$u \approx 8\,\mjb$, separate maximally near $u = 18\,\mjb$
(spread 0.185), and remain separated by $0.1$--$0.2$ over
$u = 12$--$26\,\mjb$ against errors of $0.005$--$0.03$. The gray
band is the region below the validity floor $u = 3\,\sigc =
14.3\,\mjb$ (dashed line), where the Gaussian core controls the fit,
and the green band the usable discrimination window
$u = 15$--$26\,\mjb$ of the text. The analytic
pixel-level curves (solid lines) are shown for comparison with the
peak-level Monte Carlo points, whose markers are offset slightly in
$u$ per model for legibility.} 
\label{fig:ccatxi}
\end{figure}

The unlensed $\xiu$ sweep is where CCAT is transformative
(Fig.~\ref{fig:ccatxi}). Over $u = 12$--$26\,\mjb$ the three count
models separate by $0.1$--$0.2$ in $\xi$, against simulated
bootstrap errors of $0.005$--$0.03$ from only $9.1$\,deg$^2$ of
survey: a decisive, many-sigma discrimination among functional
forms that current data cannot deliver. The physical reason is the
position of the count features relative to the confusion core: the
DPL break moves from $S_*/\sigc \approx 1.5$ at SPIRE resolution,
deep inside the Gaussian core, to $2.6$ at $15''$
(Appendix~\ref{app:counts}). What CCAT gains is not the break flux
itself, which remains below the validity floor, but the upper half of
the $0.7$\,dex slope transition that surrounds it
(Sect.~\ref{sec:theory:xi}): across the CCAT threshold ladder,
$u = 15$--$55\,\mjb$, the DPL's asymptotic $\xi$ still falls by
$0.16$, whereas across the SPIRE window, $25$--$75\,\mjb$, the same
model has already flattened to a swing of $0.04$.
The question of how the measured curve ends (decline through zero
or flatten onto a plateau) therefore becomes answerable at CCAT
resolution, and through the functional forms of the count models
(Sect.~\ref{sec:theory:reach}) it bears on the 1--10\,mJy decade
where two of the three models already conflict with the background
monopole
(Appendix~\ref{app:counts:monopole}).

The discrimination window is bounded by physics at both ends; the
simulated threshold ladder extends down to $6\,\mjb$, well below
the $3\,\sigc = 14.3\,\mjb$ validity floor, so that both boundaries are
seen. Toward high thresholds the separation
fades as all models steepen toward their cutoffs and the bright mask
approaches. Toward low thresholds the model curves converge
and cross at $u \approx 8\,\mjb \approx 1.7\,\sigc$: there the
Gaussian core, common to all models, controls the fitted shape, and
the model spread collapses to 0.006, indistinguishable even
against errors of 0.005--0.008. The usable discrimination window is
therefore $u \approx 15$--$26\,\mjb$, with its maximum leverage at
$u \approx 18\,\mjb$, and any analysis design that chases the
apparently smaller error bars at lower thresholds buys nothing: the
information is simply not there.

\subsection{The lensing modulation: a quantified requirement}
\label{sec:ccat:lensing}

\begin{figure*}
\centering
\includegraphics[width=\hsize]{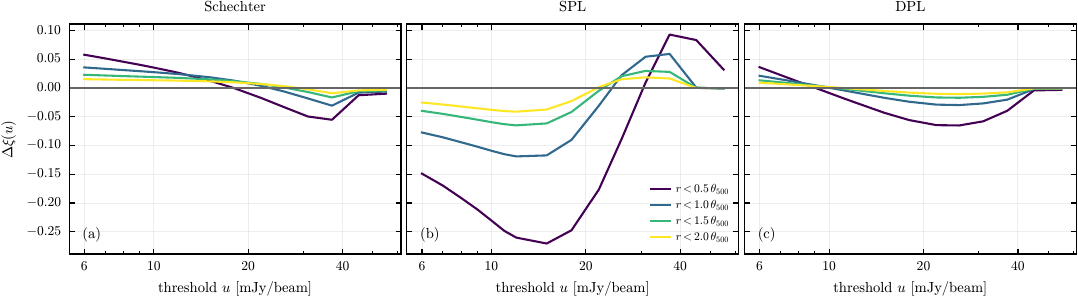}
\caption{Exact analytic $\dxiu$ of the benchmark lens
($M_{500} = 10^{15}\,\msun$ at $z=0.5$, $\theta_{500} = 3.43'$) at
CCAT resolution, per count model (\emph{(a)} Schechter, \emph{(b)} SPL,
\emph{(c)} DPL) and
aperture (color-coded as in the legend of panel b; the horizontal
line is $\Delta\xi = 0$): a deterministic single-cluster statement, free of Monte
Carlo noise. The four apertures carry area-weighted mean
magnifications $\langle\mu\rangle = 3.07$, $1.66$, $1.33$ and
$1.19$ for $r < 0.5$, $1.0$, $1.5$, $2.0\,\theta_{500}$; these are
properties of the lens alone and are common to all three panels.
The large response of the SPL in panel (b), for a model whose
count curvature $\kappa$ vanishes identically, is not a count-curvature
signal: it arises from the two scale-breaking ingredients of the exact
$\mu$-mixture, the finite endpoint at $\Scut$ (magnified to
$\mu\Scut$ in the lensed arm) and the lensed confusion core
$\sigc(\mu) = \sqrt{\mu}\,\sigc$ (Eq.~\ref{eq:sigcmu}), and its sign
reversal along the ladder is the truncation taking over at high
$u/\sigc$, which is also why the \textit{Planck} and \textit{Herschel}
optima of Table~\ref{tab:n3sigma} carry the opposite sign to CCAT's.}
\label{fig:ccatlensedtheory}
\end{figure*}

For the differential cluster signal we state a requirement 
to guide future detection efforts. 
Each instrument arm delivers three quantities: the
exact analytic $\dxiu$ of the common benchmark lens
(Fig.~\ref{fig:ccatlensedtheory}), a deterministic single-cluster
statement; the per-cluster error $\sigma_1$ from a three-arm
(lens/control/null) Monte Carlo; and the resulting number of
benchmark clusters for a $3\sigma$ detection,
$N_{3\sigma} = (3\sigma_1/\Delta\xi_{\rm theory})^2$, quoted with
the clustering-inflated variance of
Sect.~\ref{sec:sims:clustering} (Table~\ref{tab:n3sigma}). The
forecast thus rests on the analytic single-cluster signal, with the
Monte Carlo supplying only the error, a division of labor we make
explicit because the predicted signal at these thresholds is below
the Monte Carlo's own resolving power.

When the CCAT threshold ladder is extended toward the core, the
naive figure of merit improves without bound: $|\Delta\xi_{\rm
theory}|$ rises and $\sigma_1$ falls monotonically as
$u \to \sigc$, and no interior optimum exists. The
``optimal'' threshold of any such forecast is set entirely by where
the grid stops, unless a validity criterion is imposed. We impose
the floor $u \ge 3\,\sigc$, the boundary below which the fitted
shape is no longer count-dominated. Below it the clustering bias of
$\hat\xi$ switches on (Sect.~\ref{sec:sims:clustering}), the
Gaussian core controls the fit (the model-convergence point of
Sect.~\ref{sec:ccat:models} is the same boundary seen from the
discrimination side), and the pixel-versus-peak cancellation that
underwrites the $\Delta\xi$ construction degrades. With the floor
imposed, the requirement is
$N_{3\sigma} = 7.6\times10^{3}$ benchmark clusters, at
$u = 15\,\mjb$ and $r < 1.5\,\theta_{500}$, twice the number the
unextended ladder would have given, since that grid's edge sat below
the validity floor and its apparent optimum was therefore optimistic
rather than conservative.

The requirement improves by a factor $\sim 1300$ from
\textit{Planck} to CCAT (Table~\ref{tab:n3sigma}), but even CCAT's
$7.6\times10^{3}$ exceeds the number of $10^{15}\,\msun$ clusters on
the sky \citep[e.g.,][]{Holz2012}. 
Relaxing to lower masses does not obviously help, because
$\Delta\xi \propto \langle\ln\mu\rangle$ falls faster than cluster
abundance rises. The conclusion is that the cluster-lensing
$\dxiu$ is not detectable with the surveys considered here; the
calibrated upper-limit methodology of
Sect.~\ref{sec:herschel:lensed}, and the quantified requirement with
its stopping rule, are the design 
guidelines for a possible future detection. 
%
Since $\Delta\xi \propto \langle\ln\mu\rangle$, the leverage lies in
raising the aperture-averaged magnification, which means small
apertures on well-resolved, massive clusters, so \emph{resolution is
again the key factor}, both to resolve the aperture and to lower the
confusion floor inside it. The natural target is therefore not the
wide survey tier but deep, small-field observations: the CCAT
ultra-deep 4--16\,deg$^2$ tier pointed at the most massive known
clusters, 
or existing high-resolution data on cluster fields 
such as the ALMA Lensing
  Cluster Survey \citep{Fujimoto2024}.

\begin{table*}
\caption{Cluster-lensing requirement on the common benchmark lens
($M_{500} = 10^{15}\,\msun$ at $z_{\rm l}=0.5$): the number of
clusters $N_{3\sigma}$ for a $3\sigma$ detection of $\dxiu$ at the
optimal valid threshold and aperture, with clustering-inflated
errors.}
\label{tab:n3sigma}
\centering
\begin{tabular}{l c c c c}
\hline\hline
 & aperture & $u$ [$\mjb$] & $\Delta\xi_{\rm theory}$ & $N_{3\sigma}$ \\
\hline
\textit{Planck} & $r<1.0\,\theta_{500}$ & 456 & $-0.015$ & $9.9\times10^{6}$ \\
\textit{Herschel} & $r<0.5\,\theta_{500}$ & 30 & $-0.037$ & $2.8\times10^{4}$ \\
CCAT/FYST & $r<1.5\,\theta_{500}$ & 15 & $+0.015$ & $7.6\times10^{3}$ \\
\hline
\end{tabular}
\tablefoot{Columns: instrument; optimal aperture; optimal
threshold; the exact $\mu$-mixture $\PD$ prediction $\Delta\xi_{\rm theory}$
of the benchmark lens for the Schechter counts (Sect.~\ref{sec:sims:validation};
its sign reverses along the threshold ladder, Fig.~\ref{fig:ccatlensedtheory}a);
and the required cluster count. All entries impose the validity floor $u \ge 3\,\sigc$
(see text); the \textit{Planck} and \textit{Herschel} optima already
satisfy it. The \textit{Planck} row independently reproduces the
measured null of Sect.~\ref{sec:planck:lensed}. Poisson-error values
are a factor $\sim 4$ smaller in $N$; the clustering-inflated values
are quoted throughout
(Sect.~\ref{sec:sims:clustering}).}
\end{table*}

\subsection{Beam versus sensitivity: what actually buys the
improvement}
\label{sec:ccat:factorial}

Which instrument property actually buys the three-orders-of-magnitude
ladder of Table~\ref{tab:n3sigma}, and would the same gain be
available to a deeper survey at the present resolution? A two-by-two
numerical experiment answers both questions: we cross the
\{\textit{Herschel}, CCAT\} beams ($25''$, $15''$) with the
\{\textit{Herschel}, CCAT\} noise levels ($\sigN = 5.40$,
$3.10\,\mjb$) with all else fixed (common aperture
$r < 1.0\,\theta_{500}$, Schechter counts, threshold ladders matched
in $u/\sigtot$), and record $N_{3\sigma}$ at the optimal valid
threshold of each corner: $1.24\times10^4$ and $1.25\times10^4$ at
the \textit{Herschel} beam with \textit{Herschel} and CCAT noise,
and $4.0\times10^3$ and $1.66\times10^3$ at the CCAT beam. These
four values carry Poisson errors only. The clustering inflation of
Table~\ref{tab:n3sigma} is approximately common to the four corners
at matched $u/\sigtot$, so it cancels in their ratios. (The tabulated
optima of Table~\ref{tab:n3sigma}, which sit at different apertures
and thresholds, differ by a smaller factor of $3.7$.) The
factor $7.5$ between the two instruments at matched geometry
decomposes into $4.8\times$ from the beam and $1.55\times$ from
sensitivity as geometric means, but the means conceal the result:
the noise improvement is worth $0.99\times$ at the \textit{Herschel}
beam and $2.4\times$ at the CCAT beam, and the beam is worth
$3.1\times$ at \textit{Herschel} noise and $7.5\times$ at CCAT noise.
The interaction is the physically interesting part, and it has two
causes.

The first is the confusion floor. The two fiducial surveys sit at
the same place relative to it, $\sigN/\sigc = 0.68$ (SPIRE) and
$0.65$ (CCAT first-quartile weather): both are designed to integrate
down to the confusion limit and no further, which is the sensible
stopping point for any statistic of a confused map. The two
off-diagonal corners are the informative ones. At the $25''$ beam,
the CCAT noise level would take the map to $\sigN/\sigc = 0.39$,
well below a floor that the beam has already fixed; the total core
width changes by only 11\%, and the requirement does not move.
At the $15''$ beam, \textit{Herschel}'s noise level would leave the map
detector-dominated ($\sigN/\sigc = 1.13$), and the CCAT noise level
brings it back to the floor; that is the step that pays. Detector
sensitivity, which in practice means integration time, is
therefore worth buying only until $\sigN$ reaches the $\sigc$ that
the beam sets, and the beam decides how much sensitivity is worth
buying at all.

The second cause is the shape of $\dxiu$ along the threshold ladder,
and it is why the \textit{Herschel}-beam result is exactly unity
rather than merely small. Lowering the noise does reduce the
per-cluster error at $25''$ ($\sigma_1$ from $0.83$ to $0.69$, worth
a factor $1.5$ in $N_{3\sigma}$ on its own), but the ladder, matched in
$u/\sigtot$, moves the optimal threshold from $38.5$ to
$34.2\,\mjb$, toward the sign reversal of the Schechter $\dxiu$ at
this resolution (Fig.~\ref{fig:ccatlensedtheory}a shows the same
reversal at CCAT resolution), and $|\Delta\xi_{\rm theory}|$ falls
from $0.0225$ to $0.0185$, worth a factor $0.68$; the two cancel. At
$15''$ the ladder sits on the positive lobe of the curve, where
$|\Delta\xi_{\rm theory}|$ grows toward the validity floor, so the
same noise step improves the signal ($0.0159 \to 0.0204$) and the
error ($0.34 \to 0.28$) together, a factor $2.4$. Where a given
instrument's ladder lands on the $\dxiu$ curve is a property of the
count model and the beam, not of the detector, and the factorial was
run for the Schechter counts only; for the SPL and DPL the curves of
Fig.~\ref{fig:ccatlensedtheory} differ in sign structure, and the
cancellation at $25''$ would not be exact. The regime statement of
the previous paragraph is general; the precise $0.99$ is not.

The beam itself enters through two channels of comparable impact: more
independent peaks per cluster aperture ($\propto \mathrm{FWHM}^{-2}$),
and a linearly lower confusion floor ($\sigc \propto \mathrm{FWHM}$)
that admits a deeper threshold ladder where far more declustered
peaks survive. Of these, the first accounts for a factor
$(25.15/15)^2 = 2.8$ and the second, less obvious channel for the
remaining $\sim 1.7$.

The same accounting applies to the unlensed $\xiu$, and it is why 
the count-model
discrimination of Sect.~\ref{sec:ccat:models} is a resolution-driven
result rather than sensitivity-driven. The lower edge of
the discrimination window is the validity floor $3\,\sigc$, which the
beam sets and no integration time can lower: at $25''$ the DPL's
slope transition sits at $S_*/\sigc = 1.5$, inside the Gaussian core,
and a SPIRE-resolution survey of any depth would read only the
flattened end of it (a swing of $0.04$ across its window against
$0.16$ at $15''$, Sect.~\ref{sec:ccat:models}). The beam also
supplies $2.8\times$ more declustered peaks per unit area. Sensitivity
enters only through the core width $\sigtot$, which dilutes the
count-driven shape near the floor; at the CCAT beam, \textit{Herschel}'s
noise level would widen the core by 27\% and push the model-convergence
point of Fig.~\ref{fig:ccatxi} outward, but would not move the floor.
We have not simulated the unlensed factorial and quote no number for
this effect. That the CCAT window happens to fall on the upper half of
the DPL transition, where the three models separate maximally, is a
coincidence of the $350\,\mum$ counts and the $15''$ beam rather than a
property of the method; a different beam or a different band would
place the window elsewhere on the curves of Fig.~\ref{fig:ccatxi},
and the ladder should be designed against those curves. This
decomposition, rather than any single forecast number, is the
instrument-design statement of this paper.

\section{Discussion and conclusions}
\label{sec:discussion}

\subsection{The method's place among existing techniques}
\label{sec:discussion:place}

The case this paper makes rests on a three-step chain. The confusion
$\PD$ is classically treated only as a source of flux-integrated
count moments, but its tail carries a flux-resolved,
normalization-free record of the local logarithmic slope of the
source counts, recoverable as the GPD shape $\xiu$. Its lensing
modulation $\dxiu$ then recovers the local curvature of the counts:
how far $\ln(\dnds)$ bends away from a pure power law in $\ln S$.
Together, the two deliver the \emph{functional form} of $\dnds$ below
the individual-detection limit. 

The two halves of that chain are, however, unequal, by the factor
$2$--$14$ that Sect.~\ref{sec:theory:twosided} quantifies: the
unlensed sweep compares count models differing by $0.1$--$0.2$ in
$\xi$, while the lensing modulation is a curvature effect left at
$\dxiu \sim 0.01$--$0.02$ for a $10^{15}\,\msun$ cluster, with no
aperture choice able to recover it. That asymmetry (power-law response in amplitude, logarithmic in shape) is a property of
the observable rather than of the estimator, and it is what makes the
deliverable of the cluster analyses a calibrated upper limit -- plus a
quantified requirement -- rather than a detection. It is also why the
near-term return of the method lies in the unlensed count-shape
sweep, and why we present the instrument requirement for $\dxiu$ as a
design statement for future surveys rather than as a target for
existing or upcoming data.

On the relationship to existing methods, two clarifications must be made, 
because both invite overclaiming. First, at pixel level $\xiu$ is a
deterministic functional of the $\PD$, and at declustered-peak level a
functional of the same map: in either case it contains no information
the map does not, it is capped by the same information ceiling, and it
must never be combined with a full $\PD$ fit as though the two were
independent constraints; the two are strongly correlated, and
treating them as separate data would spuriously tighten errors. The
rigorous architecture for a joint analysis is a single forward model
of the counts predicting both the resolved bright-end counts and the
$\PD$ tail, fit once with the correct covariance, with $\xiu$
understood as an interpretable, robust summary of the faint-end
contribution. What the summary brings to the table is its
invariances: immunity, on the shoulder and with thresholds anchored
to the core (Sect.~\ref{sec:theory:pd}), to the map mean, the gain,
and the count normalization. This enables a freedom from the core-profile,
calibration, and noise-decomposition modeling that the full fit
cannot avoid. That advantage holds for unlensed fields just as much
as for the lensing program, which is what makes the two extractions
most valuable in comparison with each other. Full forward modeling
is the more powerful route when the instrument is perfectly
characterized, the tail sweep the more trustworthy one when it is
not, and agreement between them is itself a validation.

Second, the fluxes the statistic reads directly are those of
order the threshold, $25$--$75$\,mJy at SPIRE resolution and
$15$--$55$\,mJy at CCAT; the 1--10\,mJy decade at 350\,\mum\ is
reached through the functional forms of the count models and through
$\sigc$, and the sub-mJy population not at all: the tail is
structurally blind to the latter, whereas the CIB monopole is
exquisitely sensitive to it (Sect.~\ref{sec:theory:reach}). The two
probes are therefore nearly orthogonal in flux, and they are already
in productive tension, since two of the three count models over-produce the
measured monopole from their $S > 1$\,mJy population alone
(Appendix~\ref{app:counts:monopole}); that tension reflects our choice
of count models, which were selected to span the functional forms the
tail statistic must discriminate rather than for their consistency
with every observational constraint (Fig.~\ref{fig:counts};
Appendix~\ref{app:counts}). A joint monopole--$\xiu$ analysis
at CCAT depth is, in our assessment, the strongest available
constraint on the faint-end functional form, and the natural sequel
to this paper.

The various modeling assumptions, and in certain cases limitations, are stated throughout; we collect those here for convenience: 
\begin{enumerate}
\item \textit{Leading-order theory.} The analytic relations are
  leading-order on the shoulder and in $\ln\mu$, with all
  quantitative inference routed through exact peak-level simulation.
\item \textit{External magnification.} The magnification is an
  external input, with mass-model uncertainty to be marginalized.
\item \textit{Simplified lens model.} The lens model is a circular
  NFW halo, without ellipticity, substructure, or ray deflection
  (multiple images and arcs); this is adequate because the apertures
  deliberately exclude the strong-lensing region where these matter:
  the area at $\mu \gtrsim 5$ is a few percent of the innermost aperture
  and a fraction of a percent beyond $\theta_{500}$
  (Sect.~\ref{sec:theory:lensing}).
\item \textit{Count-model anchoring.} The count models are anchored
  to data only over 6--94.6\,mJy, with all fainter behavior an
  extrapolation.
\item \textit{Fixed source redshift.} A single source redshift
  ($z_{\rm s} = 2.0$) is assumed in the lensing predictions. Because
  the lensing efficiency $D_{\rm ls}/D_{\rm s}$ rises with $z_{\rm s}$,
  sources beyond $z = 2$ are magnified more, and those in front of it
  less, than assumed, so the sign of the resulting error in
  $\langle\ln\mu\rangle$ depends on the source redshift distribution
  behind each lens; for the highest-redshift eFEDS lenses
  ($z_{\rm l} \sim 1$), whose efficiency at $z_{\rm s} = 2$ is small,
  the assumption is more likely to under- than to overstate the
  signal.
\item \textit{Poisson simulations.} The fiducial simulations are
  Poisson (sources placed at random, not drawn from a cosmological
  simulation with clustering). Clustering effects are, however,
  bounded quantitatively, and the corresponding variance inflation is
  applied to every forecast.
\item \textit{Luminosity-independent clustering.} The clustering
  tests assume that all sources share one modulation field, whereas
  brighter CIB sources cluster more strongly \citep{Viero2013}; the
  margin of Table~\ref{tab:coretail} absorbs a large factor of
  refinement, but this is the limiting assumption of the clustering
  treatment (Appendix~\ref{app:clustering}), and the clustered
  simulations were run at the SPIRE beam only.
\end{enumerate}

\subsection{Main conclusions}
\label{sec:discussion:conclusions}

We summarize the paper's findings below, itemized by topic and
then by instrument. The headline is simple: the method performs
exactly where the beam allows it to (structurally inoperative at
\textit{Planck}'s $5'$ resolution, working as designed at
\textit{Herschel}/SPIRE, and reaching full count-model discrimination
at CCAT/FYST), while the differential cluster-lensing signal itself
remains a calibrated design target rather than a present-day
detection. The seven points below give the quantitative details behind
that headline.

\begin{enumerate}
\item \textbf{The formalism.} The peaks-over-threshold/GPD analysis
  of the $\PD$ tail delivers a flux-resolved, normalization-free
  readout of the local count slope, $\xi(u) \approx 1/(\eta(S_u)-1)$,
  and its gravitational-lensing modulation reads the local count
  curvature through the master relation
  $\Delta\xi \approx -\ln\mu\,\kappa/(\eta-1)^2$, with the amplitude
  channel carrying the classical magnification bias separately. The
  structure of both relations, the $1/(\eta-1)$ dependence and the
  proportionality of $\Delta\xi$ to $\ln\mu$ times the count curvature, is
  confirmed against exact computation and end-to-end simulation, but
  neither is quantitatively accurate at the source densities of real
  maps (Table~\ref{tab:dxisuppression}), and every numerical prediction
  in this paper is made from the exact $\mu$-mixture $\PD$ at peak level.
\item \textbf{The feasibility criterion.} A single dimensionless
  ratio, $\Scut/\sigc$ (the brightest admissible source flux over
  the confusion noise), governs whether the method can operate:
  $1.05$ at \textit{Planck} (300$''$), $12.5$ at
  \textit{Herschel}/SPIRE (25$''$), $21$ at CCAT/FYST (15$''$).
   The
  \emph{measurement} window, in which forward-simulated models are
  compared with data, additionally requires
  $\Scut/\sigc \gtrsim 3$ and closes entirely at coarse beams; it
  is void at \textit{Planck}. The narrower \emph{interpretive}
  window, over which $\xi \approx 1/(\eta-1)$ may be read
  quantitatively, requires $\Scut/\sigc \gtrsim 30$ and is empty at
  all three instruments: at present-day resolution the analytic
  relations can orient the interpretation but cannot supply a number,
  which is why every model curve in this paper is simulated.
\item \textbf{Clustering robustness.} Source clustering leaves the
  tail shape unbiased to within $\pm0.01$ above $u \approx 3\,\sigc$
  (a $-0.02$ shift below, decaying through zero), as predicted by the single-big-jump
  principle and verified by direct clustered simulation; but it
  inflates exceedance-count scatter by $\sim 2\times$ in rms. This reconciles our approach with the clustering bias
  reported for classical $\PD$ count inference by \citet{Wang2026},
  which lives in the blended core that the tail statistic discards.
\item \textbf{\textit{Planck} at $5'$: the resolution floor.} The
  machinery operates on a real confusion-limited $\PD$, the first
  EVT characterization of one, but the shoulder is structurally
  absent at this beam, and the measured $6$--$7\sigma$ positive
  excursion of $\hat\xi(u)$ is a measurement of the bright and
  clustered sky, not of the faint counts, so we quote no count-model
  ranking from this arm. The absolute background does yield one
  result no differential measurement can: the single power law
  over-produces the 857\,GHz monopole by $5.4\times$ and is excluded
  outright. The cluster-differential $\dxiu$ behind 363 PSZ2 clusters
  is null at every measurable aperture ($95\%$ limits
  $|\Delta\xi| < 0.52$--$0.61$ against a prediction three orders of
  magnitude below), and the one mildly significant negative offset is
  attributed by angular-size splits to a sky systematic (wrong sign
  in mass for either lensing or cluster dust).
\item \textbf{\textit{Herschel}/SPIRE at $25''$: the measurement.}
  The first count-sensitive measurement of $\xi(u)$ in the confusion regime: 
  $\hat\xi$ rising from $0.141\pm0.009$ at $u=25\,\mjb$ to
  $0.48\pm0.10$ at $75\,\mjb$ on the unmasked map, with a
  parameter-free forward simulation of the Schechter model
  reproducing the level and run of the bright-masked variant ($+0.037$
  against $+0.043$ at $u=25\,\mjb$; $\chi^2/\nu = 2.7$) and a covariance-correct ranking of Schechter, then
  DPL, then SPL. The rise itself is the arm's most instructive
  result: no intrinsic count model can produce it, and forward
  simulation attributes it to the strongly lensed bright population
  (identified as lensed through the $500\,\mum$ samples; $+0.0102$ of
  the $+0.011$\,mJy$^{-1}$ by which the measured slope
  exceeds the truncated model's). The imprint of the lensed bright
  population is therefore already detected in the $\PD$ tail, in
  the unlensed channel, and the statistic locates the transition from
  intrinsic to lensing-dominated counts on its own. Behind 110 eFEDS
  clusters the differential $\dxiu$ is a clean null with all null
  tests passing, giving the first observational limits on the lensing
  modulation, $|\Delta\xi| < 0.067$--$0.37$ (95\%).
\item \textbf{CCAT/FYST at $15''$: where the regime opens.} The three
  count models separate by $0.1$--$0.2$ in $\xi$ against errors of
  $0.005$--$0.03$ from $\sim\!9$\,deg$^2$, within a discrimination
  window bounded by physics at both ends
  ($u \approx 15$--$26\,\mjb$; models converge at $\approx 8\,\mjb$,
  maximal separation at $\approx 18\,\mjb$), so the functional form
  of the sub-confusion counts becomes distinguishable. The
  cluster-lensing forecast, by contrast, has no interior optimum
  (it improves monotonically toward the confusion core) and is
  therefore defined only with the explicit validity floor
  $u \ge 3\,\sigc$, at which a $3\sigma$ detection would require
  $\approx 7.6\times10^{3}$ clusters of $10^{15}\,\msun$: beyond any
  survey considered here, and stated as a designable requirement
  rather than softened.
\item \textbf{Instrument design.} \emph{Beam before sensitivity:} at
  $25''$, reducing the detector noise by a factor $1.7$ buys nothing
  for the Schechter counts ($0.99\times$); at $15''$, the identical improvement buys
  $2.4\times$. Angular resolution both multiplies the independent
  peaks and linearly lowers the confusion floor, and detector
  sensitivity pays only until the survey reaches that floor, so the
  beam decides how much sensitivity is worth buying. The same
  accounting, not separately simulated, makes the count-model
  discrimination a resolution result: the $3\sigc$ floor, which no integration time lowers,
  decides whether the count features fall inside the window.
\end{enumerate}

Three practical legacies accompany these conclusions. The calibrated
differential pipeline (common-mode filtering, declustering,
lensed-arm threshold construction, covariance-correct combination,
simulation-calibrated nulls) transfers to any future survey once the
filter forward model is rebuilt for the new instrument. The
statistical machinery for correlated thresholds
(Appendix~\ref{app:stats}) should apply to other threshold-swept
statistics, well beyond this paper's. And the design rules (the
$\Scut/\sigc$ feasibility criterion, the $3\sigc$ validity floor,
beam before sensitivity) give the next instrument, and the next
analysis, their benchmark.

Looking forward, a near-term task would be to derive a joint
  monopole--$\xi(u)$ constraint on the 1--10\,mJy decade at CCAT
  depth, where two of the three current models demonstrably fail. In the longer term, we can explore
  deep, high-resolution fields on the most massive clusters for the
  differential lensing signal, e.g., from the ALMA Lensing
  Cluster Survey or a future CCAT survey on massive clusters.
  Once the counts are pinned down in such fields, the formalism
  can be run in reverse: with the count slope known, the amplitude
  channel applied to large cluster ensembles becomes a magnification,
  and hence mass, probe.

A different route to the cluster-lensing signal is also worth mentioning, 
because our own results point to it.
Equation~(\ref{eq:sigcmu}) shows that the dominant response of the
lensed $\PD$ is not a change of tail shape but a broadening of the
confusion core, $\sigc(\mu) = \sqrt{\mu}\,\sigc$, which reaches
$75\%$ in the innermost aperture of
Table~\ref{tab:dxisuppression} against the $\dxiu \sim 0.01$ of the
shape channel. A forward model of the \emph{full} lensed $\PD$ ---
Eq.~(\ref{eq:mumixture}) evaluated over the whole distribution rather
than above a threshold, and predicted from $n_0(S)$ and an
externally supplied $\mu(\theta)$ rather than fitted with a flexible
analytic family --- would use that response and the exceedance rate alongside the
tail, and should be considerably more sensitive than $\dxiu$ alone. Whether it would also be more \emph{accurate} is the
open question. Drawing its power from the blended core, such an
analysis forfeits the invariances that motivate the tail statistic
(Sect.~\ref{sec:theory:xi}) and becomes exposed to the absolute
calibration, to cluster dust (Sect.~\ref{sec:planck:dust}), and to
the clustered-CIB bias that \citet{Wang2026} report for core-based
count inference --- a bias that, behind clusters, is correlated with
the signal rather than random. Establishing that trade quantitatively
will be the task for a future paper.

\begin{acknowledgements}
We thank Benjamin Magnelli for discussions; Thomas Reiprich for providing comments to the preprint-version; 
and Dominik Rhiem, Christopher L\"{o}rler, Andreas Kundik, and Alina Kubica for their earlier work on related projects. 
We thank the CCAT-prime collaboration for sharing internal results from the forecast study of the upcoming 850\,GHz survey.  
This paper uses Planck Third Party Products (CIB maps by \citet{Lenz2019}) from the Legacy Archive for Microwave Background Data Analysis (LAMBDA), database part of the High Energy Astrophysics Science Archive Center (HEASARC). HEASARC/LAMBDA is a service of the Astrophysics Science Division at the NASA Goddard Space Flight Center. 
We also make use of data from the Herschel Extragalactic Legacy Project (HELP), which is a European Commission Research Executive Agency funded project under the SP1-Cooperation, Collaborative project, Small or medium-scale focused research project, FP7-SPACE-2013-1 scheme, Grant Agreement Number 607254. 
The eFEDS galaxy cluster catalogue is based on data from eROSITA, the soft X-ray instrument aboard SRG, a joint Russian-German science mission supported by the Russian Space Agency (Roskosmos), in the interests of the Russian Academy of Sciences represented by its Space Research Institute (IKI), and the Deutsches Zentrum f\"{u}r Luft- und Raumfahrt (DLR).  
We acknowledge funding from the Collaborative Research Center 1601 (SFB 1601), funded by the Deutsche Forschungsgemeinschaft (DFG, German Research Foundation); and from the Verbundprojekt 05A2023 -- D-MeerKAT III, funded by the German Federal Ministry of Education and Research (BMBF).

\end{acknowledgements}

\begin{spacing}{0.85}
{\small
\noindent\textit{AI-usage disclaimer:}
This paper was prepared with extensive use of
AI assistants, specifically the Claude Opus- and Fable-class models (Anthropic)
operating in agentic, tool-using mode. The AI wrote the analysis and figure-generation
code, produced the figures, performed numerical consistency checks, 
helped to refine the paper's presentation, and handled the 
\LaTeX{} formatting. The research questions, the theoretical framework and the
interpretation of the results are the authors', as is the verification of every
quantitative claim reported here. The authors take full responsibility for the
work, including the originality of the content and the accuracy of the results.
}
\end{spacing}

\vspace{5mm}
\begin{spacing}{0.85}
{\small
\noindent\textit{Data availability:} The entire codebase for this paper is available at the public GitHub repository \href{https://github.com/kmbasu/GPD-Analysis-of-CIB}{GPD-Analysis-of-CIB}, with the Zenodo DOI \href{https://doi.org/10.5281/zenodo.22947949}{10.5281/zenodo.22947950}. 
This repository includes the full simulation suites, \textit{Planck} and \textit{Herschel}/SPIRE analysis scripts, the number-count fits, and the script that generates the plots. 
}
\end{spacing}

\bibliographystyle{aa}
\bibliography{final_refs}


\begin{appendix}
\section{Count-model fits at 350 \texorpdfstring{\mum}{um}}
\label{app:counts}

\subsection{Data}

All count models are fitted to the 350\,\mum\ differential counts of
\citet[Table~3, ``All'' column]{Bethermin2012}, which combine
three techniques over complementary flux ranges: stacking on 24\,\mum\
priors in GOODS-N (2.1--4.2\,mJy) and COSMOS (6.0--16.8\,mJy), and
directly resolved counts in COSMOS (23.8--133.7\,mJy). The table
gives Euclidean-normalized counts $Y \equiv (\dnds)\,S^{2.5}$ in
Jy$^{1.5}$\,sr$^{-1}$; we convert to $\dnds$ in
mJy$^{-1}$\,deg$^{-2}$ (the conversion factor, combining the unit
changes of flux and solid angle, is $9.633\,Y/(S/{\rm mJy})^{2.5}$,
verified with \texttt{astropy.units} rather than derived by hand),
with fractional errors carried over unchanged since the bin centers
are fixed.

Two exclusions define the fitted subset, and both are physical
rather than cosmetic. The brightest point (133.7\,mJy) is excluded
because above $\sim\!100$\,mJy the observed 350\,\mum\ counts are
dominated by strongly lensed sources \citep{Negrello2007,Lima2010b,Bethermin2012,Bethermin2012b}
 and no longer trace the intrinsic $\dnds$ that any of the three models describes.
This is the same physics that sets $\Scut$ throughout this paper, and
whose imprint Sect.~\ref{sec:herschel:rise} detects in the map. The
three GOODS-N stacked points are excluded because they are
methodologically distinct (stacking in a single small field, with
its own completeness and cosmic-variance systematics) and because
they visibly pull the fits away from an otherwise excellent
description of the remaining nine points: with them removed the
Schechter $\chi^2/\nu$ improves from 1.36 to 0.19 while its
parameters move by more than their formal errors
($\alpha$: $-1.28 \to -1.89$; $S_*$: 13.6 $\to$ 19.0\,mJy). The
final fits therefore use $N = 9$ points over $S = 6.0$--$94.6$\,mJy,
common to all three models. After these exclusions, the faint end below 6\,mJy has no
observational anchor in the fit at all, and all faint-end behavior
of the models is extrapolation, constrained externally only by the
background monopole (Appendix~\ref{app:counts:monopole}).

\subsection{Models and fitting methodology}

The three functional forms are, with $x \equiv S/S_*$,
\begin{align}
\text{Schechter:}\quad & n_0 = \frac{n_*}{S_*}\,x^{\alpha}\,
  \mathrm{e}^{-x} , \label{eq:nsch}\\[2pt]
\text{SPL:}\quad & n_0 = N_0 \,(S/S_0)^{\beta} , \label{eq:nspl}\\[2pt]
\text{DPL:}\quad & n_0 = \frac{\phi_*}{S_*}
  \bigl[x^{\alpha} + x^{\beta}\bigr]^{-1} , \label{eq:ndpl}
\end{align}
the SPL pivot fixed at $S_0 = 2.2$\,mJy following
\citet{Patanchon2009} and the DPL taken with $0 < \alpha < \beta$
\citep[the form of][]{Fujimoto2024}. Everything the tail statistic
sees follows from the local slope and curvature of
Eq.~(\ref{eq:etakappa}), which for these three forms are
\begin{equation}
\begin{array}{lcc}
\text{model} & \eta(S) & \kappa(S) \\[3pt]
\hline
\rule{0pt}{12pt}
\text{Schechter} & -\alpha + x & -x \\[4pt]
\text{SPL} & -\beta & 0 \\[4pt]
\text{DPL} &
 \dfrac{\alpha x^{\alpha} + \beta x^{\beta}}{x^{\alpha} + x^{\beta}} &
 -\dfrac{(\beta-\alpha)^{2}}{4}\,\mathrm{sech}^{2}\dfrac{t}{2}
\end{array}
\label{eq:etamodels}
\end{equation}
with $t \equiv (\beta-\alpha)\ln x$. The sign conventions of
Eqs.~(\ref{eq:nsch})--(\ref{eq:ndpl}) follow the source papers and
are not uniform: the fitted Schechter $\alpha$ and SPL $\beta$ are
negative while the DPL slopes are positive, so that all three models
give $\eta > 0$ as Eq.~(\ref{eq:etamodels}) requires
(Table~\ref{tab:countfits}). Three consequences organize the
whole comparison. The SPL has $\kappa \equiv 0$, so it is the
scale-free fixed point of Sect.~\ref{sec:theory:lensing} and shows no
lensing shape response at all. The Schechter curvature grows without
bound as $x \to \infty$, which is what drives $\xi$ down toward zero.
And the DPL's $\eta$ is a logistic in $\ln S$: its transition from
$\alpha$ to $\beta$ has a $10$--$90\%$ width of
$2\ln 9/[(\beta-\alpha)\ln 10] = 1.91/(\beta-\alpha)$\,dex, fixed by
the slope difference and \emph{not} a free parameter of this family.
For the fit below that is $0.67$\,dex, so the ``break'' is a ramp
spanning $5.4$--$25.3$\,mJy rather than a step, which is the reason
Sect.~\ref{sec:theory:xi} describes the DPL fingerprint as a broad
ramp onto a plateau. A sharper break would require the
one-parameter generalization
$n_0 \propto x^{-\alpha}[1 + x^{\Delta}]^{-(\beta-\alpha)/\Delta}$,
of which Eq.~(\ref{eq:ndpl}) is the $\Delta = \beta-\alpha$ case; we
do not adopt it, because nine points spanning
$6$--$94.6$\,mJy cannot constrain $\Delta$, and because at
$\Delta \gg \beta-\alpha$ the break would fall below the validity
floor of every instrument considered here and the DPL would lose its
shape signature entirely.

The Schechter and SPL fits minimize the standard weighted $\chi^2$
with the published uncertainties taken at face value
(\texttt{absolute\_sigma}), so that the reduced $\chi^2$ remains
informative about relative goodness of fit rather than being
normalized away. The DPL, with four parameters against nine points,
has a shallow, elongated $\chi^2$ valley in $(\alpha, S_*)$ and is
fitted as a maximum-a-posteriori problem: a Gaussian prior on
$\log_{10} S_*$ enters as a pseudo-residual in a nonlinear
least-squares solver, the internal parametrization
$(\alpha, \beta{-}\alpha, \log_{10}S_*, \log_{10}\phi_*)$ enforces
$\alpha < \beta$ by construction, and the quoted $\chi^2/\nu$ is
computed from the data term alone so that it remains comparable to
the unregularized fits. In the final (GOODS-N-excluded) fit the
prior is loose and the break is data-driven: $S_*$ moves to
$11.70 \pm 1.89$\,mJy with both slopes well determined. The adopted
parameters of all three models are collected in
Table~\ref{tab:countfits}.

\begin{table}[h]
\caption{Adopted 350\,\mum\ count models: functional forms of
Eqs.~(\ref{eq:nsch})--(\ref{eq:ndpl}), best-fit parameters with
$1\sigma$ errors, and reduced $\chi^2$ with $\nu = N - N_{\rm par}$
degrees of freedom. Sign conventions differ between rows: the Schechter
$\alpha$ and the SPL $\beta$ are signed logarithmic slopes, whereas the DPL
$\alpha, \beta$ are positive slope magnitudes.}
\label{tab:countfits}
\centering
\begin{tabular}{l l c}
\hline\hline
Model & Best-fit parameters & $\chi^2/\nu$ \\
\hline
Schechter & $\alpha = -1.890 \pm 0.186$ & 0.19 \\
          & $S_* = 19.01 \pm 2.76$\,mJy & \\
          & $n_* = 7014 \pm 2156$\,deg$^{-2}$ & \\
SPL       & $\beta = -3.279 \pm 0.049$ & 12.6 \\
          & $N_0 = (1.053\pm0.120)\times10^{5}$\,mJy$^{-1}$deg$^{-2}$ & \\
DPL       & $\alpha = 1.082 \pm 0.507$, $\beta = 3.928 \pm 0.149$ & 3.9 \\
          & $S_* = 11.70 \pm 1.89$\,mJy & \\
          & $\phi_* = (1.33\pm0.37)\times10^{4}$\,deg$^{-2}$ & \\
\hline
\end{tabular}
\tablefoot{SPL pivot fixed at $S_0 = 2.2$\,mJy. The DPL $\chi^2/\nu$
uses the nominal $\nu = N - 4$; the loose prior on $S_*$ makes the
effective value marginally smaller. The relative quality of the fits
is itself informative: an independently adjustable bright-end power
law (DPL) does not substitute for a true exponential cutoff on these
data.}
\end{table}

The models are used exactly as fitted, with no renormalization. This
choice was tested rather than assumed: the like-for-like comparison
of predicted and measured confusion widths at \textit{Herschel} agrees to
7--9\%, inside the fits' own $\pm 15\%$ normalization uncertainty
(Appendix~\ref{app:matchedfilter}). Two truncations are applied
throughout the paper: $\Scut = 100$\,mJy (the lensing-domination
boundary above) and $S_{\min} = 1$\,mJy. The latter is a modeling
choice, not a property of any model. For the Schechter and
DPL the confusion variance converges at the faint end and the choice
is immaterial ($\lesssim 2\%$ in $\sigc$); for the SPL, whose second
moment diverges as $S_{\min} \to 0$, $\sigc$ is set by the cut
rather than by the data: at \textit{Planck}'s beam the SPL
$\sigc$ changes by 50\% between $S_{\min} = 0.1$ and 1\,mJy.
SPL-derived core widths are therefore never quoted as model
predictions anywhere in this paper.

\subsection{The background monopole, and two falsifications of the
SPL}
\label{app:counts:monopole}

The count models are anchored to data only over 6--94.6\,mJy, so
their behavior below that range is extrapolation and is constrained
externally. The strongest such constraint is the absolute background,
\begin{equation}
q_1 = \int_{S_{\min}}^{\infty}\! S\,n_0(S)\,\mathrm{d}S ,
\label{eq:monopole}
\end{equation}
which for our three fits, extrapolated to $S_{\min} = 1$\,mJy,
predicts an 857\,GHz monopole of $0.906$ (Schechter), $3.58$ (SPL)
and $0.602$ (DPL) MJy\,sr$^{-1}$. The measured values are the FIRAS power-law fit evaluated at 857\,GHz, $0.66$\,MJy\,sr$^{-1}$ \citep[with $\sim\!30\%$
amplitude uncertainty;][]{Fixsen1998} and the more precise
$0.576 \pm 0.034$\,MJy\,sr$^{-1}$ of \citet{Odegard2019}; the factors
below are quoted against the FIRAS value, and are $15\%$ larger against
the \citet{Odegard2019} value ($6.2\times$ and $1.6\times$ in place of
$5.4\times$ and $1.4\times$). The single
power law over-produces the background by a factor $5.4$ and is
excluded as a global count model by Eq.~(\ref{eq:monopole}) alone,
independently of anything the tail statistic says. It survives in
this paper only as a local description of the 6--94.6\,mJy counts,
which is the capacity in which the tail statistic probes it, and its
extrapolation-dependent quantities ($\sigc$, the core width) are
never quoted as model predictions.

The Schechter model's more moderate $1.4\times$ over-production
(with the $\sim\!30\%$ uncertainty of its fitted normalization $n_*$
alone, Table~\ref{tab:countfits}, an excess of only $1$--$2\sigma$, which
is why it is called a tension rather than an exclusion throughout) is
itself informative. Because $S\,n_0(S)$ is spread nearly evenly over
every decade of flux for these slopes, while the confusion variance
$q_2$ converges at the faint end, the two integrals weight the counts
almost orthogonally, and the Schechter excess arises already from
the $S > 1$\,mJy population, before any faint-end extrapolation. The
tension therefore sits in the 1--10\,mJy decade: the same decade that
carries a third of $q_2$, and the one the tail statistic targets
(Sect.~\ref{sec:theory:reach}). The count models, fitted to resolved
and stacked counts, demonstrably fail somewhere in exactly the decade
this method is designed to measure, which is less an embarrassment
than the motivation for measuring it.

A second, fully independent falsification of the SPL comes from the
flux-cut dependence of the confusion noise. \citet{Nguyen2010}
measured the confusion noise $\sigc$ of the SPIRE
350\,\mum\ maps as a function of the pixel flux cut, obtaining the
ratio $\sigma_{\rm c}({<}19\,{\rm mJy})/\sigma_{\rm c}(\infty) =
0.73$. The Schechter and DPL models predict 0.76 for this ratio;
the SPL predicts 0.95 (at $S_{\min} = 0.1$\,mJy; 0.89 at 1\,mJy).
This test involves no absolute normalization and no beam model
(the ratio cancels both) and is fully independent of the monopole
argument. Both tests fail the SPL and pass the other two models.

\subsection{Count features and instrument resolution}

The detectability of a count feature in $\xiu$ is set by its flux
relative to the confusion core, $k_* = S_*/\sigc$, where $\sigc$ is
computed from the model's own counts (for the DPL,
$7.58\,\mjb$ at $25''$ and $4.52\,\mjb$ at $15''$, against $7.98$
and $4.76$ for the Schechter, all at $S_{\min} = 1$\,mJy). For the
DPL break
at 11.7\,mJy: $k_* \approx 1.5$ at SPIRE resolution ($25''$), deep
inside the Gaussian core, inaccessible, against $k_* \approx 2.6$
at $15''$, entering the usable window. Because $\sigc$ is linear in
the beam FWHM at fixed counts, the product $k_* \times \mathrm{FWHM}
\approx 39''$ is constant across beam sizes, which turns
the criterion into a resolution requirement: an instrument in the
$10$--$12''$ class would place this particular break comfortably in
the fitting window, the $15''$ of CCAT/FYST places it at its inner
edge (Sect.~\ref{sec:ccat:models}), and SPIRE-class resolution
cannot reach it at all.

One regime caveat accompanies the DPL specifically. Computed from
its own fitted counts, its source density per beam at the \textit{Herschel}
and CCAT beams ($\Nbeam \approx 2$ and $0.7$) sits below the
$\Nbeam \gtrsim 5$--$7$ boundary at which a fully Gaussian confusion
core forms, so its $\PD$ occupies a slightly different physical
regime from the Schechter's at the same beam. The three model curves
are nonetheless directly comparable everywhere in this paper,
because they are compared through the identical peak-level
simulation pipeline (the regime difference is forward-modeled,
not assumed away), but analytic core-based reasoning should not be
applied to the DPL configurations without this caveat in mind.

\section{Beam, declustering, and the usable threshold window}
\label{app:beam}

This appendix collects the operational machinery connecting the
idealized relation of Eq.~(\ref{eq:xislope}) to a measurement on a
beam-smoothed, noisy map: which distributions are actually involved,
where the usable threshold window lies, what declustering does and
does not accomplish, and why the common alternative, deconvolution, would be a mistake
for this measurement.

\subsection{Two distributions, three correlation effects}

Two distinct distributions and three distinct correlation effects
run through every analysis in this paper, and conflating them
produces subtle errors, so we fix the vocabulary here first. The \emph{pixel} distribution is the $\PD$ proper: the
one-point marginal of the smoothed field, the object of
Eq.~(\ref{eq:condon}). The \emph{peak} distribution is the marginal
of declustered local maxima, a different random variable with a
different marginal, related to the pixel distribution through the
correlation structure of the field. The three correlation effects
are: (A) serial dependence of neighboring pixels within one
beam, which declustering removes by construction; (B) the
peak-selection effect, by which the marginal of maxima
differs from the marginal of pixels even after declustering (this
is a change of estimand, not a bias, and it is why peak-level model
curves are required); and (C) shared-realization covariance
between nested threshold sets, which no amount of declustering can
remove and which is the subject of Appendix~\ref{app:stats}.
Keeping (A), (B), and (C) separate (one removed, one modeled, one
propagated) is the organizing principle of the pipeline.

\subsection{The usable window and a design rule}

The measurable $\hat\xi(u)$ of a confusion-limited map interpolates
between three regimes. Below $\sim\!3\,\sigc$ the Gaussian core
controls the fit: the fitted shape is negative (a Gaussian tail has
an effective endpoint on any finite sample) and climbs as the
threshold rises, the ``penultimate'' regime of extreme-value
theory, in which the asymptotic GPD has not yet been reached and the
fitted $\hat\xi$ reflects the crossover rather than the counts. On
the shoulder the fitted shape tracks the local count slope, which is
the science regime. Above $\sim\!\Scut/10$ the truncation or mask
depresses $\hat\xi$, a fall that measures the mask, not the counts.
This upper bound is calibrated rather than asserted. Holding the
beam at $20''$ and varying only $\Scut$ over $30$--$400$\,mJy, the
depression of $\hat\xi$ relative to the effectively untruncated
($400$\,mJy) case is, to good accuracy, a function of $u/\Scut$ alone
(Table~\ref{tab:truncbias}):

\begin{table}[h]
\caption{Truncation bias $\delta\hat\xi_{\rm trunc}$ of $\hat\xi$, calibrated at a $20''$ beam
with $\sigN = 0$ on the illustrative DPL of
Fig.~\ref{fig:beamxi}, against the $\Scut = 400$\,mJy reference.}
\label{tab:truncbias}
\centering
\begin{tabular}{r r r r r r}
\hline\hline
$u/\Scut$ & 0.025 & 0.05 & 0.10 & 0.15 & 0.20 \\
\hline
$\delta\hat\xi_{\rm trunc}$ & $-0.01$ & $-0.02$ & $-0.05$ & $-0.10$ & $-0.18$ \\
\hline
\end{tabular}
\tablefoot{The bias reaches the scale of the count-model separations
($0.1$--$0.2$) by $u \approx \Scut/5$; $\Scut/10$ is the level at
which it becomes comparable to a typical measurement error.}
\end{table}

\noindent
so that the \emph{interpretive} window, over which
Eq.~(\ref{eq:xislope}) may be read quantitatively, is
Eq.~(\ref{eq:window}), $3\,\sigc \lesssim u \lesssim \Scut/10$,
non-empty only for $\Scut/\sigc \gtrsim 30$. The wider \emph{measurement} window of the data chapters is defined in Sect.~\ref{sec:theory:beam}.

The design rule is to verify before an analysis is attempted that the
measurement window is non-empty, which requires
$\Scut/\sigc \gtrsim 3$: at \textit{Planck}'s $300''$ beam
($\Scut/\sigc = 1.05$) it is not, and no amount of sky area can
recover it (Sect.~\ref{sec:planck:dilution}). The number of sources
per beam, $\Nbeam$, plays a complementary diagnostic role (it
controls whether the faint-slope branch of the counts is observable
at all) and is reported alongside every configuration in this
paper. The low-threshold boundary reappears in the CCAT forecast as
the validity floor $u \ge 3\,\sigc$
(Sect.~\ref{sec:ccat:lensing}): the same physics, seen from the
forecasting side.

\subsection{Declustering: rationale, mechanics, and the tail}

\begin{figure*}
\centering
\includegraphics[width=\hsize]{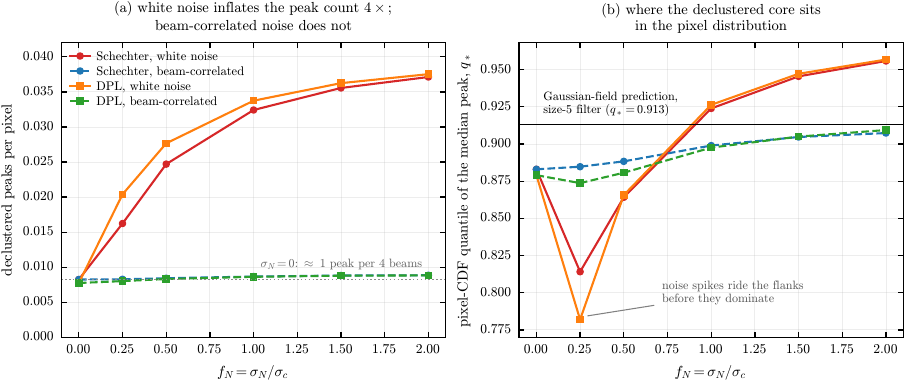}
\caption{The two measurements underlying the declustering
conventions of this paper, as a function of the noise-to-confusion
ratio $f_N = \sigN/\sigc$. Each noise mode is run for two count
models, the Schechter (circles) and the illustrative DPL of
Fig.~\ref{fig:beamxi} (squares): white noise with solid lines,
beam-correlated noise with dashed lines. \emph{(a)} White pixel noise inflates the
retained declustered peak count by a factor $\approx 4$, because
isolated noise spikes survive the local-maximum test; beam-correlated
noise leaves it at the $\sigN = 0$ level of $\approx 1$ peak per four
beam solid angles (dotted horizontal line). All simulations in this paper therefore use
beam-correlated noise. \emph{(b)} The location of the declustered
core within the pixel distribution, expressed as the pixel-CDF
quantile $q_*$ of the median peak, against the
Gaussian-random-field prediction $q_* = 0.913$ for a one-FWHM (five-pixel, ``size-5'')
maximum filter (solid horizontal line;
Appendix~\ref{app:gaussianpeaks}). The dip at $f_N \approx 0.25$ in the
white-noise legs is the regime in which noise spikes populate the
flanks of the distribution before they come to dominate it.}
\label{fig:noisedecl}
\end{figure*}

We decluster for three reasons, not one: to restore the independence
that the GPD likelihood assumes (effect A); to prevent a single
bright source from entering the fit many times through its
contiguous pixels; and to make the exceedance count an interpretable
number (peaks per beam solid angle) rather than a pixel-scale
artifact. The implementation retains local maxima separated by at
least one beam FWHM, via a maximum filter with the corresponding
window correction calibrated on simulations.\footnote{The
procedure is a two-dimensional declustering analogue  
of the one-dimensional POT literature
(\citealt[ch.~5]{Coles2001}; \citealt{FerroSegers2003}): a pixel is retained if it
equals the maximum of a square window of side one beam FWHM
(\texttt{scipy.ndimage.maximum\_filter}), the same rule being applied
to the \textit{Planck} and \textit{Herschel} maps and to every
simulation. No published image-declustering recipe is followed; the
window correction and the peak density it implies are calibrated on
the simulations of Fig.~\ref{fig:noisedecl}.}

Three properties of this operation matter downstream.
\emph{First}, declustering is a rank operation on the smoothed
field: it depends only on the ordering of pixel values, so it
commutes with any monotonic rescaling of the flux axis. This is what
makes the entire declustering machinery common-mode between lensed
and control arms, up to the flux-scale caveat below.
\emph{Second}, the declustered peak density of a confusion-limited
map is approximately one peak per four beam solid angles (not one per
beam), and exceedance budgets must be computed accordingly; the
naive $A/\Omega_{\rm beam}$ estimate is optimistic by a factor of
several. \emph{Third}, in the far tail, declustering acts as a pure
rate thinning with a closed-form constant: the ratio of pixel
to peak exceedances approaches
$\mathcal{C}(u) \to \xi\,\Omega_{\rm beam}/A_{\rm pix}$, independent
of threshold, because each tail peak is one source whose profile
places a calculable pixel area above any level it exceeds. We log
$\mathcal{C}(u)$ alongside every measured curve: its
flatness in $u$ is a cheap, fit-free diagnostic that the
window sits on the power-law plateau, available before any GPD is
fitted, and its departure from flatness is an early warning that the
window has strayed into the core or the mask ceiling.

The peak-selection effect (B) has a compact analytic form. Because a
pixel is retained only if it exceeds the $n_{\rm win}-1$ others
within its declustering window, and those neighbors are
approximately exchangeable, the peak marginal is the corresponding
order statistic of the pixel marginal,
\begin{equation}
p_{\rm pk}(D) \;\propto\; p_{\rm pix}(D)\,
\bigl[F_{\rm pix}(D)\bigr]^{\,n_{\rm win}-1},
\label{eq:peakmarginal}
\end{equation}
with $F_{\rm pix}$ the pixel CDF and $n_{\rm win}$ the effective
number of independent pixels per declustering window, a calibrated
quantity rather than one derived from an independence assumption (distinct from the effective number of independent thresholds $n_{\rm eff}$ of Appendix~\ref{app:stats}). Equation~%
(\ref{eq:peakmarginal}) exhibits both regimes at once: in the core
$F_{\rm pix} \ll 1$ and the bracket crushes the distribution, while
in the far tail $F_{\rm pix} \to 1$ and the peak density is the pixel
density merely rescaled, which is the rate-thinning limit
$\mathcal{C}(u)$ above, seen from the other side. It is a change of
estimand, not a bias.

Operationally we calibrate the resulting shift per configuration as a
quantile $q_*$ (never as a multiple of $\sigc$, since the shift is
not Gaussian-universal), with the exact Gaussian-field theory of
Appendix~\ref{app:gaussianpeaks} supplying the reference in the
core. Two boundaries of validity are respected throughout: the
calibration describes the peak core and is never
extrapolated into the tail (where the rate-thinning constant
governs instead, with different physics), and the net
pixel-versus-peak difference in the fitted shape (a median of
$0.03$--$0.06$, reaching $0.09$ at SPIRE and $0.22$ at CCAT
resolution at the ends of the threshold ladder) is the reason all
model curves in this paper are simulated at peak level
(Sect.~\ref{sec:sims:validation}). Measured against the
asymptotic $1/(\eta-1)$ rather than the pixel $\PD$, the
same gap is $0.3$--$0.5$ (Fig.~\ref{fig:fingerprintspeak}).

\subsection{Instrument noise and declustering}

The interaction of noise with declustering depends on the noise
correlation length, not only its amplitude. White pixel noise
decorrelates neighboring pixels and multiplies the retained peak
count several-fold (nearly all of the additions being noise spikes),
which corrupts any peak-level statistic. Beam-shaped noise, with
correlation length equal to the beam, adds no spurious peak
population. All simulations in this paper therefore use
beam-correlated noise, matching the matched-filtered data whose
noise correlation length is the beam scale by construction
(Appendix~\ref{app:matchedfilter}).

\subsection{Why not deconvolve}

A natural-seeming alternative to declustering is iterative source
extraction, CLEAN-like peak finding with subtraction, followed
by inference on the recovered sources. For this measurement it would
be a mistake three times over. It changes the estimand: the
recovered catalog depends on the extraction prior, and at
$\Nbeam \gg 1$ source densities the ``sources'' are
prior-dominated blends, rather than physical objects. It forfeits
the universality that makes the POT construction attractive: the
GPD limit applies to threshold exceedances of the observed field,
not to the output of a nonlinear, flux-dependent iterative
algorithm. And lastly, decisively for the differential argument, it
breaks the common-mode cancellation: the effective operator of an
iterative extraction depends on the field brightness, which differs
between lensed and control arms by construction, so the two
arms would no longer share their systematics. A fixed, linear,
common filter plus a rank-based declustering preserves exactly the
properties the differential design relies on.

\subsection{Lensing and the flux scale}
\label{app:beam:lensing}

Under magnification the dimensionless machinery above (the
declustering convention, the quantile $q_*$, the peak density, the
filter) is common-mode and cancels in $\dxiu$ to 1--3\%: lensing
changes the counts, not the beam, the pixel grid or the declustering
rule, so only the mild change in the shape of the field's correlation
structure survives. In simulations at a $15''$ beam with a uniform
aperture-averaged magnification $\mu = 1.60$ (an annulus around a
$4\times10^{14}\,\msun$ halo at $z_{\rm l} = 0.3$), the peak quantile
$q_*$ moves by $-0.006$ (Schechter) and $-0.015$ (DPL) between the
control and lensed arms, and the declustered-peak density by $-1.2\%$
and $-3.1\%$, which is the origin of the figure quoted in
Sect.~\ref{sec:theory:lensing}. The flux scale does not, by
Eq.~(\ref{eq:sigcmu}), so the two arms have their usable windows at
different absolute fluxes. The operational consequence is a
convention: the threshold ladder of a differential measurement is
built from the \emph{lensed} arm's window, the narrower of the
two, and never from the control arm, as applied in
Sects.~\ref{sec:herschel:lensed} and \ref{sec:ccat:lensing}.

\section{Peak statistics of two-dimensional Gaussian random fields}
\label{app:gaussianpeaks}

The calibration of the peak-selection shift (Appendix~\ref{app:beam})
rests on the exact extreme-value statistics of a smoothed
two-dimensional Gaussian random field, a subject with a classical
literature in other corners of astronomy: ocean-wave statistics
\citep{LonguetHiggins1957}, the general theory of level crossings
and maxima of random fields \citep{Adler1981}, and the peaks of
cosmological density and CMB fields \citep{BondEfstathiou1987}. Both
relevant limits of our maps are fields of this class: the confusion
core of a beam-smoothed sky at high $\Nbeam$, and beam-shaped
instrument noise.

For a stationary isotropic Gaussian field of variance $\sigma_0^2$
the relevant descriptors are the spectral moments and the two shape
parameters built from them,
\begin{equation}
\sigma_j^2 = \int\!\frac{\mathrm{d}^2\ell}{(2\pi)^2}\,
\ell^{2j}\,P(\ell) ,
\qquad
\theta_c = \frac{\sigma_1}{\sigma_2} ,
\qquad
\gamma = \frac{\sigma_1^{2}}{\sigma_0\sigma_2} ,
\label{eq:specmom}
\end{equation}
$\theta_c$ being the coherence scale and $\gamma$ the spectral
narrowness. Our two Gaussian limits, beam-shaped instrument noise and the
confusion core at high $\Nbeam$, are both white noise smoothed by a
Gaussian beam of width $\sigma_{\rm b}$, for which
$P(\ell) \propto \mathrm{e}^{-\ell^2\sigma_{\rm b}^2}$ and
Eq.~(\ref{eq:specmom}) evaluates in closed form,
\begin{equation}
\sigma_j^2 = \frac{j!}{4\pi\,\sigma_{\rm b}^{\,2j+2}}
\qquad\Longrightarrow\qquad
\theta_c = \frac{\sigma_{\rm b}}{\sqrt{2}} ,
\qquad
\gamma = \frac{1}{\sqrt{2}} .
\label{eq:specgauss}
\end{equation}
Two exact results follow. The total density of local maxima is
\begin{equation}
n_{\rm pk} = \bigl(8\sqrt{3}\,\pi\,\theta_c^{2}\bigr)^{-1} ,
\label{eq:npk}
\end{equation}
which with Eq.~(\ref{eq:specgauss}) and
$\Omega_{\rm beam} = 2\pi\sigma_{\rm b}^{2}$ gives
$n_{\rm pk}\,\Omega_{\rm beam} = 1/(2\sqrt{3}) = 0.289$: one peak per
$3.46$ beam solid angles, against the $\approx\!1$ per $4$ measured
on the confusion field (Appendix~\ref{app:beam}). Direct maximum-filter
counts on simulated Gaussian-smoothed white noise reproduce
Eq.~(\ref{eq:npk}) to $0.3\%$ over $\sigma_{\rm b} = 3$--$6$ pixels;
the fiducial simulations adopt the measured confusion-field rate,
one declustered peak per four beam solid angles, as their peak
budget. Second, the density of maxima above a high level
$\nu = D/\sigma_0$ is
\begin{equation}
n_{\rm pk}({>}\nu) \;\simeq\;
\frac{\nu\,\mathrm{e}^{-\nu^{2}/2}}{4\,(2\pi)^{3/2}\,\theta_c^{2}} ,
\label{eq:npknu}
\end{equation}
so maxima are heavier-tailed than pixels by a factor $\propto\nu$,
the Gaussian-limit form of Eq.~(\ref{eq:peakmarginal}).

The same peak statistics fix the reference
quantile used in Appendix~\ref{app:beam}: defining $\nu_*$ by
$n_{\rm pk}({>}\nu_*) = n_{\rm pk}/2$, the median peak sits at the
pixel-CDF quantile
\begin{equation}
q_* = \Phi(\nu_*) ,
\label{eq:qstar}
\end{equation}
with $\Phi$ the standard normal CDF. Because the median lies at
$\nu_* \approx 1.3$, where the high-level asymptote of
Eq.~(\ref{eq:npknu}) is not yet accurate, $\nu_*$ is evaluated
numerically from the exact peak statistics of the smoothed field
with the one-FWHM maximum-filter window of the pipeline, which gives
$q_* = 0.913$ ($0.90$ for the population of all maxima), the
prediction plotted against the measured $0.88$--$0.91$ in
Fig.~\ref{fig:noisedecl}.
Transfer to the real, non-Gaussian confusion field is made at the
level of quantiles: Eq.~(\ref{eq:qstar}) supplies the \emph{shape} of
the pixel-to-peak mapping and one number per configuration is
calibrated on the end-to-end simulation. These are calibrations of
the peak core; they mis-predict the peak CDF at the
several-percent level outside it and are never extrapolated into the
power-law tail, whose declustering behavior is governed instead by
the rate-thinning constant of Appendix~\ref{app:beam}.

\section{Matched-filter treatment of the SPIRE data}
\label{app:matchedfilter}

\subsection{Why filter at all}

The unfiltered SPIRE map cannot be declustered meaningfully. Its
peak population is a mixture of two components, sky structure and
correlated instrument-noise structure, and the mixture is directly
visible in the pixel statistics (the lag-one autocorrelation values of Sect.~\ref{sec:herschel:data}). A declustering pass on
the unfiltered map would therefore return a peak sample of
indeterminate composition, and no peak-level tail statistic could be
interpreted. The purpose of the matched filter in this analysis is
accordingly \emph{not} noise suppression: the filtered map retains a
substantial instrument-noise component
($\sigN = 5.40\,\mjb$ against the beam confusion width
$\sigc = 7.98\,\mjb$, i.e.\ $\sigN/\sigc = 0.68$; in like-for-like
robust widths measured on the maps themselves the ratio is $1.43$
before filtering, $9.3$ against $6.5\,\mjb$, and $1.04$ after it,
$5.4$ against $5.2\,\mjb$). Matched
filters are often credited with rendering maps ``sky-dominated'';
on either convention this one leaves an instrument floor of order
two-thirds to unity of the confusion, which must be carried
explicitly in every forward model. What the filter removes is the
noise correlation structure, and that is what makes
one-peak-per-beam declustering, and hence peak-level extreme-value
analysis, applicable.

\subsection{The operator, and self-filtering of cutouts}

The filter is the confusion-weighted matched filter of
\citet{Chapin2011}, as delivered with the HELP products
\citep{Shirley2019}: a kernel $K$ (101$\times$101 pixels at
$8''$) constructed from the beam and the confusion power spectrum,
whose effective point-source response has a Condon-equivalent
FWHM of 25.15$''$, which is the beam value used throughout this
paper. The delivered operator is the inverse-variance-weighted
convolution
\begin{equation}
D_{\rm filt} = \frac{(d/\sigma^2) \ast K}{(1/\sigma^2) \ast K^2},
\qquad
\sigma_{\rm filt} = \bigl[(1/\sigma^2) \ast K^2\bigr]^{-1/2},
\label{eq:mfilt}
\end{equation}
with $d$ the signal map and $\sigma$ the delivered per-pixel error
map; a plain unweighted convolution $d \ast K$ is \emph{not} the
operator (it reproduces the delivered map only to correlation
$0.992$). We re-implemented Eq.~(\ref{eq:mfilt}) and verified it against the
delivered filtered map at a correlation of unity to ten decimal
places, with residual
rms $1.5\times10^{-6}\,\mjb$ (machine precision). Because the
kernel has finite support ($13.5'$), filtering a cutout that has
been padded by the kernel half-width and then trimming is
identical to filtering the full map, and we verified this too: the
maximum difference is $7\times10^{-14}\,\mjb$. Filtering unpadded
cutouts, by contrast, corrupts a $6.7'$ border annulus (rms
$0.45\,\mjb$) and is never done. The analysis therefore
self-filters padded cutouts from the primary signal map (the per-pixel error map varies by 11\% between cutouts, Appendix~\ref{app:matchedfilter:caveats}).

\subsection{Noise budget}

The filtered noise is characterized along two independent routes:
from the delivered per-pixel error and exposure maps, and from the
variance-versus-inverse-exposure regression of the map itself,
which separates the exposure-scaled instrument term from the
exposure-independent confusion floor. The routes agree on
$\sigN = 5.40\,\mjb$ and a confusion width consistent with the
adopted count models at the 7--9\% level. One methodological lesson
from this comparison is recorded because it is easy to repeat
elsewhere: a Condon-style \emph{rms} confusion prediction must
never be compared against a \emph{MAD-derived} measured width. On a
distribution as skewed as the confusion $\PD$ (skewness
$\approx 1.7$ here) the rms-to-MAD ratio is 1.25, and the mismatch
masquerades as a 25\% ``over-prediction'' of the confusion noise:
an artifact of estimator choice, not physics. All width comparisons
in this paper are like-for-like.

\subsection{Caveats carried}
\label{app:matchedfilter:caveats}

Two properties of the filter are carried as explicit caveats rather
than corrected. The matched-filter noise varies by 11\%
cutout-to-cutout across the mosaic (range $3.8$--$5.5\,\mjb$), making
the filter formally non-stationary and the observed $\PD$ a mixture
over noise levels; the induced core broadening is estimated at well
below a percent and neglected in the forward model, with the
per-cutout values retained in the data products. And the filter's
point-source flux response is 1.043 relative to unit normalization
when evaluated on an idealized Gaussian PSF. A cross-match against
the HELP XID+ catalog \citep{Hurley2017}, whose 350\,\mum\ fluxes
are PSF-fitted on the unfiltered image at prior positions and are
therefore independent of the filter, shows that the excess is real:
for 132 isolated XID+ sources of 60--400\,mJy the self-filtered peak
is $1.04 \pm 0.01$ times the catalog flux (median with bootstrap
error; the raw-image peak is $0.973 \pm 0.007$ times it, the
deficit being the pixel-integrated PSF at $8''$ pixels, so the
filtered-to-raw gain is $1.061 \pm 0.007$, a median of per-source
ratios rather than a ratio of the two medians), consistent across
flux bins ($1.03$, $1.02$, $1.07$ for 60--90, 90--140 and
140--400\,mJy) and rising to $1.06$ under stricter isolation cuts. The flux axis of Sect.~\ref{sec:herschel} is therefore
stretched by $\approx 4\%$ relative to the unit-response models. At
the measured $\mathrm{d}\xi/\mathrm{d}\ln u$ this is worth
$|\Delta\xi| \lesssim 0.013$ (comparable to the statistical error
at the lowest threshold and below it elsewhere); it is common-mode
in the differential measurement of Sect.~\ref{sec:herschel:lensed},
and repeating the covariance-weighted model comparison of
Sect.~\ref{sec:herschel:unlensed} with the model curves evaluated at
$u/1.04$ leaves the ranking unchanged ($\chi^2 = 23.4$, $112$ and
$396$ for Schechter, DPL and SPL, against $30.1$, $138$ and $472$).
We therefore report it as a known systematic rather than rescale
the maps.

\subsection{Forward model of the filtered map and of the bright mask}
\label{app:matchedfilter:forward}

Two properties of the data pipeline are not reproduced by simulating
the sky and the instrument noise through the same Gaussian beam and
truncating the injected counts at $\Scut$. Both are forward-modeled
here, and Table~\ref{tab:forwardconventions} quantifies what each
changes.

\emph{The filter's correlation structure.} In the filtered map the
sky response is the PSF convolved with the kernel $K$ and the
instrument noise is white noise convolved with $K$ alone; because $K$
has negative sidelobes (its effective solid angle is $504$\,arcsec$^2$
against $717$ for the $25.15''$ Gaussian), both are less correlated
than a beam-correlated field. The lag-one autocorrelation of the
filtered data is $0.72$ against $0.87$ for the Gaussian-beam
simulation, and the declustered-peak density is $0.374$ per beam
against $0.255$: a $32\%$ shortfall of the simulated peak density,
which the peak-level curves inherit. The simulator therefore
generates the beam-smoothed sky plus white noise on a padded grid,
applies $K/\sum K^2$ (the HELP operator with uniform weights,
Eq.~\ref{eq:mfilt}), divides by the kernel's point-source
response $R = 1.0426$ on the Gaussian PSF (the value of Appendix~\ref{app:matchedfilter:caveats},
recovered here independently) so that a unit source still peaks at
unity, and crops the padding; the white rms is set so that the
filtered noise has the measured $5.40\,\mjb$. This reproduces the
data's lag-one autocorrelation ($0.78$), pixel rms ($9.6$ against
$9.0\,\mjb$), peak-core width ($8.7$ against $8.1\,\mjb$) and peak
density ($0.359$ per beam, $4\%$ below the measured value). Relative
to the Gaussian-beam peak-level simulation, the filtered
truncated-counts curves rise by $+0.01$ to $+0.04$ inside the
comparison window and are unchanged above it (the level shift quoted
in Sect.~\ref{sec:herschel:rise} is measured against the
Gaussian-beam analytic pixel model instead, and has the opposite
sign); the slope decomposition of Sect.~\ref{sec:herschel:rise}
is unaffected at the $10^{-4}$\,mJy$^{-1}$ level.

\emph{Mask versus truncation.} Removing the pixels above
$100\,\mjb$ from the map (with a one-FWHM growth) is not the same
operation as removing the sources above $100$\,mJy from the injected
counts. The mask imposes a hard endpoint in \emph{map} units, whereas
the truncation leaves a soft edge smeared by confusion and
instrument noise, and sources of $100$--$130$\,mJy whose peaks fall
below the mask level survive it. A simulation that contains the
observed bright population (the power law of Sect.~\ref{sec:herschel:rise},
identical for every model) and is then masked exactly as the data
gives a Schechter curve $0.01$ below the truncated-counts curve at
$25\,\mjb$, $0.09$ below at $50\,\mjb$, and $0.4$ below at
$75\,\mjb$, where the masked measurement itself falls to
$-0.45 \pm 0.10$ (Fig.~\ref{fig:herschelxi}): the plunge of the masked
data above the comparison window is the mask, and only a masked
simulation reproduces it. The masked area is $0.09$--$0.13\%$ of the
field per model against $0.10\%$ measured. The fiducial model curves of
Fig.~\ref{fig:herschelxi} therefore carry the observed bright
population, the filter forward model, and the data's own baseline
removal, mask, declustering and edge exclusion, with the simulated
peaks shifted so that their median coincides with the measured
peak-core centroid ($-1.6$, $-1.0$ and $-5.0\,\mjb$ for Schechter, DPL
and SPL; without the shift the $\chi^2$ are $24$, $120$ and $273$ and
the ranking is the same). Table~\ref{tab:forwardconventions} gives the
covariance-correct $\chi^2$ under the four combinations of the two
conventions: the ranking is the same in every one, the two effects
partly cancel for the Schechter model, and the filter widens the
separation.

\begin{table}[h]
\caption{Covariance-correct $\chi^2$ of the bright-masked
$\hat\xi(u)$ ($u = 25$--$50\,\mjb$, 11 thresholds, Hartlap factor for
400 replicates) against the three count models under the four
forward-model conventions; 400 maps per model.}
\label{tab:forwardconventions}
\centering
\setlength{\tabcolsep}{4pt}
\begin{tabular}{l l r r r}
\hline\hline
noise & bright end & Schechter & DPL & SPL \\
\hline
beam   & truncated        & 27.7 & 108.0 & 315.4 \\
beam   & observed, masked & 56.1 & 106.8 & 385.9 \\
filter & truncated        & 23.0 & 179.7 & 459.1 \\
filter & observed, masked (fiducial) & 30.1 & 138.4 & 472.4 \\
\hline
\end{tabular}
\tablefoot{``Beam'': sky and noise through the same Gaussian beam;
``filter'': the matched-filter forward model. The first row is the
Gaussian-beam convention evaluated in the present pipeline. All
rows use the same source realizations.}
\end{table}

\section{Statistical methodology for correlated thresholds}
\label{app:stats}

\subsection{Threshold covariance}

The values of $\hat\xi(u_i)$ at different thresholds are fitted to
nested exceedance subsets of the same peaks and are therefore
strongly correlated: effect (C) of Appendix~\ref{app:beam}, the one
that declustering cannot remove. The correlation has two
components with different origins. The \emph{nesting} component is
present even for perfectly independent peaks, because the
exceedances above a higher threshold are a subset of those above a
lower one; it is predicted analytically by
$\rho \approx \sqrt{N_{\rm exc}(u_2)/N_{\rm exc}(u_1)}$ (for
$u_2 > u_1$), and we audit every bootstrap covariance against this
floor: a matrix falling below it indicates too few
replicates. The \emph{shared-realization} component arises because
all thresholds are measured on the same finite sky, and it appears
as a long-range correlation floor that i.i.d.\ peak resampling
would destroy, one of two reasons the bootstrap must resample
structures, not peaks (below).

The effective number of independent thresholds is defined from the
eigenvalues $\{\lambda_i\}$ of the threshold correlation matrix
$\mathsf{R}$ as the participation ratio
\begin{equation}
n_{\rm eff} \;=\; \frac{\bigl(\sum_i \lambda_i\bigr)^{2}}
                       {\sum_i \lambda_i^{2}}
            \;=\; \frac{N_u^{2}}{\sum_{ij} \mathsf{R}_{ij}^{2}} ,
\label{eq:neff}
\end{equation}
the second form following from $\sum_i\lambda_i = N_u$ for a
correlation matrix of $N_u$ thresholds. It equals $N_u$ for
uncorrelated thresholds and $1$ for perfectly correlated ones.
Measured on our data, adjacent thresholds correlate at
$\rho \approx 0.9$ (\textit{Herschel}) and $0.7$--$0.9$
(\textit{Planck}), giving $n_{\rm eff} \approx 2.2$ of 21
(\textit{Herschel}) and $2.8$ of 10 (\textit{Planck}); the
trace-based alternative $N_u^2/\sum_{ij}\mathsf{R}_{ij}$ gives $1.6$
and $2.2$, and we quote the two as a range. A corollary worth
internalizing: a many-point $\hat\xi(u)$ curve from one map is
close to one measurement, not many, and an unlucky
realization displaces the whole curve coherently, producing
exactly the visually ``systematic-looking'' offsets that invite
physical over-interpretation. We quote $n_{\rm eff}$ alongside
every multi-threshold result in this paper.

Model comparison uses
$\chi^2 = \alpha_{\rm H}\,\mathbf{d}^{\mathsf{T}}
\hat{\mathsf{C}}^{-1}\mathbf{d}$, with $\mathbf{d}$ the vector of
measured-minus-model $\hat\xi(u_i)$ over the $p$ thresholds retained,
$\hat{\mathsf{C}}$ its bootstrap covariance, and
$\alpha_{\rm H} = (N - p - 2)/(N - 1)$ the \citet{Hartlap2007}
debiasing factor for a covariance estimated from $N$ bootstrap
replicates. Because the replicates resample a finite number of
cutouts, the independent information in the covariance is bounded by
the cutout count, not the replicate count; with $N = 103$ cutouts and
$p = 11$ thresholds the factor is $0.88$ against $0.97$ for $N = 400$,
and Sect.~\ref{sec:herschel:unlensed} quotes both. The factor is exact
for a Wishart-distributed sample covariance from independent Gaussian
realizations; bootstrap replicates of cutouts are neither, so the two
conventions bracket the correction rather than fix it. At \textit{Planck}
(1600 cutouts) the distinction is immaterial. Single combined numbers are never formed by
inverse-variance averaging across thresholds; instead the identical
weighted mean is applied inside every bootstrap replicate
and the replicate scatter is quoted, which propagates the full
covariance without ever inverting it. The distinction is not
pedantry: naive quadrature is wrong here by up to a factor $2.3$ and,
importantly, not uniformly in one direction. The leading
principal component of the $\hat\xi(u)$ error budget, carrying
55--72\% of the variance depending on the arm, is a coherent
vertical shift of the whole curve; model pairs that differ mainly
by such a shift have their naive separations deflated by the
correct treatment, while pairs whose curves cross (differing in a
direction orthogonal to the noisy mode) have them sharpened.
Because the full-covariance statistic draws part of its power from
low-variance modes whose Gaussianity the bootstrap cannot
guarantee, we quote it together with $n_{\rm eff}$ and treat a
ranking as established only where the per-threshold comparison shows
it as well (Sect.~\ref{sec:herschel:unlensed}).

\subsection{Null Monte Carlo and estimator calibration}

Every quoted significance is calibrated against an end-to-end null:
simulated skies with the measured Gaussian budget and truncated
Poisson counts, processed through the identical cutout, filter,
decluster, threshold, and fit pipeline, with matched peak counts.
The null serves two purposes. First, it replaces asymptotic GPD
sampling theory in the small-sample regime. For $\xi > -1/2$ the
maximum-likelihood estimator is asymptotically normal with
\begin{equation}
n\,\mathrm{Cov}(\hat\xi,\hat\sigma) \;\longrightarrow\;
\begin{pmatrix}
(1+\xi)^{2} & -\sigma(1+\xi)\\
-\sigma(1+\xi) & 2\sigma^{2}(1+\xi)
\end{pmatrix},
\label{eq:gpdvar}
\end{equation}
so that $\mathrm{Var}(\hat\xi) \simeq (1+\xi)^{2}/n$
for $n$ exceedances \citep{Smith1984}; the $(1+\xi)$ factor is the
one carried through the per-cluster significances of
Table~\ref{tab:dxisuppression}. Equation~(\ref{eq:gpdvar}) is an
asymptotic statement, and at the exceedance counts of the
\textit{Planck} cluster apertures it is optimistic: there
$\hat\xi$ additionally carries a positive bias of order $1/n$ and a
strongly right-skewed sampling distribution
\citep{HoskingWallis1987, Smith1985}, both of which the null
propagates automatically where Eq.~(\ref{eq:gpdvar}) cannot. Second,
it defines the reference against
which ``excess tail structure'' is meaningful, making explicit what
the null does and does not contain (the \textit{Planck} null
contains neither the bright extension nor clustered/cirrus
non-Gaussianity, which is precisely how the origin of the measured
excursion was diagnosed in Sect.~\ref{sec:planck:unlensed}).

\begin{figure*}
\centering
\includegraphics[width=\hsize]{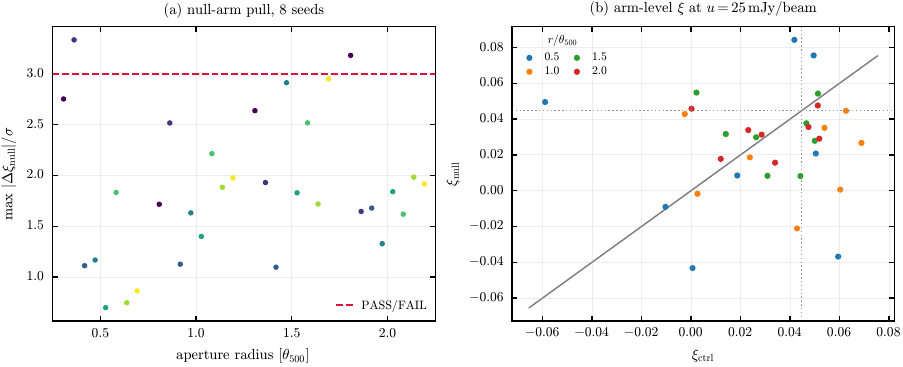}
\caption{Calibration of the null-test statistic on a seed ensemble
(eight independent realizations of the \textit{Herschel}-arm lensed
Monte Carlo). \emph{(a)} The distribution of the maximum null pull
over 9 thresholds $\times$ 4 correlated apertures: each point is
the maximum $|\Delta\xi_{\rm null}|/\sigma$ over the nine thresholds
at one aperture (abscissa) for one seed, color-coded by seed (the
eight seeds run from dark purple to yellow, the points of one seed
being offset horizontally for legibility); the red dashed line is
the pass/fail threshold at a pull of 3. The observed
per-seed maxima over all $9 \times 4$ tests (mean 2.55, two of eight exceeding 3) are the expectation for
a maximum over many correlated tests, and the aperture attaining the
maximum differs between seeds. \emph{(b)} Arm-level $\xi$ of the two arms of the
control-versus-control null test at the single threshold
$u = 25\,\mjb$ (null arm against control arm), one point per seed
and aperture (colors \emph{here} encode aperture, as in the legend); the gray diagonal marks
equality and the dotted lines mark the independent unlensed
reference value in both coordinates. The single-seed
excursion stands out as an ordinary unlucky control-arm draw.}
\label{fig:seedensemble}
\end{figure*}

One methodological refinement emerged from this work and is worth
recording for reuse (Fig.~\ref{fig:seedensemble}). A null test
defined as the maximum pull over many correlated thresholds
and apertures does not have the intuitive reference distribution:
for 9 thresholds $\times$ 4 strongly correlated apertures, a
maximum near 2.5--3$\sigma$ is the expectation, not a
warning sign. When a single fiducial realization of the \textit{Herschel}
simulation null showed a maximum pull of 3.18, we calibrated the
statistic on an ensemble of eight independent seeds rather than
re-rolling once: the ensemble gave mean maximum pull 2.55 with two
of eight seeds exceeding 3, the identity of the ``failing''
aperture moving between seeds (the signature of a maximum over
correlated tests, not of a defect attached to any geometry), and
the arm-level diagnosis identified the single outlying arm as an
ordinary $1$--$2\sigma$ draw of the control ensemble. Null tests of
this kind should be reported as calibrated distributions, not as
single pass/fail verdicts; we adopt this convention throughout.

\subsection{Bootstrap design}

Uncertainties are bootstrap estimates over the largest exchangeable
units: cutouts for the unlensed analyses (peaks within a cutout
share a baseline fit, a noise level, and a cirrus realization;
i.i.d.\ peak resampling would both underestimate the shared
variance and sever the long-range covariance floor above), and
cluster/control pairs for the differential analyses, so that
sky systematics cancel within every replicate rather than on
average. Full replicate matrices are retained for every statistic,
so that any derived quantity, including the threshold covariance
itself, can be recomputed without re-running a pipeline. Two
guard rules complete the design. Points near the minimum-exceedance
floor are treated as conditionally biased, not merely noisy: the
survival cut selects realizations that happened to draw a heavy
tail, so surviving fits near the floor are biased high; they are
excluded from combined statistics wherever a science window is
defined (Sects.~\ref{sec:planck:lensed} and
\ref{sec:herschel:unlensed}), and enter the inverse-variance
combinations of Sect.~\ref{sec:herschel:lensed} only with the small
weight their errors give them (Fig.~\ref{fig:herscheldxi}). And no undersampled
cluster-aperture $\hat\xi(u)$ curve is ever presented on a fine
threshold grid (the 21-point unlensed \textit{Herschel} ladder of
Fig.~\ref{fig:herschelxi} is well sampled):
with adjacent-threshold correlations of 0.9, a handful of
thresholds carries the same information without the illusion of
independent structure.

One implementation detail interacts with the first of those guard
rules. For speed, the simulation scans thinned any pooled peak
sample above a fixed size by uniform random sub-sampling. Uniform
thinning is unbiased in principle (the retained sample is an
i.i.d.\ draw from the same marginal), but it becomes a bias
wherever it pushes a threshold down toward the minimum-exceedance
floor, precisely because of the conditioning effect above. In one
diagnostic the thinned arm returned $\Delta\xi = +0.32$ against an
analytic $-0.02$, with the wrong sign, purely from this mechanism.
All scans reported in this paper are run unthinned; with the cap in
place, the shape estimates move by $\le 0.03$ across the
model-comparison windows while the error bars inflate by $15$--$40$\%
at \textit{Herschel} and by about a factor two at CCAT, so the cap's
effect is almost entirely on the uncertainties.

\section{Numerical tests of source clustering}
\label{app:clustering}

This appendix collects the quantitative results behind
Sect.~\ref{sec:sims:clustering}: the amplitude-free core--tail
contrast, the calibration of the clustered simulations, the direct
measurement of the clustering shift of $\hat\xi(u)$, and the
inflation of exceedance-count scatter.

\subsection{The core--tail contrast}

Writing the clustering corrections to the second cumulant of the
$\PD$ (Sect.~\ref{sec:sims:clustering}) in ratio to the Poisson
terms, the core and tail corrections scale as
$\Rcore \propto \bar w\,\Nbeam\,\langle S\rangle^2/\langle
S^2\rangle$ and $\Rtail \sim \bar w\,\Nbeam({>}u)$, with $\bar w$
the beam-scale mean angular correlation. Their ratio is independent
of $\bar w$ entirely,
\begin{equation}
\frac{\Rcore}{\Rtail} \;=\; \frac{2\,q_1^2}{q_2\,N({>}u)} ,
\label{eq:coretail}
\end{equation}
with $q_k = \int S^k n_0\,\mathrm{d}S$. This is the useful form, because
it survives any uncertainty in how strongly the sources actually
cluster. Table~\ref{tab:coretail} evaluates it on the adopted Schechter counts at the SPIRE
350\,\mum\ beam, over the \textit{Herschel} fitting window.

\begin{table}[h]
\caption{The amplitude-free core--tail contrast
(Eq.~\ref{eq:coretail}) over the \textit{Herschel} fitting window.}
\label{tab:coretail}
\centering
\begin{tabular}{r r r r}
\hline\hline
$u$ [$\mjb$] & $N({>}u)$ [deg$^{-2}$] & $\Nbeam({>}u)$ & $\Rcore/\Rtail$ \\
\hline
25 & 537 & $3.0\times10^{-2}$ & 123 \\
40 & 122 & $6.7\times10^{-3}$ & 541 \\
50 & 51  & $2.8\times10^{-3}$ & 1\,290 \\
75 & 7.2 & $4.0\times10^{-4}$ & 9\,170 \\
\hline
\end{tabular}
\tablefoot{Columns: threshold, identified with the source flux
$S_u = u$ of the tail; integral source density above $u$; sources per
beam above $u$; and the core-to-tail clustering-response ratio of
Eq.~(\ref{eq:coretail}). Two to four orders of magnitude of protection for the
tail relative to the core, with no clustering amplitude assumed
anywhere; $q_1$ and $q_2$ are integrated from the paper's
$S_{\min} = 1$\,mJy (extrapolating to 1\,$\mu$Jy would raise the
ratios sevenfold through the faint-end $q_1$ that
Appendix~\ref{app:counts:monopole} shows to be over-predicted).}
\end{table}

\subsection{Calibration of the clustered simulations}

For direct simulation an absolute clustering amplitude is needed.
We adopt a power-law angular spectrum
$C_\ell^{\rm clust} = q_2\,(\ell/\ell_{\rm eq})^{-1.2}$ with
$\ell_{\rm eq} = 2000$: the slope matches both measured CIB spectra
\citep{Addison2013} and the Fourier transform of the
$w(\theta) \propto \theta^{-0.8}$ correlation function adopted by
\citet{Wang2026} for SPIRE data (the slope fixed from optical and
radio surveys, the scale $\theta_0$ fitted), and the amplitude (clustering
power equal to shot power at $\ell = 2000$) introduces no free
parameter, is self-consistent for any count model, and reproduces
the WebSky 857\,GHz CIB spectrum \citep{Stein2020} across two
decades of $\ell$. With this normalization the clustering
correction to the beam-scale confusion variance has the closed form
$\Rcore = \Gamma(0.4)\,(\ell_{\rm eq}\sigma_{\rm b})^{1.2}$
(Gaussian beam of width $\sigma_{\rm b}$), which evaluates to 0.111
at the $20''$ simulation beam, 0.146 at SPIRE's $25.15''$, and
2.86 at \textit{Planck}'s $5'$: clustering adds $\sim 15\%$ to
the confusion variance at \textit{Herschel} resolution and nearly
quadruples it at \textit{Planck}'s, consistent with the measured
non-Poisson core of Sect.~\ref{sec:planck:data}. Clustered skies
are realized as lognormal-modulated Cox processes with this input
spectrum; the realized spectra reproduce the input at the percent
level, and the unclustered mode reproduces the fiducial pipeline
bit for bit.

\subsection{The clustering shift of \texorpdfstring{$\hat\xi(u)$}{xi(u)}}

Table~\ref{tab:clusteringxi} gives the direct comparison of
$\hat\xi(u)$ measured on clustered and Poisson skies of identical
counts (96 maps per arm, SPIRE-beam configuration, declustered
peaks, paired seeds). The shift is real, negative, and confined to
the core region: $-0.02$ at $u \approx \sigc$, decaying
monotonically through zero by $u \approx 3\,\sigc$, and
unresolvable above, exactly the pattern predicted by the
single-big-jump argument, since the pair term adds exceedances
preferentially at moderate amplitudes. The nested thresholds of the
table are strongly correlated (Appendix~\ref{app:stats}) and constitute
roughly two independent numbers, not eight. The protection is
positional, not absolute: a $-0.02$ shift would be comparable to
model separations if it sat inside the fitting window; it sits
below it, and in $\dxiu$ it is common-mode besides.

\begin{table}[h]
\caption{Measured clustering shift of the declustered-peak
$\hat\xi(u)$: lognormal-clustered versus Poisson skies of identical
counts (96 maps per arm; SPIRE-beam configuration,
$\sigc = 8.06\,\mjb$ for the clustering-test convention of
$S_{\min} = 0.1$\,mJy, within 1\% of the $7.98\,\mjb$ of
Table~\ref{tab:ladder}).}
\label{tab:clusteringxi}
\centering
\setlength{\tabcolsep}{4pt}
\begin{tabular}{r r r r r r}
\hline\hline
$u$ & $u/\sigc$ & $\hat\xi$ Poisson & $\hat\xi$ clustered &
$\Delta\xi$ & signif.\ [$\sigma$]\\
\hline
 5 & 0.6 & $-0.062$ & $-0.080$ & $-0.019$ & 11.3 \\
10 & 1.2 & $-0.013$ & $-0.033$ & $-0.020$ &  9.7 \\
15 & 1.9 & $+0.014$ & $-0.003$ & $-0.017$ &  6.6 \\
20 & 2.5 & $+0.027$ & $+0.014$ & $-0.012$ &  3.8 \\
25 & 3.1 & $+0.022$ & $+0.020$ & $-0.002$ &  0.5 \\
30 & 3.7 & $+0.014$ & $+0.018$ & $+0.003$ &  0.7 \\
40 & 5.0 & $-0.025$ & $-0.013$ & $+0.012$ &  1.5 \\
50 & 6.2 & $-0.063$ & $-0.052$ & $+0.011$ &  0.9 \\
\hline
\end{tabular}
\tablefoot{Thresholds $u$ in $\mjb$. ``signif.''\ is the per-threshold significance of the
clustered-minus-Poisson difference; thresholds are nested and
strongly correlated ($\sim$2 independent numbers). The shift is
confined below $u \approx 3\,\sigc$, outside the fitting windows
used in this paper.}
\end{table}

\subsection{Where clustering does bite: the exceedance-count
scatter}

The realization-to-realization scatter of exceedance counts is
measured by the Fano factor $F = \mathrm{Var}(N)/\mathrm{E}[N]$ in
cells, compared between clustered and Poisson skies
(Table~\ref{tab:clusteringfano}). Two technical points govern the
reading. First, the declustered Poisson process is sub-%
Poissonian ($F < 1$) because declustering imposes a minimum
separation (a hard-core process), so the meaningful comparison
is clustered-versus-Poisson at fixed threshold, never $F$ against
unity. Second, the source-density modulation $\delta$ that drives
the over-dispersion depends on the first moment $q_1$ of the
counts, which two of our fitted models over-predict against the measured
CIB monopole (Appendix~\ref{app:counts:monopole}); anchoring the
modulation to the measured monopole (the physically correct
choice) strengthens it by 2.7$\times$ in rms over the model's own
$q_1$. With that anchoring, exceedance-count error bars derived
from Poisson simulations are too small by a factor
$\approx 2$ in rms at the floor of the fitting window, falling
to $\sim 1.5$ at its top. The measured
excess exceeds the naive counts-in-cells prediction by an order of
magnitude, because a local overdensity not only adds sources but
lifts the local confusion pedestal, pushing additional peaks over a
fixed threshold: the count responds to $\delta$ with an
effective index above unity. The factor-4 variance inflation is a
conservative proxy: the count
over-dispersion does not propagate one-to-one into the fitted shape,
whose realization-to-realization scatter in the same clustered
simulations is inflated by only $1.0$--$1.15$ over $u = 15$--$50\,\mjb$,
so the count-based factor is an upper bound on the shape-error
inflation and the requirements of Table~\ref{tab:n3sigma} are
conservative in this respect.

\begin{table}[h]
\caption{Over-dispersion of declustered exceedance counts: Fano
factors in 21.3$'$ cells (96 maps), Poisson versus clustered skies
with the source-density modulation anchored to the measured CIB
monopole.}
\label{tab:clusteringfano}
\centering
\setlength{\tabcolsep}{4pt}
\begin{tabular}{r r r r r}
\hline\hline
$u$ & $u/\sigc$ & $F$ Poisson & $F$ clustered &
error bars small by \\
\hline
 5 & 0.6 & 0.45 & 2.04 & $2.12\times$ \\
10 & 1.2 & 0.61 & 3.10 & $2.25\times$ \\
15 & 1.9 & 0.75 & 3.88 & $2.28\times$ \\
20 & 2.5 & 0.84 & 4.22 & $2.24\times$ \\
25 & 3.1 & 0.90 & 4.07 & $2.13\times$ \\
30 & 3.7 & 0.89 & 3.71 & $2.05\times$ \\
40 & 5.0 & 0.98 & 2.85 & $1.71\times$ \\
50 & 6.2 & 1.02 & 2.13 & $1.45\times$ \\
\hline
\end{tabular}
\tablefoot{Thresholds $u$ in $\mjb$. $F < 1$ for the Poisson sky is real: declustering
imposes a minimum peak separation (hard-core process). The final
column is the rms inflation applicable to any
Poisson-simulation-derived uncertainty.}
\end{table}

\subsection{Standing caveats}

Three limitations bound the realm of validity of these tests. The
implementation assumes luminosity-independent clustering (all
sources share one modulation field), whereas the observed
flux dependence of CIB halo occupation \citep{Viero2013} implies
brighter sources cluster more strongly; the two-to-four orders of
magnitude of margin in Table~\ref{tab:coretail} absorb a large
factor of such refinement, but this is the limiting assumption, not
a detail. The numerical tests were run at a $25''$-class beam;
at \textit{Planck}'s $5'$ beam the core correction is an order of
magnitude larger ($\Rcore = 2.9$) and, while the tail argument is
beam-independent, the tests have not been repeated there, one
reason the \textit{Planck} chapter never leans on Poisson
simulations for its conclusions. And the tail correction depends on
the pair correlation at sub-beam separations, where the power-law
extrapolation is least trustworthy; this affects the absolute
$\Rtail$ but not the amplitude-free ratio of
Eq.~(\ref{eq:coretail}), which is why that form is the one quoted.
\end{appendix}

\end{document}